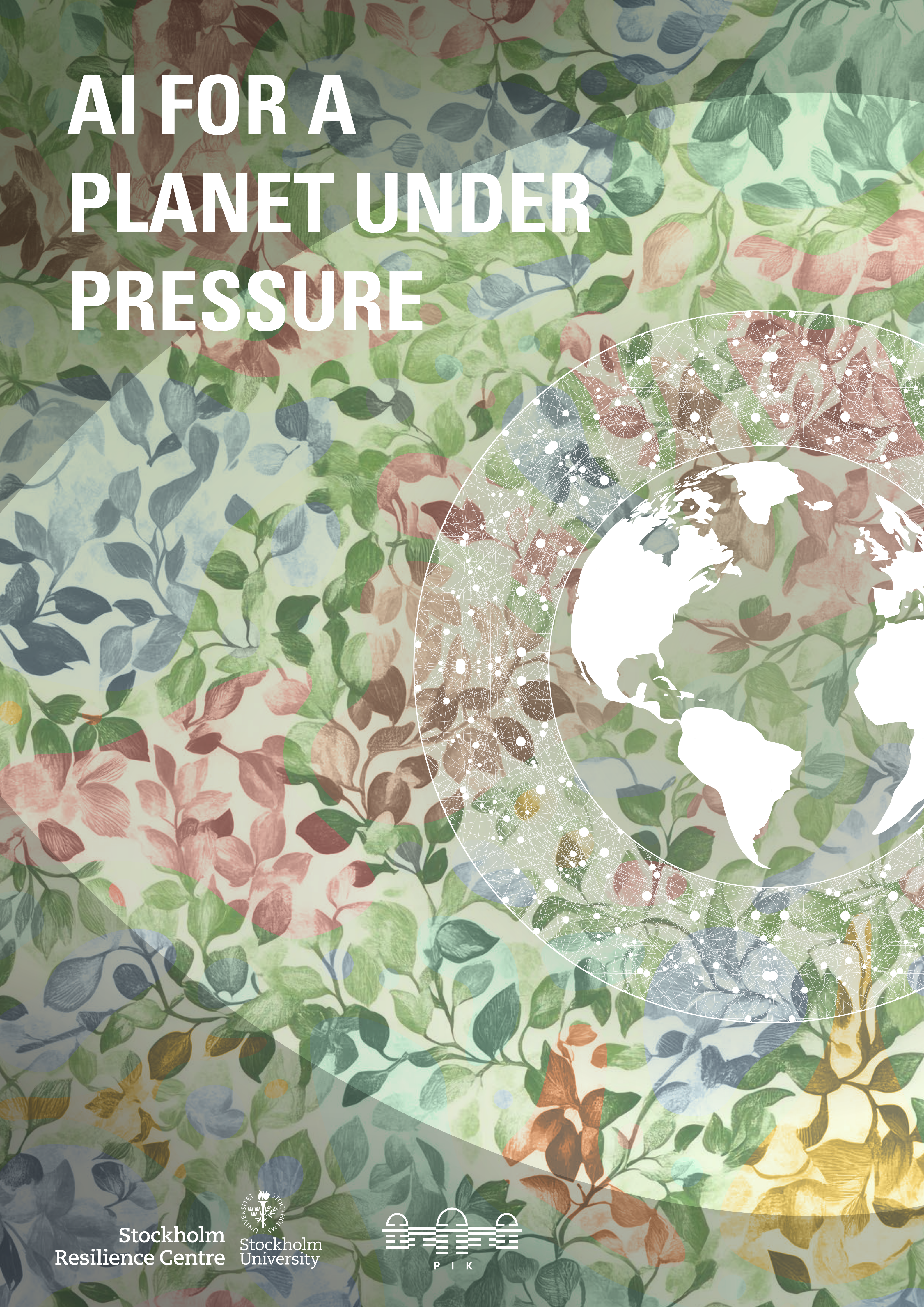
AI FOR A PLANET UNDER PRESSURE
Stockholm Resilience Centre
Stockholm University
PIK

**Title:** AI for a Planet Under Pressure

**Authors:** Victor Galaz,[1,2*] Maria Schewenius,[1,2,6*] Jonathan F. Donges,[1,3,5] Ingo Fetzer,[1,2,7] Erik Zhivkoplias,[1] Wolfram Barfuss,[8,9,10] Louis Delannoy,[1,11,12] Lan Wang-Erlandsson,[1,3] Maximilian Gelbrecht,[3,13] Jobst Heitzig,[3] Jonas Hentati-Sundberg,[14] Christopher Kennedy,[15] Nielja Knecht,[1] Romi Lotcheris,[1] Miguel Mahecha,[16] Andrew Merrie,[1,24] David Montero,[16] Timon McPhearson,[1,15] Ahmed Mustafa,[15] Magnus Nyström,[1] Drew Purves,[4] Juan C. Rocha,[1] Masahiro Ryo,[17,18] Claudia van der Salm,[4] Samuel T. Segun,[19] Anna B. Stephenson,[20] Elizabeth Tellman,[21] Felipe Tobar,[22] Alice Vadrot[23]

*Editors

**Affiliations:**
1. Stockholm Resilience Centre, Stockholm University, Sweden
2. Beijer Institute of Ecological Economics, Royal Swedish Academy of Sciences, Sweden
3. Potsdam Institute for Climate Impact Research (PIK), Member of the Leibniz Association, Germany
4. Google DeepMind, United Kingdom
5. Max Planck Institute of Geoanthropology, Germany
6. Faculty of Engineering and Sustainable Development, University of Gävle, Sweden
7. Bolin Centre for Climate Research, Stockholm, Sweden
8. Transdisciplinary Research Area Sustainable Futures, University of Bonn, Germany
9. Center for Development Research, University of Bonn, Germany
10. Institute for Food and Resource Economics, University of Bonn, Germany
11. Global Economic Dynamics and the Biosphere, Royal Swedish Academy of Sciences, Sweden
12. Swedish Centre for Impacts of Climate Extremes (climes), Uppsala University, Sweden
13. School of Engineering and Design, Technical University of Munich, Germany
14. Swedish University of Agricultural Sciences, Sweden
15. Urban Systems Lab, New York University, USA
16. Institute for Earth System Science and Remote Sensing, Leipzig University, Germany
17. Brandenburg University of Technology Cottbus-Senftenberg, Germany
18. Leibniz Centre for Agricultural Landscape Research (ZALF), Germany
19. Global Center on AI Governance, Canada
20. High Meadows Environmental Institute, Princeton University, USA
21. Nelson Institute for Environmental Studies, University of Wisconsin-Madison, USA
22. Department of Mathematics, Imperial College London, United Kingdom
23. Department of Political Sciences, University of Vienna, Austria
24. Planethon, Sweden

**Suggested citation for the full report:**
Galaz, V. and M. Schewenius (eds, 2025). *AI for a Planet Under Pressure.* Stockholm Resilience Centre, Potsdam Institute for Climate Impact Research. Stockholm. Report. Online: http://arxiv.org/abs/2510.24373.

**Suggested citation for individual chapters in report (e.g.):**
Wang-Erlandsson, L., N. Knecht, R. Lotcheris, I. Fetzer. “Securing Freshwater for All” in Galaz, V. and M. Schewenius (eds, 2025). *AI for a Planet Under Pressure.* Stockholm Resilience Centre, Potsdam Institute for Climate Impact Research. Stockholm. Report.

**ISBN:** 978-91-89107-61-8
**e-book ISBN:** 978-91-89107-62-5

**Graphics and layout:** Jerker Lokrantz/Azote

November 2025

---

The production of this report was supported by additional funding from the Beijer Institute of Ecological Economics (Royal Swedish Academy of Sciences), and Google.org through ClimateIQ at the Urban Systems Lab at New York University (USA).

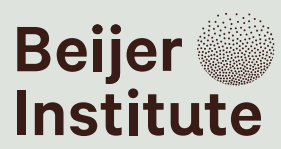

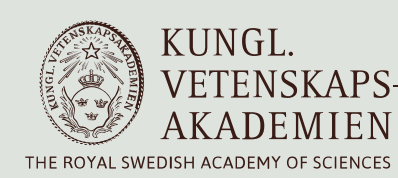

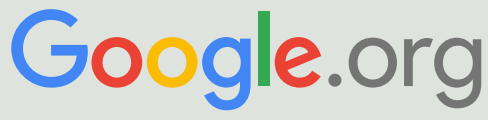

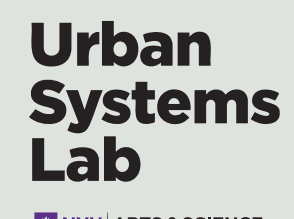

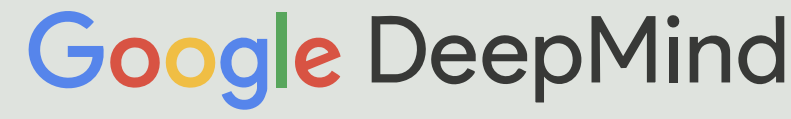

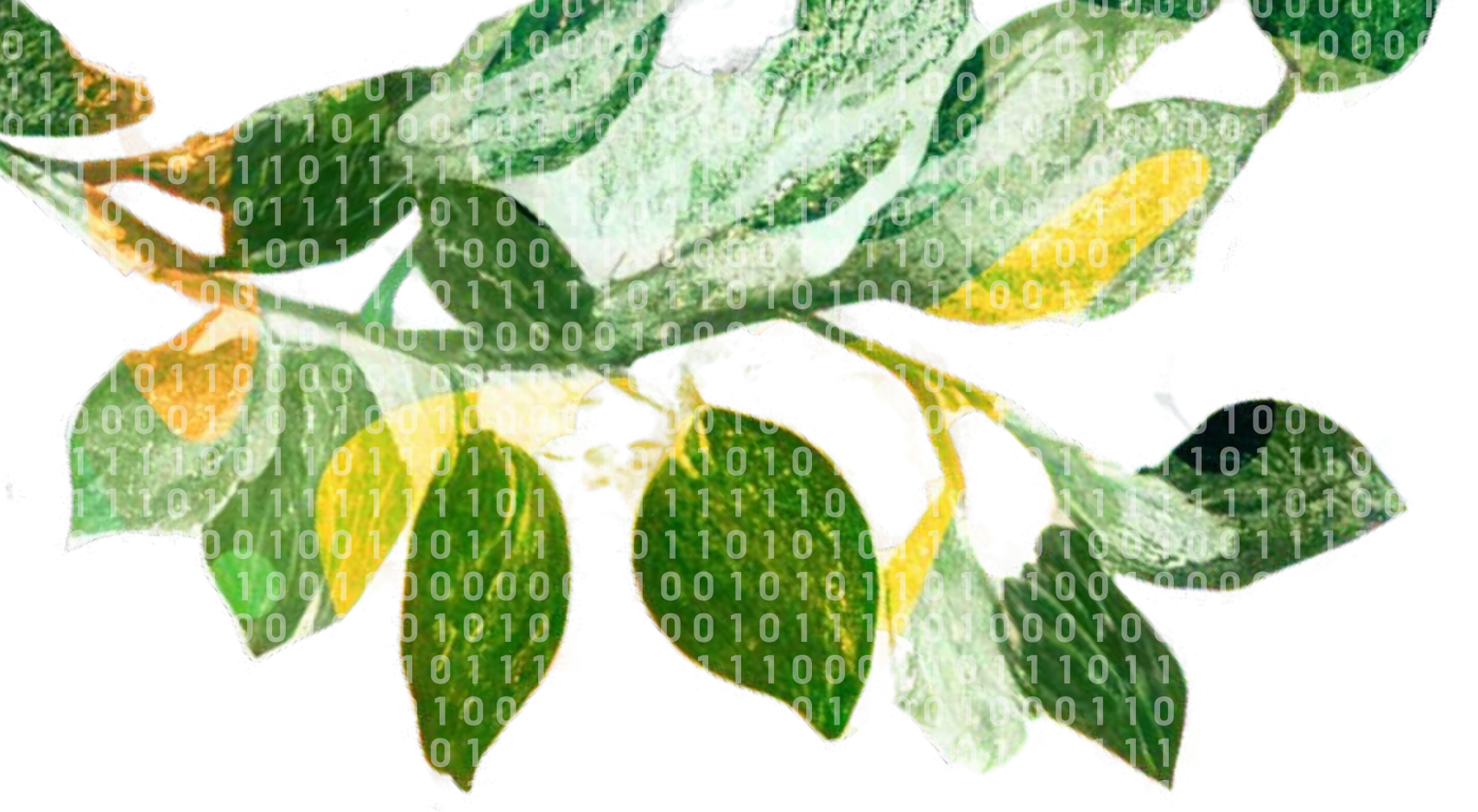

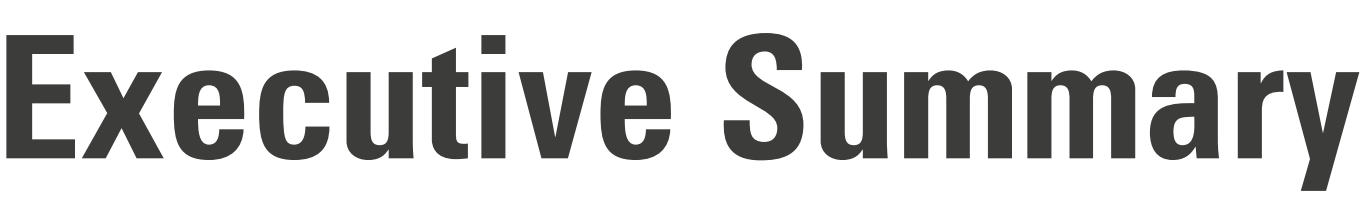

# Executive Summary

Artificial intelligence (AI) is already driving scientific breakthroughs in a variety of research fields, ranging from the life sciences to mathematics. This raises a critical question: can AI be applied both responsibly and effectively to address complex and interconnected sustainability challenges? These challenges include climate change, biodiversity loss, ocean acidification, and other transformations of our living planet. Each of these poses serious risks to societal stability, human health and well-being, and the ability of present and future generations to thrive within planetary boundaries. Here, science plays a fundamental role in two ways. **First**, by helping accelerate changes and innovations that move us closer to a just and safe future for all. **Second**, by making sure that AI is developed and used in ways that neither exacerbates inequalities, nor increases planetary pressures.

This report explores the potential and limitations of using AI as a research method in eight issue areas: *Preparing for a Future of Interconnected Shocks, Understanding a Complex Earth System, Stewarding Our Blue Planet, Securing Freshwater for All, Enhancing Nature's Contributions to People, Prospering on an Urban Planet, Improving Sustainability Science Communication,* and *Collective Decisions for a Planet Under Pressure*. Together, these areas reflect a range of complex research challenges where people, nature, and climate interplay. By taking such an integrated approach, our work stands out within the growing landscape of "AI for Science," and brings together "AI for Climate" and "AI for Nature" initiatives.

Our results build on iterated expert dialogues and assessments, a systematic AI-supported literature overview including over 8,500 academic publications, and expert deep-dives into each specific issue area. In conclusion, we show that: (**1**) AI offers vast potential to accelerate progress across the sustainability sciences. (**2**) AI can sharpen our decision-making and clarify complex environmental challenges for researchers and the public alike. (**3**) However, realizing this promise requires careful navigation of the risks, including AI's own environmental footprint, inherent biases, and the challenge of unequal access. (**4**) Despite these hurdles, responsible and ethical applications of AI in sustainability research are not just a possibility—they are an urgent necessity. (**5**) Pioneering these uses can unlock the breakthroughs we need to build a more sustainable future.

Advancing from the potential of AI to its responsible application in research that benefits both people and the planet will require a balance between urgency and innovation, and a commitment to ethical and inclusive practices. Our concluding recommendations—addressed to researchers, policymakers, funding agencies, and philanthropic organizations—outline practical steps for achieving this balance. Leveraging powerful compute (i.e., computational power) and new methodologies, advancements in AI enable the discovery of patterns and relationships that were previously too complex, or too interdisciplinary, to tackle. As experience has shown, AI has the potential to open up new, exciting areas of inquiry for sustainability research. It's an opportunity the world cannot afford to miss. It is now up to sustainability researchers, companies, technology entrepreneurs, and decision-makers to carefully shape AI development in the service of a just and safe future.

**Use of AI in Writing This Report**

L.W-E. used Grammarly for editing the Freshwater chapter. N.K. has used ChatGPT to rephrase and shorten parts of the draft version of the Freshwater chapter. E.Z. used Gemini 2.5 Flash for grammar and syntax checks of all writing contributions to the report chapters; GitHub Copilot assisted with script programming to improve the visual appeal of the plots in the literature review and conclusion chapters; and used DeepSeek-R1-7b for analysis of the collected literature, as described in the corresponding chapter. A.V. used Gemini for search of relevant literature (together with Scopus) and editing. L. D. used ChatGPT (4o) for spell checking and rewording individual sentences, upon which he reviewed and edited the content. W.B. used Grammarly for language editing. M.R. used ChatGPT to refine the academic language. M.S. used Copilot in Microsoft 365 to revise some of the sentences in the full report to improve readability. C.S. used Gemini to provide input to the summary of the full report, and worked with V.G. and other chapter authors with the report summary, conclusions, and recommendations.

**Funding**

Workshop, travel costs, research assistance time, and design and lay-out of this report, has received funding from Google DeepMind. V.G. would like to acknowledge funding from the Marianne and Marcus Wallenberg Foundation, Google.org, and the Beijer Institute of Ecological Economics (Royal Swedish Academy of Sciences). M.S. acknowledges financial support from Stockholm Resilience Centre, the Beijer Institute of Ecological Economics, and Google.org. L.W-E. acknowledges funding from Formas (2022-02089, 2023-00321), and the IKEA Foundation. J.R. acknowledges support from Vetenskapsrådet grant 2022-04122. I.F. acknowledges funding from the IKEA Foundation. A.V. acknowledges funding from the European Research Council (ERC), Grant agreement No 101124903 – TwinPolitics – ERC-2024-CoG. M.M. acknowledges funding from ML4Earth by DLR. R.L. acknowledges funding from Formas (2022-02089). L.D. thanks the Erling-Persson Family Foundation for funding. M.G. acknowledges founding from the Volkswagen Foundation under the Freigeist program. W.B. acknowledges financial support received from the Cooperative AI Foundation. T.M. acknowledges financial support from Google.org. J.H-S. acknowledges funding from Marcus and Marianne Wallenberg foundation 2018-0093, and Vetenskapsrådet 2021- 03892. M.R. gratefully acknowledges support from the project “KI und Citizen Science gestütztes Monitoring von zertifizierten Biodiversitätsprojekten (KICS-Zert)” (16LW0441), funded by BMFTR, Germany. A.S. received support from William H. Miller III, the Princeton University Dean for Research, the High Meadows Environmental Institute, and the Army Research Office under grant number W911NF2410126.

**Competing Interests**

D.P. and C.S. are both employed by Google DeepMind. F.T. has received funding from Google Research between 2020 and 2024 through the programmes: Latin America Research Awards (LARA), AI for Social Good, and the Award for Inclusion Research (AIR). T.M., A.M., V.G. and M.S. receive financial support from Google.org. V.G. has also engaged as consultant for Klarna and Milkywire for the program “AI for Climate Resilience”. The workshop and travel costs, research assistance time, and design and lay-out of this report, has received funding from Google DeepMind.

**Thank You**

Emelie Elfvengren (Stockholm Resilience Centre, Stockholm University), has provided helpful research assistance in the compilation of this report. Fredrik Moberg (Stockholm Resilience Centre, Stockholm University, and Albaeco) has proofread and commented on an early version of the full report. Elinor Kruse (Google DeepMind) contributed to reviewing the Taxonomy chapter and the final version of the report. Marcus Lundstedt (Stockholm Resilience Centre, Stockholm University) has provided helpful advice in discussions about the report title, key messages, and Executive Summary. Professional proofreading has been done by The Content Creation Company (CCC), UK.

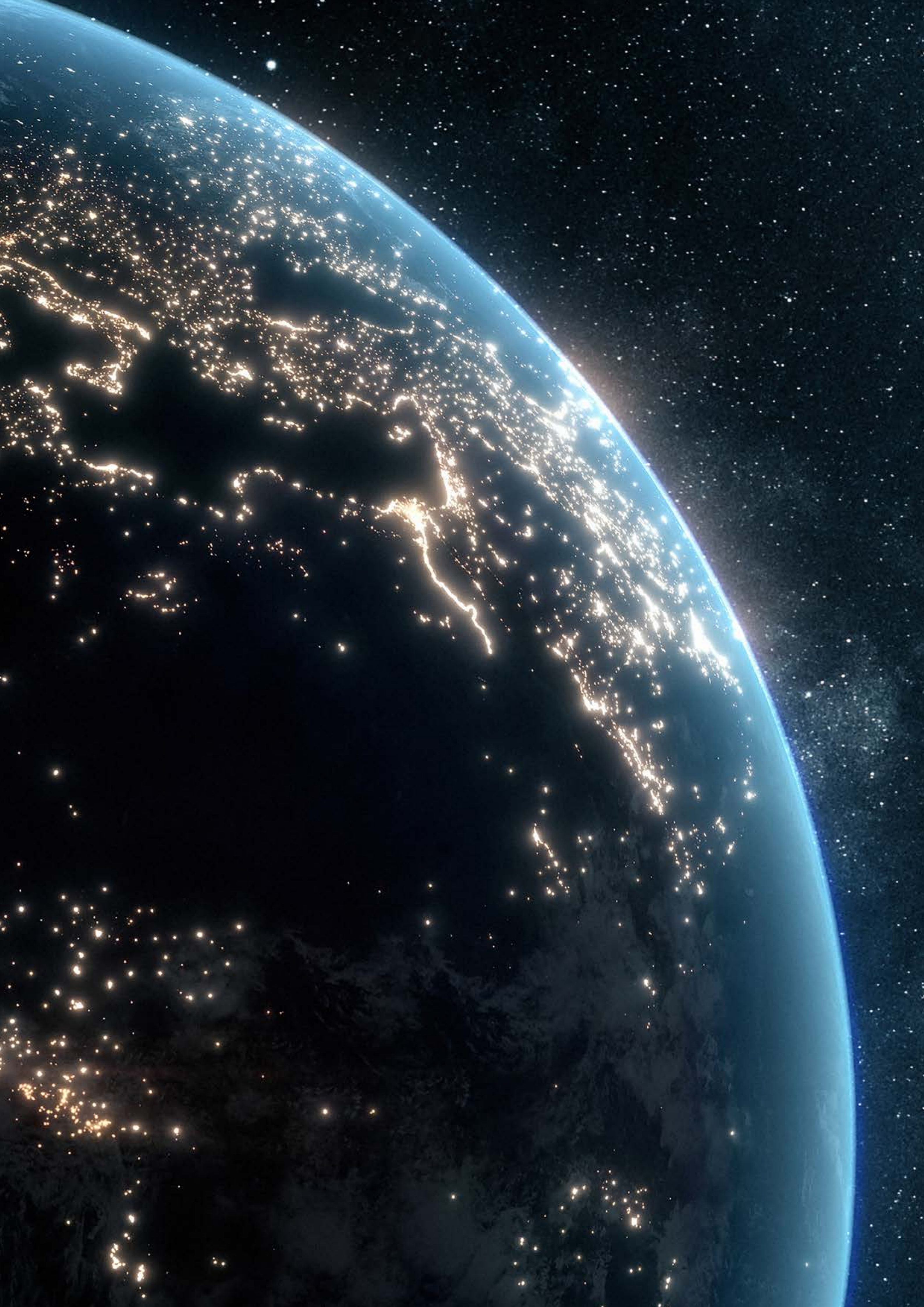

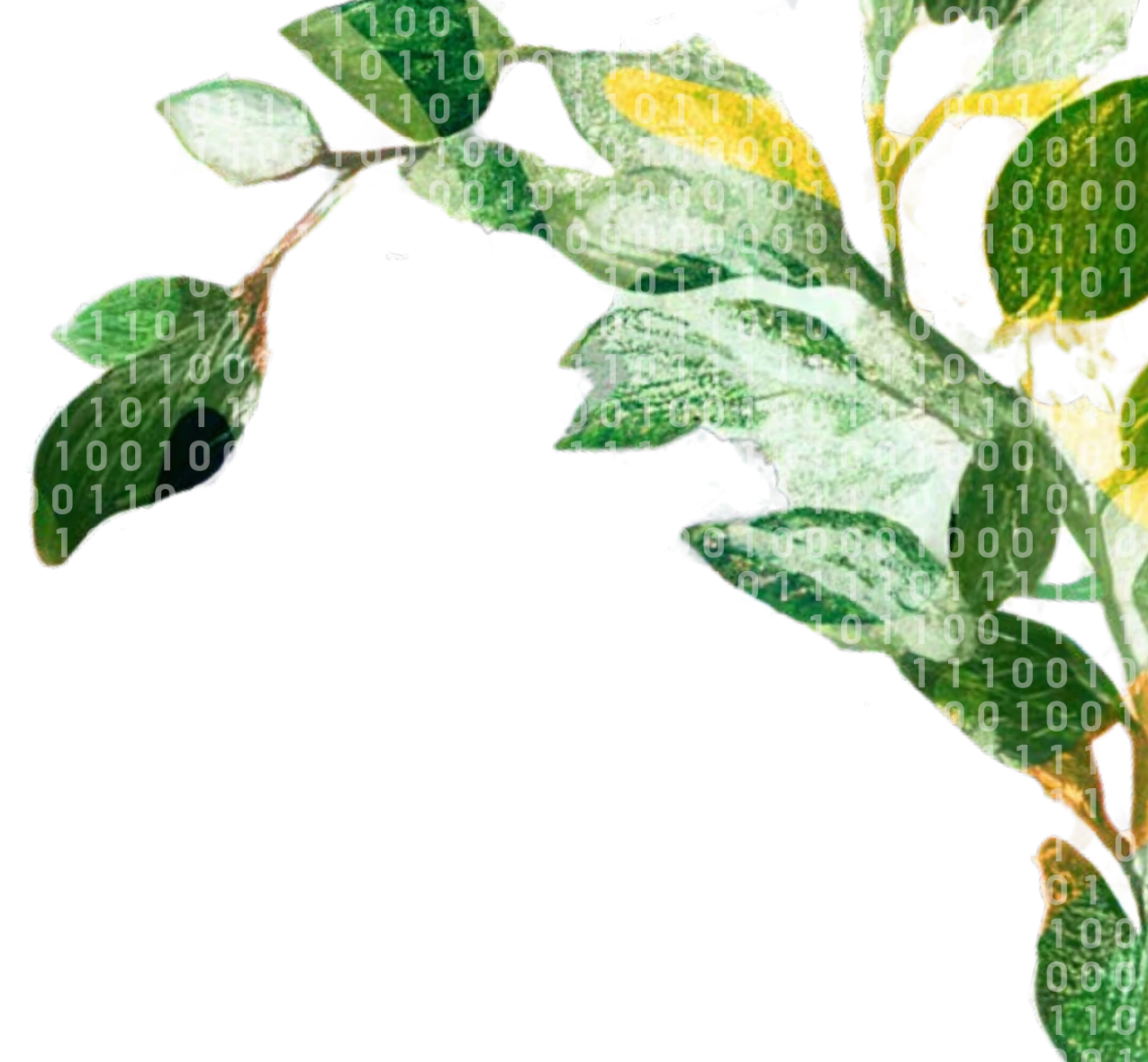

# Introduction

Our living planet and the climate system are changing at an unprecedented pace. The years 2023 and 2024 were characterized by unprecedented warming across the globe.[1] Ocean heat reached its highest observed levels in 2024,[2] causing severe ecological pressure on marine life and putting communities who depend on the ocean for their food security, way of life, and livelihoods at increased risk.[3] Children born this decade are expected to experience an unprecedented number of extreme weather events during their lifetime.[4] The continued loss of ecosystems and biodiversity,[5] and the transgression of several planetary boundaries—the estimated safe operating space for Earth System stability and human prosperity—pose severe risks to society.[6] The time for forceful and evidence-based action that secures a sustainable future for all is now.

These profound changes on the planet are unfolding at the same time as another important global shift: a period of rapid technological change and disruption. Sensors, autonomous systems, artificial intelligence (AI), and other technological advances are on the verge of disrupting multiple aspects of societies and economies. Interestingly, advances in AI are also contributing to novel scientific breakthroughs.[7] Examples include mathematics and computer sciences,[8] the prediction of protein structures,[9] and the life sciences,[10] to mention only a few.

Could uses of AI in science also help us better understand and tackle the complex social, economic, and ecological repercussions of a changing planet? This report focuses on this particular question with a focus on issues related to the sustainability sciences. This field is inherently inter- and transdisciplinary; integrates the analysis of humans, climate, and nature; and strives to inform people's and societies' ambitions to secure a sustainable future for all.[11,12] As we discuss here, AI offers untapped and under-explored potential to accelerate this type of research.

## Our Ambition

This report explores the potential of AI as a set of research methods in a number of issue areas associated with the sustainability sciences. At the core of the sciences is the understanding that the biosphere, with resilient ecosystems and the services they provide, is vital for human health and prosperity, and that human stewardship is key to supporting ecosystem resilience.[12] Technology, traditionally encompassing built and engineered environments, but now ever more digital technologies, is an integral part of these social-ecological-technological systems,[13] with far-reaching impact. As we discuss in this report, AI is at breakneck speed becoming increasingly engrained in shaping how we understand and interact with the world around us and, indeed, our living planet.

Our work complements previous attempts to explore uses of AI for climate change and related issues (e.g., Earth System forecasting[14] and biodiversity monitoring[15]). Our approach is deliberately broader than that of other synthesis reports. We do not focus on direct applications of AI or prescribe specific implementation strategies for the public or private sectors.[16,17] Instead, we emphasize the systemic, cross-cutting implications of AI for sustainability research. We choose areas where science and real-world action meet, in an attempt to lay a foundation for an "AI for Sustainability Science" agenda.

Our broad systems perspective is certainly more demanding. However, we argue that it's urgently needed to build the capacity required to tackle the system dynamics of interconnected sustainability challenges, especially in light of the accelerating pace of planetary change.

## Our Approach

This report is the result of a collaboration that started with a two-day workshop hosted in Stockholm in mid-January 2025, with >20 scientists from various scientific backgrounds, including ecology, sustainability, social, Earth System, and computer science. Many have experience from work in NGOs, private companies, and public universities of applying and developing AI methods in their research. Our work has been structured in the following way:

a. Editorial team started with the identification of issue areas (November and December 2024)

b. Systematic collection and AI-augmented analysis of relevant literature for each issue area (November 2024 to June 2025)

c. Domain expert analysis and redrafting of issue areas, partly informed by the identified literature analysis (January 2025 to August 2025)

d. Iterated discussions with chapter author teams about content of identified issue areas (February 2025 to June 2025)

e. Collective grading, and analysis of potential and gaps (July 2025 to August 2025)

Domain expert discussions and writing have been structured around three main subareas, summarized in Fig. 1: AI for data collection and generation; AI-assisted predictive modeling; and AI-assisted decision-making.

## Limitations

A large part of this report centers around eight broad issue areas. The areas were preselected by the editorial team and discussed, and reformulated after discussions with contributing participants of the workshop and co-authors. The issue areas have been chosen because of their inter- and transdisciplinary features, as well as their importance to people in a time of rapid change for our living planet and climate system. It is important to acknowledge that the issue areas and questions could have been framed differently, and additional ones could have been included. Such changes might have influenced the outcomes and insights presented. The analysis also relies heavily on the composition of invited co-authors and their expertise. We hope, however, that the approach chosen here can inspire further analyses by others.

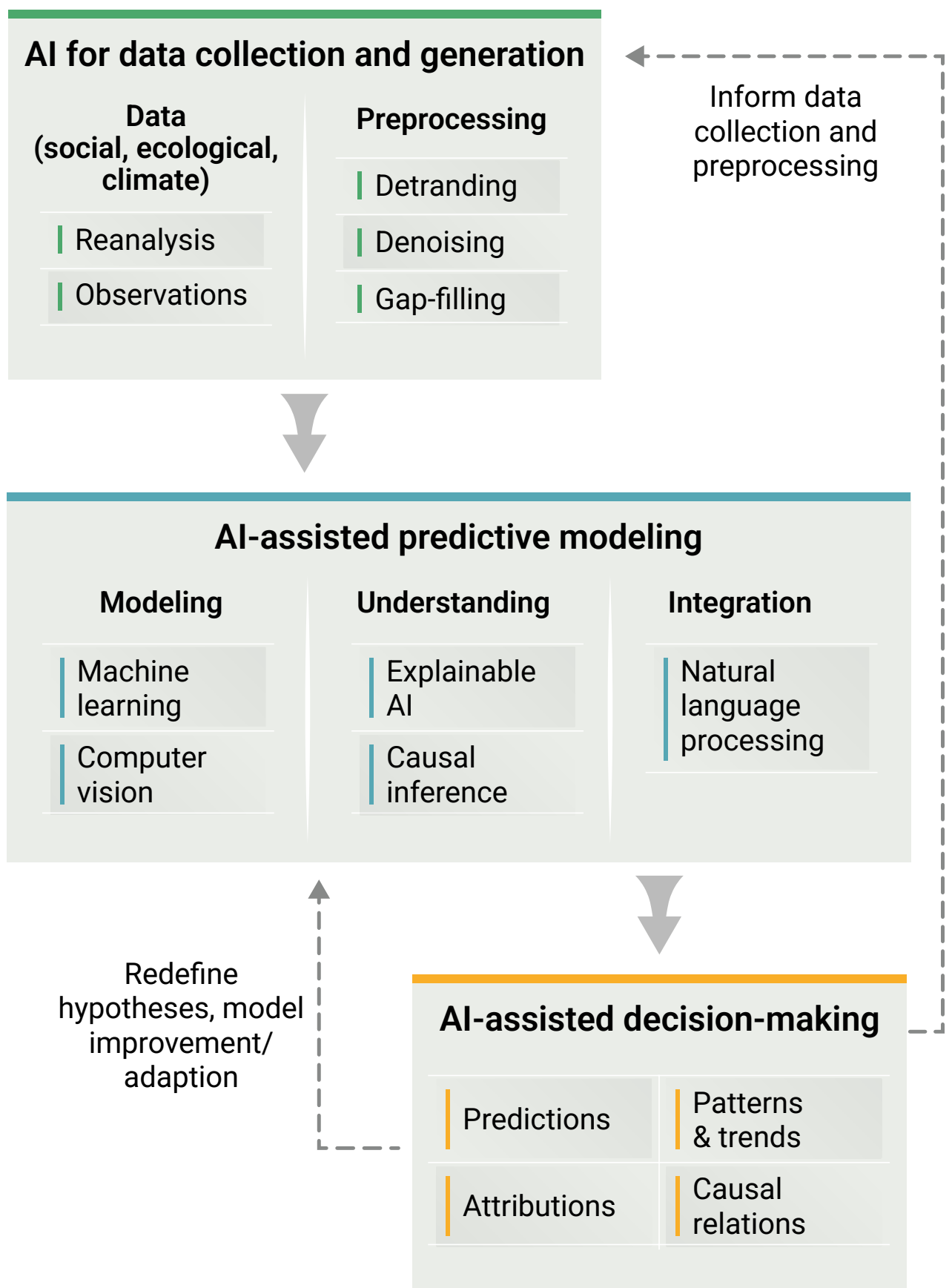

Fig. 1. Flowchart illustrating AI-assisted research processes. Adapted from Camps-Valls et al., 2025, p. 2.[18]

## Potential of Work

Why is this report needed? We believe that while AI has much to offer to the sustainability sciences, it has so far been unclear how the potential could be fulfilled. A number of reports have explored the potential of AI for science in general,[43–45] but few have addressed issues related to people, climate, nature, and their interactions. This is surprising considering the importance of a stable climate system and a resilient living planet for human development and well-being, economic prosperity, and innovation.

The work presented here should be of interest to:

- **Sustainability researchers**, by offering an overview of uses of AI for sustainability research in various areas
- **AI developers in the private sector** looking to contribute to research and development in areas of urgent importance to society
- **Public agencies** interested in assessing how AI may be used in their work to understand sustainability challenges
- **Research agencies and philanthropies** looking to assess emerging areas where novel applications of AI methods could help drive scientific breakthroughs

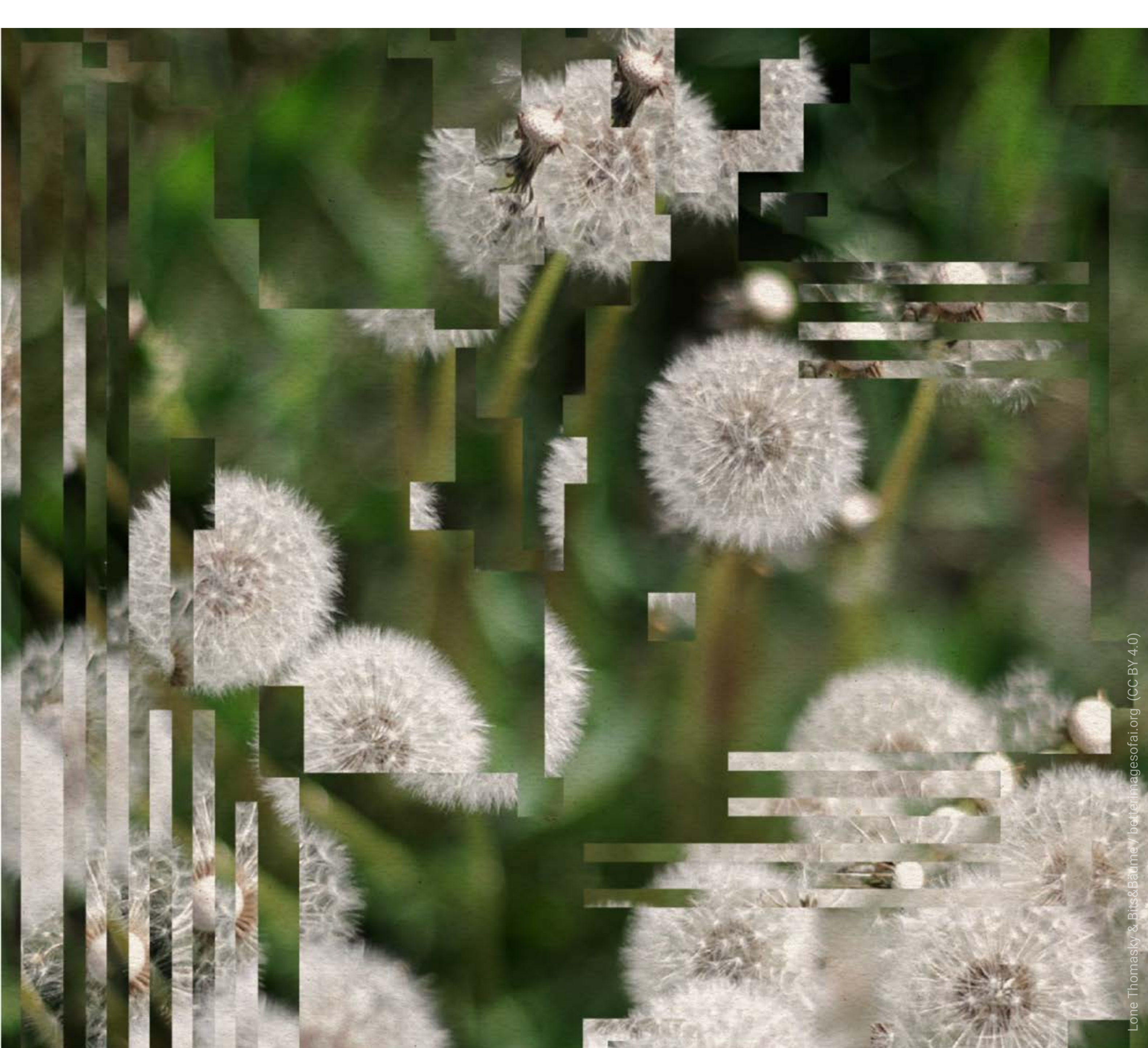

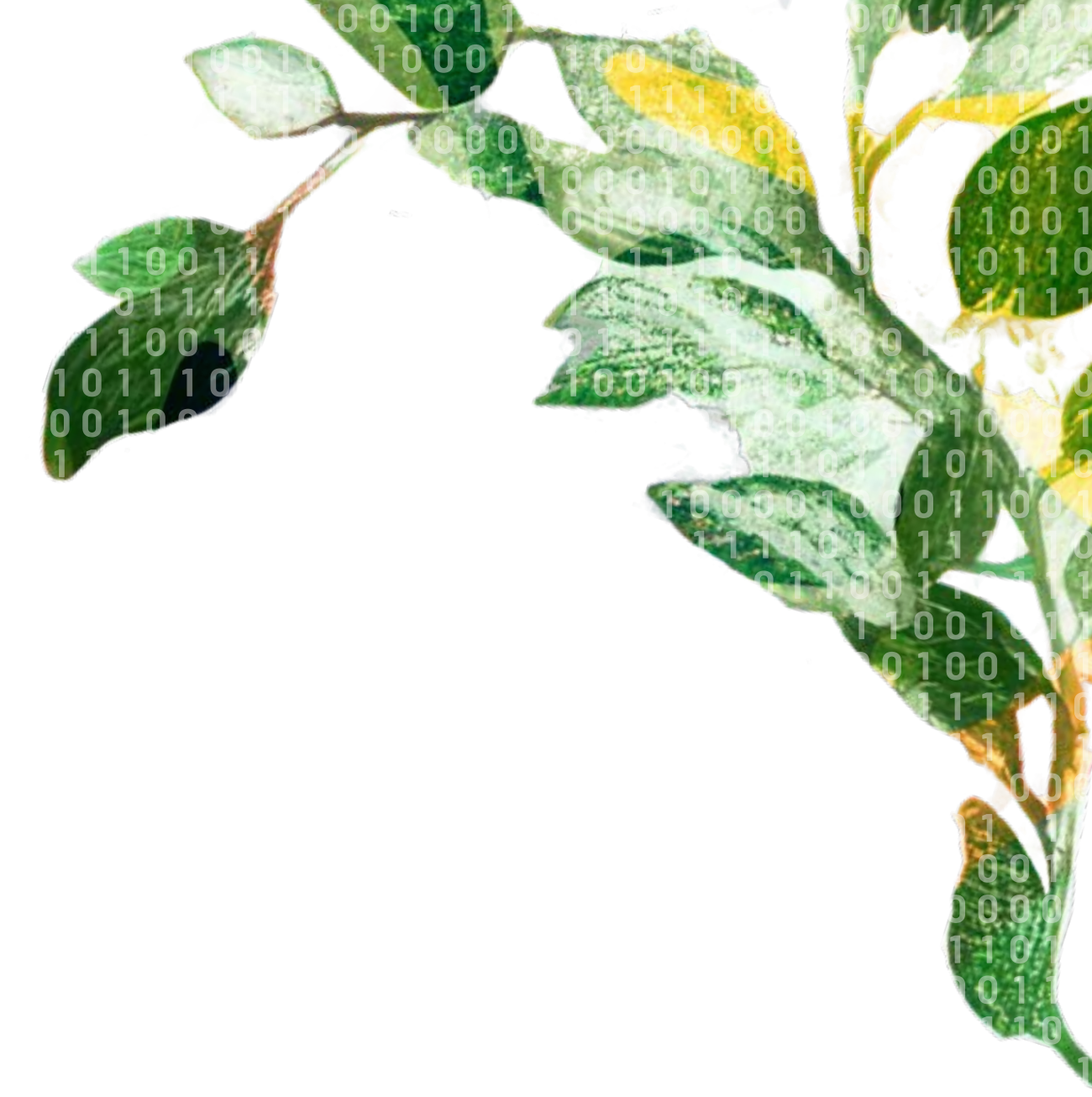

# Sustainability Risks and Allocative Harms

Authors: Samuel T. Segun, Victor Galaz, Maria Schewenius

*AI offers numerous opportunities for climate and sustainability research and action, but which are the most pertinent risks from a sustainability perspective?*

Artificial intelligence (AI), autonomous systems, and associated technologies are increasingly being presented as disruptive technologies with the ability to fundamentally change research, society, and industry. How to develop such technologies responsibly, and how to best monitor and govern possible risks, has led to a growing body of scientific literature. This includes research on AI ethics,[1] law,[2] sociology,[3] allocative harms that affect vulnerable social groups,[4–6] and the political economy of an increasingly digitized and data-driven world[7,8]—just to mention a few.

A number of sustainability risks and ethical challenges emerge from increased uses of AI.[9] We discuss some of these below but would like to emphasize that the list is not exhaustive. Sustainability scientists, large research programs, and research centers planning on using AI in their work need to consider each of these risks, while also looking into other potential ones, and mitigate harmful effects.

## Material, Carbon, and Water Footprint

As the use of AI as a consumer product increases—for example in digital services like chatbots, and the generation of synthetic content like videos and text—so does the ecological and climate footprint of its underpinning infrastructure. This includes the amount of e-waste, such as obsolete specialized processors (graphics processing units [GPUs], tensor processing units [TPUs]), data center infrastructure (servers, hard drives, networking equipment), and end-of-life AI devices (robots, autonomous vehicles, smart sensors).

E-waste represents a growing risk as it is known to release toxic substances like lead, mercury, and cadmium into the environment. These pollutants infiltrate soil and water sources, and can lead to significant ecological degradation and health risks for communities near e-waste disposal sites.[10] Communities affected by e-waste pollution often lack adequate legal protection or avenues for recourse against corporations responsible for harm to the environment and, indeed, living beings.

The impacts of AI training and use are also associated with notable increases in energy use and resulting $CO_2$ emissions.[11,12] Calculating the resulting emissions has proved challenging,[13] but a number of attempts have been made in recent years to assess the carbon footprint of individual AI models, company level emissions, and future estimates at the country level.[11] Whether the longer-term emissions of AI development and use will be net positive remains a contested issue. Some analyses point to potential climate net benefits, provided that AI development and use is matched by targeted investments and supportive climate policies.[14,15] Others instead point at secondary and rebound effects, which could lead to growing climate harms despite increased efficiency over time.[16,17]

There is also a growing understanding and concern regarding the freshwater use associated with AI development and use. Such uses emerge as the result of onsite server cooling (to avoid server overheating) and offsite electricity generation (i.e., through cooling at thermal power and nuclear plants, and expedited water evaporation caused by hydropower plants).[18] Recent estimates show that the growing use of AI is likely to lead to increased water use in many parts of the world, at times even in areas that already face water scarcity.[19]

The expansion of infrastructure needed for AI compute could thus increase the risk of tensions over water rights, land use, and access to rare earth minerals.[20] This is an issue that requires further attention, especially considering marginalized and underserved populations that are already disproportionately impacted by climate change-related events such as extreme weather and natural hazards, and resource scarcity resulting from increased energy and water use.

## Algorithmic Bias and Hallucinations

The risks and impacts of possible algorithmic biases and their allocative harm[21,22]—that is, when opportunities or resources are withheld from certain people or groups—have gained considerable attention in the last few years. Algorithmic biases of this sort can have a number of sources,[23] and could emerge in the sustainability domain in the following ways:[24]

**Training data bias** could emerge if AI systems are designed with poor, limited, or biased datasets. For example, the use of AI based on deep learning (DL) for precision and decision support can help smallholder agrarian communities adapt to a changing climate. However, algorithmic biases created by data gaps could also put the same communities at serious risk if the system's recommendations are flawed or are not validated properly with local knowledge and expert opinion.

**Transfer context bias** could emerge when AI systems designed and trained on one ecological, climate, or social-ecological context are applied to a different, incorrect, or unintended context. While the training data and the resulting model may be developed and suitable for the initial social-ecological situation (say, a large industrial farm in a data-rich context), using it in a different setting (e.g., a small farm) could lead to flawed, biased, and inaccurate results. AI systems built on historical ecological conditions may also fail as ecosystems across land and seascapes shift in unexpected and sometimes irreversible ways.

Even if both the training data and the context in which the algorithm is used are appropriate, their application can still lead to **interpretation bias**. In this type of bias, an AI system might be working as intended by its designer. The user, however, might not fully understand its utility or might try to infer a different meaning that the system might not support.

**Bias and the risks of harm can also be found in new forms of AI**. Large language models (LLMs) have demonstrated a well-documented ability to rapidly summarize, tailor, and communicate climate and environmental information (see later chapters). However, their potential to create hallucinations—that is, inaccurate or misleading information, for instance false facts or citations—remains a concern.[25,26] LLMs can generate malfunctioning code; refer to or cite non-existent material like legal briefs and articles; produce factually incorrect information; introduce subtle inaccuracies, oversimplifications or biased responses; and make common-sense reasoning failures.

**Hallucinations** are not only caused by the technical features of LLMs, but also evolve as the result of how humans interact with such AI models through prompting and feedback.[27] Several private and public–private initiatives have evolved in recent years to address these issues through technical research and development, and safety protocols (examples in Appendix 1).[28–30] The current lack of universal industry standards, as well as agreed-upon and regulated best practices, continues to pose challenges to mitigating harmful uses of LLMs, however.[31] In some extreme cases, LLMs have been used to plagiarize or

**Most and least studied countries in AI for the sustainability sciences**

Fig. 2. The number of occurrences of the ten most frequently mentioned countries, along with a selection of the least frequently mentioned ones (chosen to illustrate overall geographical distribution), including affiliations of first and second authors, across the >8,500 studies on AI and the sustainability sciences included in our literature review. Large world economies dominate, such as the USA, China, India, and countries in Western Europe. Map produced by Maria Schewenius.

produce low-quality scientific papers and academic books, risking the erosion of integrity and public trust in science.[32,33] Such risks should not be taken lightly as sustainability scientists strive to experiment and use AI for research.

## Data Gaps and Bias

Growing volumes of environmental, social, and ecological data are a fundamental prerequisite for the application of AI in, for example, forest biodiversity monitoring[34] and climate modeling.[35] Environmental and ecological data have well-known limitations, however, both in their temporal coverage, geographical spread, and information about biological and ecosystem diversity.[36–39] The rapid growth of data from cell phones and satellites offers vast opportunities to map and respond to social vulnerabilities, such as poverty and malnutrition. Meanwhile, it has become increasingly evident that solutions driven by so-called "big data" and AI analysis can be skewed, since the most disadvantaged populations tend to be the least represented in emerging digital data sources.[40] Indeed, as indicated by the results of the literature analysis conducted in preparation for this report (see AI for the Sustainability Sciences—A Literature Review), most research sites and researchers in the field of AI for the sustainability sciences are associated with high- to higher-medium-income countries (e.g., China, the USA, and countries in Europe) (Fig. 2).

Several of the countries that are the least represented in AI for the sustainability sciences research are also projected to experience some of the most profound changes related to the global issues discussed in this report. For example, Niger, which was not observed in the dataset, has one of the most rapidly increasing populations in the world. It's projected to double from 26 million people in 2024 to 56 million by 2054.[41] Tuvalu and the Maldives are examples of low-lying islands that are particularly vulnerable to extreme sea-level rise, but had zero and two occurrences, respectively.[42]

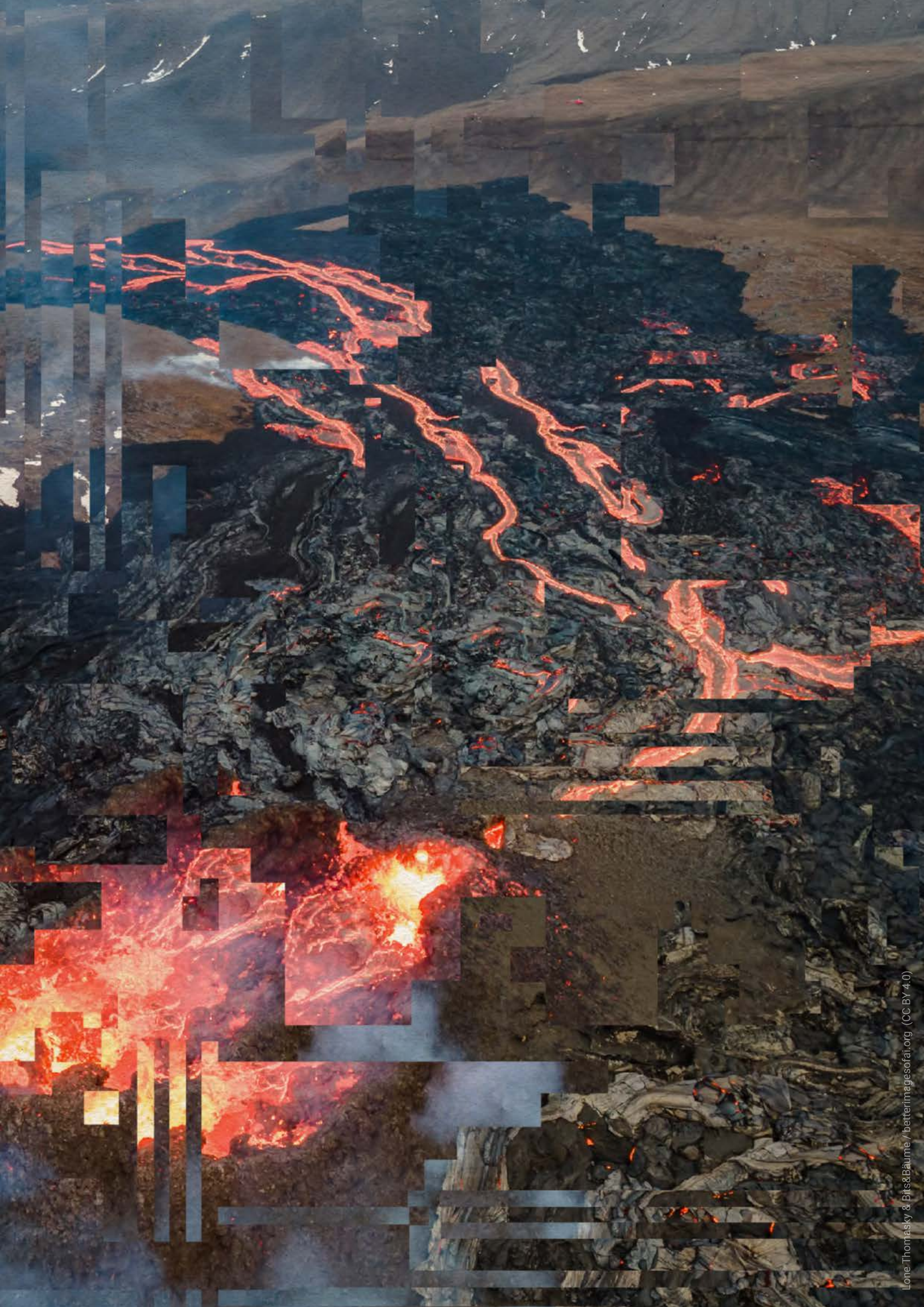

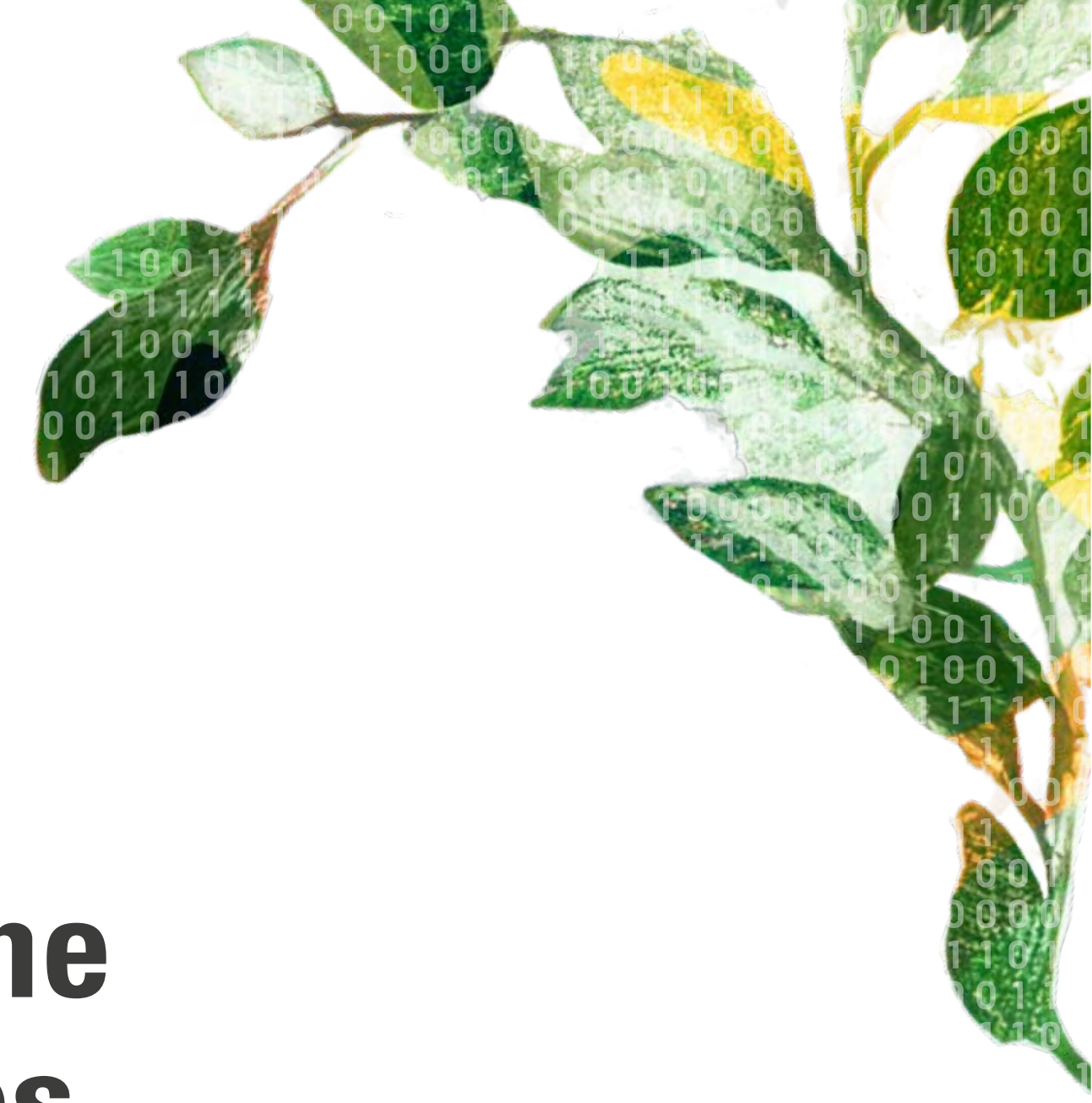

# A Taxonomy of AI for the Sustainability Sciences

Authors: Claudia van der Salm, Drew Purves, Felipe Tobar, Erik Zhivkoplias

## Introduction

Artificial intelligence (AI) has made significant contributions across scientific disciplines, with growing applications to sustainability challenges. This section introduces a taxonomy of AI tools to establish the technical context for the application areas examined in this report. The methods are presented in a way that is consistent with broad data-driven research, placing particular emphasis on current applications, emerging trends, and prospective opportunities in sustainability.

The interdisciplinary nature of AI in sustainability often leads to subcomponents fitting into multiple categories, highlighting the inherent fluidity and interconnectedness of this domain. Therefore, the taxonomy presented in this chapter is structured to reflect both AI tools and the tasks they address in an intertwined manner. Rather than treating methods and applications as separate categories, our presentation is intended to provide a clear and intuitive account of key techniques and their use in sustainability. This perspective aims to provide a clearer understanding of how AI is being effectively applied in diverse sustainability challenges.

It is widely accepted that the mere definition of AI is a matter of open discussion. Different disciplines and application domains consider AI as a collection of methods with widely different levels of automation, data-processing abilities, or even perceived cognition.[1,2] In this taxonomy, we consider a general but pragmatic definition of AI as the field encompassing machine learning paradigms and methods (both classic and modern), symbolic approaches, and hybrid simulation techniques. This way, we focus on methodologies that fuse, possibly to a different extent, domain expertise with techniques for knowledge discovery to address relevant challenges involving (usually large) datasets and/or simulations.

This chapter is organized as follows. We will first detail the different categories of learning settings within **machine learning (ML)**. We will then walk through various well-established AI methods, starting with **classical ML**, which relies heavily on classical statistical algorithms and is characterized by methods developed before the deep learning (DL) era. This is followed by **deep learning**, a methodology relying on deep neural networks. Within deep learning, we will cover **focused DL**, **domain foundational DL**, **agentic LLMs**, **generative AI**, and **deep reinforcement learning**. Next, we cover **symbolic AI**, which relies on explicitly representing knowledge through symbols, rules, logic, and ontologies. Following this is **hybrid simulation AI**, which is presented as an extension of any of the previous AI methods, combined with process-based simulation models.

## Machine Learning

AI systems are regarded as capable of high-level data processing, sometimes comparable—or in some respects superior—to that of humans. To build an AI system, a human expert could include an exhaustive list of rules so that the system knows how to operate in every possible situation. Since this is often unfeasible due to the prohibitively large number of required rules, many of which are unknown, the predominant contemporary paradigm for endowing a system with AI is **ML**. Formally, ML is a field in the intersection of computer science and mathematics that allows machines (or software) to learn to solve problems without being explicitly programmed to do so.[3] This distinction is exemplified by the sharp contrast in methodology between the *Deep Blue system*[4]—an AI chess player that was explicitly programmed with an exhaustive list of rules and famously beat the world champion—and DeepMind's breakthrough *AlphaGo Zero*[5]—part of the modern, more effective, and scalable paradigm of ML where the system learns and develops its own "strategies" from data and experience.

Within ML, different types of learning settings cater for a wide range of data-driven challenges. These are mainly:

- Supervised learning: where the relationship between a dependent (output) and one or more independent (input) variables is discovered based on available data. For instance, predicting the precipitation levels at a given city over the next year (regression), or determining if one of the generators in a wind farm is faulty (classification).
- Unsupervised learning: where relationships and patterns among samples in a dataset are discovered. For instance, grouping households by energy consumption patterns (clustering), or identifying the main deforestation trends from satellite climate observations (dimensionality reduction).
- Reinforcement learning: where an agent learns to make sequential decisions through trial and error to maximize long-term rewards, guided by a specific objective in a particular environment. For instance, optimizing the operation of a smart grid to balance renewable energy supply with fluctuating demand.
- Generative AI: where models learn to create new data resembling the properties of real data. For instance, the generation of synthetic climate scenarios using historical data to evaluate the impact and effectiveness of new policies.

In general, most real-world applications fall into one of these learning settings, or a combination of them, and identifying the appropriate setting is a necessary step toward successfully deploying an ML pipeline. Next, we present a subset of well-established ML models, which can be used (almost) interchangeably across the learning settings described above.

**Classical ML** refers to algorithms developed before the deep learning (DL) era (see next), which mainly rely on a limited number of so-called "hand-crafted features"—that is, transformations of the data following expert knowledge to aid the solution of a learning problem. The features are used as inputs to statistical models such as support vector machines (SVMs[6]), decision trees and random forests,[7] k-nearest neighbors,[8] and linear or logistic regression. These methods are computationally inexpensive and interpretable. They are particularly well-suited for small to medium-sized datasets and when expert domain knowledge is available. As such, classical ML methods have been extensively employed in the natural sciences over the last few decades, and their use will continue, including as components in more complex AI pipelines where appropriate.[9]

One significant application where established classical ML techniques are employed to address sustainability challenges is environmental forecasting, exemplified by the National Oceanic and Atmospheric Administration's (NOAA) *Harmful Algal Bloom (HAB) Operational Forecast System*.[10] This system relies on classical ML methods, and is trained on historical data for select features, including water temperature, nutrient levels, salinity, wind conditions, and satellite-derived chlorophyll concentrations. These models then predict the probability of HAB events in specific coastal regions, such as the Gulf of Mexico, bordered by the US, Mexico, and Cuba, or Lake Erie, located on the international boundary between Canada and the US.[11] Such forecasts are critical for public health advisories, fisheries management, and protecting coastal economies, directly contributing to environmental and human well-being.[12]

## Deep Learning

**Deep learning (DL)** is a methodology that leverages a general class of mathematical models referred to as artificial neural networks (NNs) to represent relationships among data. NNs are a collection of interconnected layers comprising elementary processing units referred to as neurons. The NN architecture is a simplified representation of the physical structure of biological neural networks found in animals' brains, where, by processing data through successive layers, the NN extracts information relevant to the task at hand. The name deep learning reflects the incorporation of an increasingly large number of layers in the NN architecture, where, by going deeper through the stack of layers, the data processing becomes more complex.

Deep NNs are capable of identifying intricate patterns that are beyond the reach of classical ML; however, this is only possible with sufficiently large training datasets and computational resources. Where these are available, NNs can process massive and unstructured inputs, such as images, audio, video, text, and even graphs (as those representing user interactions in a social network). As such, DL has dramatically improved the state-of-the-art in speech recognition, object detection, and specialized domains such as drug discovery and genomics.[13] Particular DL architectures widely used in practice are convolutional neural networks (CNNs),[13] recurrent neural networks,[14] autoencoders,[15] and transformers.[12]

The most direct, and currently the most common, way to apply the power of DL in sustainability science is to define a specific task, and then train a DL model for that task using just the data that the practitioner judges are needed. We refer to this approach as **focused DL**. For instance, NASA's *NeMO-Net project*[1] employs this focused DL approach, training a CNN to analyze high-resolution satellite and airborne fluid-lensing imagery of coral reef environments. The core methodology for this model was specifically designed and trained to delineate and classify coral reef extent, composition, and health status.[16] This work directly supports coral reef conservation and broader marine biodiversity sustainability.

In contrast to the narrow scope considered by focused DL, **domain foundational DL** addresses a more general learning setting within an application domain, where several particular (or narrow) tasks can be cast as special cases. That is, one model supports many tasks. The success of this approach builds on the increasing volume of multimodal datasets and computational resources. Foundational models are first pre-trained on massive general-task datasets. Subsequently, one of a variety of methods can be used to fine-tune the foundational model to enable a specific downstream task. The benefit of this approach is that, compared to focused DL, the end user can achieve the same performance with much less data (or alternatively, can achieve greater performance for the same amount of data). In addition, the user will generally need much less computing power and may not require specialist DL skills.[17]

The domain foundational approach is exemplified by *Aurora*, a model pre-trained on over one million hours of global geophysical data, including forecasts, analysis, reanalysis, and climate simulations.[18] This task-agnostic pre-training allows *Aurora* to learn a general-purpose representation of the geophysical aspects of Earth System dynamics. Adapting *Aurora* for downstream tasks provided state-of-the-art performance in air quality prediction, ocean wave modeling, tropical cyclone tracking, and high-resolution weather simulation, with greatly reduced computational requirements compared to traditional methods.[18] We are also witnessing the increasing emergence of foundational general vision models[19] and specialist models tailored for remote sensing.[20]

An application that has been at the center of AI since its early years is the processing of natural (or human) language, when early statistical language models were developed to process and generate text in restricted domains. Currently, **large language models (LLMs)** are the de facto resource for natural language processing. These models began as domain foundation models for human language, enabling downstream tasks such as generation, summarization, and translation. Later, they were adapted to allow for repeated rounds of conversation and to control tools that execute instructions generated by the same LLM; these are known as agents or **agentic LLMs**. For instance, agentic LLMs can now search on the web, produce and compile code, or assess tabular data to, for example, forecast the dynamics of the Earth System.[18]

LLMs, therefore, have a wide variety of practical uses in sustainability by improving complex workflows involving data handling, processing, and visualization. An additional, but lesser discussed, potential use of LLMs is as core engines of analysis, prediction, and forecasting, where they could supplement, or in some cases replace, the classical ML or DL models used currently. An example of this approach is the project *PandemicLLM*, which implements fine-tuned LLaMA-2 models to reformulate real-time pandemic forecasting (demonstrated with COVID-19) as a text reasoning problem.[21] Diverse data streams—including textual public health policies, textual and sequential genomic surveillance information, textualized spatial data, and epidemiological time series (which are partly textualized and partly encoded by an RNN)—are integrated into structured textual prompts via an "AI–human cooperative design."[21] The composite information is then processed to predict hospitalization trends, outperforming previous models against the tests reported in the publication.[21]

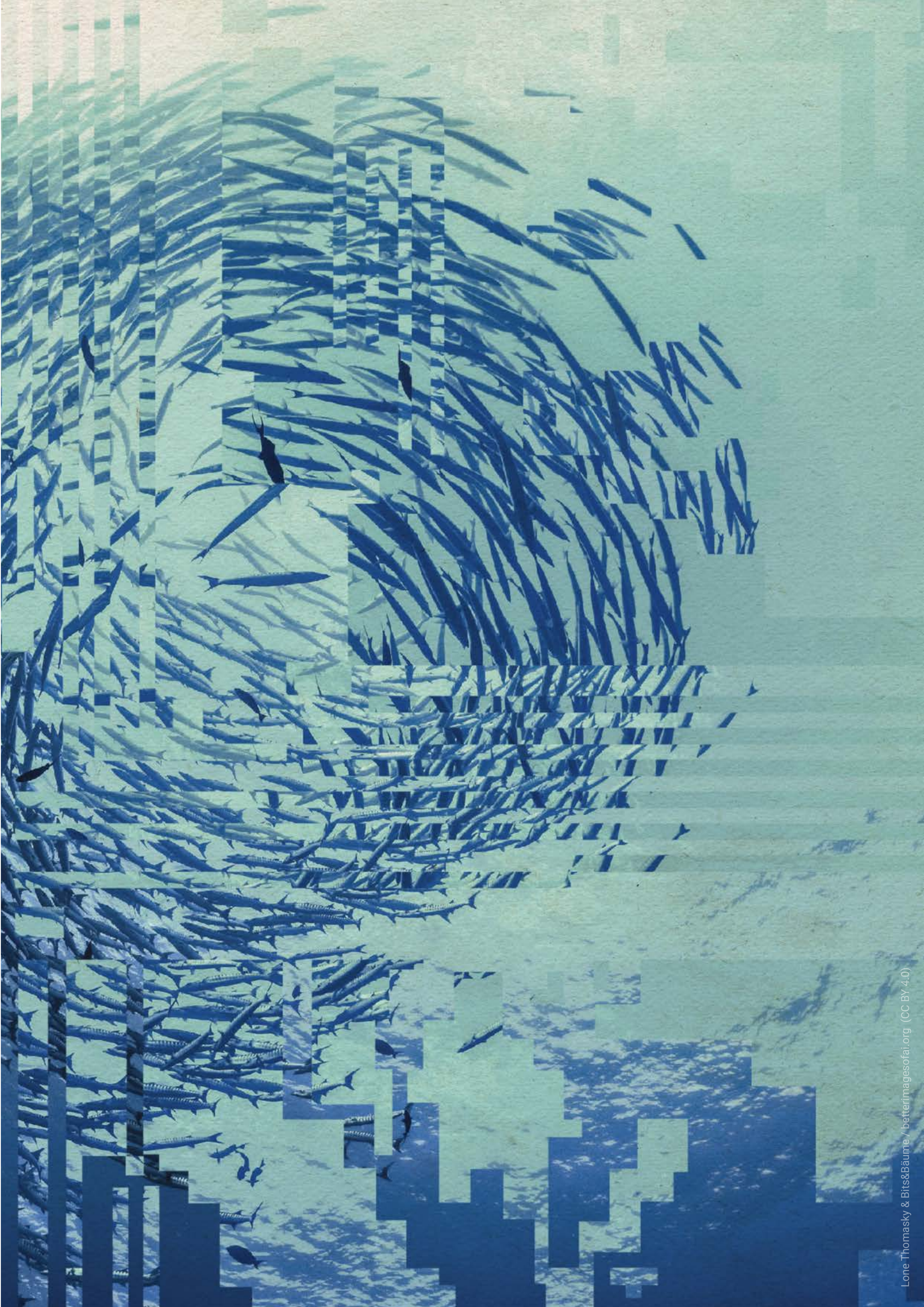

**Generative models** are AI systems capable of producing new data, such as text, images, or audio, and represent a paradigm shift in both AI model architectures and the tasks they perform. In ML, generative AI is technically defined as models that learn to approximate, either explicitly or implicitly, the underlying probability distribution of a dataset to generate new, synthetic data samples that resemble the training data. For this report's focus, however, a more practical distinction is helpful. The distinguishing feature of generative AI is its perceived ability to create, rather than just analyze, complex information. Non-generative (also known as discriminative) AI methods receive inputs that, especially in DL, can be large and information-rich, then produce outputs that are smaller and simpler than the inputs. For example, an image can be the input, and a label the output. By contrast, generative models produce outputs that are large and complex, and reproduce the characteristics of the complex data on which they have been trained. These models can generate outputs that are as complex as their inputs (e.g., video frame in, video frame out), or even more complex than their inputs (e.g., text label in, video out). At the time of writing, the overwhelming majority of generative AI models under discussion in the scientific community are powered by DL, hence why we have included them within the Deep Learning section here.

Interestingly, to work with multimodal data, many current generative AI models incorporate LLMs alongside state-of-the-art models for image generation, such as diffusion models,[22] with all of these having deep NNs as building blocks. These multimodal generative systems have reached, and even exceeded, human-level performance in tasks such as coding, writing, and illustration, making them an attractive tool for professionals across domains. A catalyst for the adoption of these models by the wider, non-specialist community is their free or inexpensive availability (e.g., *ChatGPT*, *Gemini*, *Claude*, and *Mistral*).[23] Potential barriers to the adoption of generative AI include their tendency to sometimes "hallucinate" (that is, provide realistic, believable, but incorrect outputs) and to sometimes produce biased outputs (reflecting biases in their training data). Thus, at the time of writing, practitioners need to put in place careful quality control systems to ensure the appropriate use of generative AI systems.

As outlined above, reinforcement learning (RL) is an ML technique where an agent learns to make sequential decisions through trial and error to maximize long-term rewards guided by a specific objective in a particular environment. **Deep reinforcement learning (DRL)** refers to the use of deep NNs within an RL framework. For example, traditional RL often involves the development of policy models (which generate a suggested action in a given circumstance) and/or value functions (which predict the rewards that would result from alternative actions). Both policies and value functions can be replaced with DL models, creating RL systems that can maximize rewards in more complex environments. RL and DRL address dynamic optimization problems, characterized by the need to make sequential decisions where the resulting feedback guides each choice. This makes RL/DRL a powerful tool for addressing sustainability challenges, particularly in applications focused largely on energy and transportation, as noted in the review by Zuccotto et al. (2024).[24] Importantly, there are many classical optimization methods that are separate from RL and which are better suited to static decision problems, where a one-time decision ought to be made given a fixed set of information. These methods, which include, for example, linear programming, share some properties with classical ML and some properties with symbolic AI (see below).[25]

Among other emerging applications of DRL, Google DeepMind's application of deep RL to reduce energy consumption in their data centers represents a notable real-world success.[26] By training an RL agent on historical sensor data (temperatures, power, pump speeds, etc.), the system learned to dynamically adjust cooling system operations (e.g., fan speeds, chiller settings) to minimize energy use while maintaining safe operating temperatures. This led to signifi-cant energy savings, directly contributing to environmental sustainability by reducing the carbon footprint of large-scale computing infrastructure.

## Symbolic AI/Knowledge-Based Systems

Distinct from ML, symbolic AI represents knowledge explicitly through symbols, rules, logic, and ontologies.[1] Classical examples include expert systems, semantic networks, and logic programming. Symbolic AI systems can still interface with large datasets, but these datasets must first be structured and mapped to the system's formal ontology or knowledge representation. The systems then apply explicit, human-defined rules and logical principles to the structured data (by contrast, ML models can infer implicit patterns and statistical relationships directly from raw data). Fundamental to the history of AI, symbolic methods remain highly relevant, particularly for tasks requiring transparency, reproducibility, explainability, and the direct encoding and integration of established domain expertise or complex scientific models, as seen in advanced decision support systems.

The *OECD QSAR Toolbox*, for example, aids in assessing chemical hazards, aiming to reduce animal testing.[27] The toolbox embeds expert knowledge and regulatory guidelines as structured rules and decision workflows. This symbolic AI approach guides users through complex chemical data integration and hazard prediction, translating scientific and regulatory information into actionable assessments. It thus supports sustainable chemical management worldwide by promoting safer chemical design and protecting human and environmental health.

## Hybrid Simulation AI

Simulation models are built on mechanistic understanding of dynamic systems. By analogy, civil engineers routinely use simulations of bridges and other structures, based on Newtonian physics, to simulate the response of the structure to various forces, such as wind, and iterate on the design until appropriate levels of safety have been reached. This explicit, mechanistic approach contrasts with classical ML and DL, both of which learn empirical patterns from data. A key advantage of simulation models is that, in principle, they allow for prediction and inference in previously unseen scenarios (e.g., unprecedented $CO_2$ levels and climate extremes). In contrast, the predictive ability of classical ML and DL is not guaranteed when used with new data that are statistically different from the training set (known as the out-of-distribution setting).

Process-based simulation (or mechanistic) methods for Earth System components include atmospheric physics models used to drive traditional weather forecasting, which are used alongside ocean physics models within general circulation models (GCMs) to predict climate change.[28] Other examples include models of hydrology, ecology (e.g., global vegetation[29] and atmospheric chemistry[18]), and a variety of economic and socioeconomic models.[30]

However, the predictive ability of simulation models depends on a correct understanding and encoding of the mechanistic rules. Many systems in sustainability science are characterized by large numbers of interacting processes, some of which are not understood sufficiently to enable accurate mechanistic modeling. For example, agricultural yields emerge from an interaction among plants, soils, hydrology, weather, and indeed people. Even systems that are at first glance more purely physical, such as atmospheric and oceanic dynamics, are subject to a "long tail" of additional factors that are hard to identify and specify *a priori*. It can therefore be valuable to combine simulation modeling with ML, in methods known as, for example, physics-informed ML, knowledge-guided ML, scientific ML, surrogate modeling (emulation), and ML-based parameter estimation, to name a few.[9]

To date, most hybrid simulation-AI methods have employed classical ML methods, but we are seeing the emergence of methods based on DL. *GraphCast*, for instance, trained a graph neural network on historical weather reanalysis data (which themselves are outputs of physics-informed models) to forecast global weather.[31] The related project *NeuralGCM* adopted a more tightly integrated hybrid approach: DL was used to replace or augment specific computationally intensive parameterizations (e.g., for cloud processes) within the structure of an otherwise traditional GCM.[32] In both cases, the hybrid physics-informed-AI approach allowed for improved accuracy at a reduced computational cost (see also the reference to *Aurora* earlier in this chapter).

## Conclusion

We hope that the above taxonomy and examples will be helpful to those with an interest in the current, and potential future, landscape of AI as applied to the sustainability sciences. Before concluding, we add the following caveats.

**First**, in several places we have contrasted previous innovations with more recent ones. For example, we contrasted classical ML with the more recent development of DL, and we contrasted the original use of LLMs (focusing narrowly on human language) with the more recent development of multimodal agentic LLMs. The caveat here is that we do not mean to imply that the more recent developments will replace the previous ones. There will continue to be problems where a more traditional or classical approach is the best choice, and where moving to a more sophisticated approach is methodological overkill. It is therefore very reasonable to expect that all of the paradigms and methods we described above will continue.

**Second**, to illustrate our taxonomy, we sought representative, focused use cases that illustrated the particular method under discussion. For example, we used *Aurora* to illustrate the concept of domain foundational DL. The caveat here is that many, potentially most, solutions in sustainability science will involve combinations of the methods we listed above. For example, one might use DL to extract data from raw sensor observations, which are then passed into a hybrid simulation-AI modeling stage, producing a model that leverages RL to find an optimal decision … and all of this might be coded up with the help of an agentic LLM!

**Third**, AI is a rapidly evolving field. We hope that our taxonomy will remain relevant, but it is likely to become relatively less and less comprehensive through time, as new methods and paradigms appear, which themselves can be hybridized with existing methods. For example, we outlined hybrid simulation AI as an overall approach, but there are many different ways that simulations and AI can be hybridized, and it may be that some of these will come to constitute new paradigms that could sit alongside other recent paradigms, such as deep generative AI.

**Fourth**, we chose to focus on the fundamentals and uses of different methods, rather than their relative environmental footprint. The caveat here is that some recent methods have, at the time of writing, significantly higher environmental costs than others. For example, overall, DL methods currently tend to require more compute (and hence energy) than analogous classical ML methods; whereas, some uses of domain foundational models, which are based on DL, require very little compute at the point of use, compared to focused DL. We therefore encourage all potential users to assess the marginal environmental costs of any AI application in sustainability, adopting solutions only where the likely sustainability benefits outweigh the likely environmental footprint.

**Fifth and finally**, we stress that responsible uses of AI seek to empower and augment, rather than replace, human expertise, including but not limited to scientific expertise. Human creativity and expertise, and human values, are still needed to select the relevant problems to address, select the appropriate models and datasets, evaluate model predictions rigorously, and develop those predictions into policy. Carried out correctly, AI has great potential to not only increase the productivity of sustainability scientists, but also to lead to scientific conclusions and predictions that would not be possible without AI—or without humans.

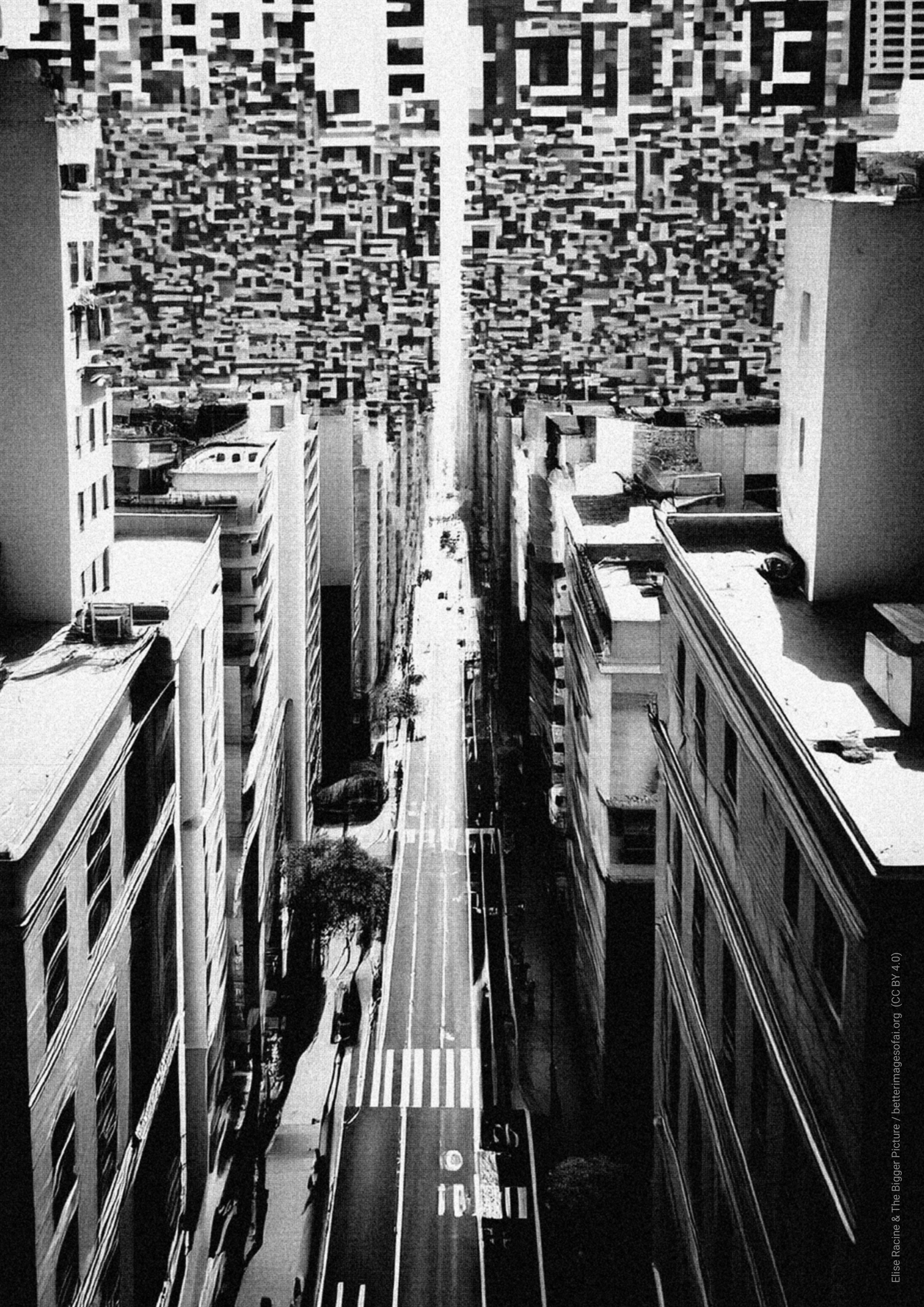

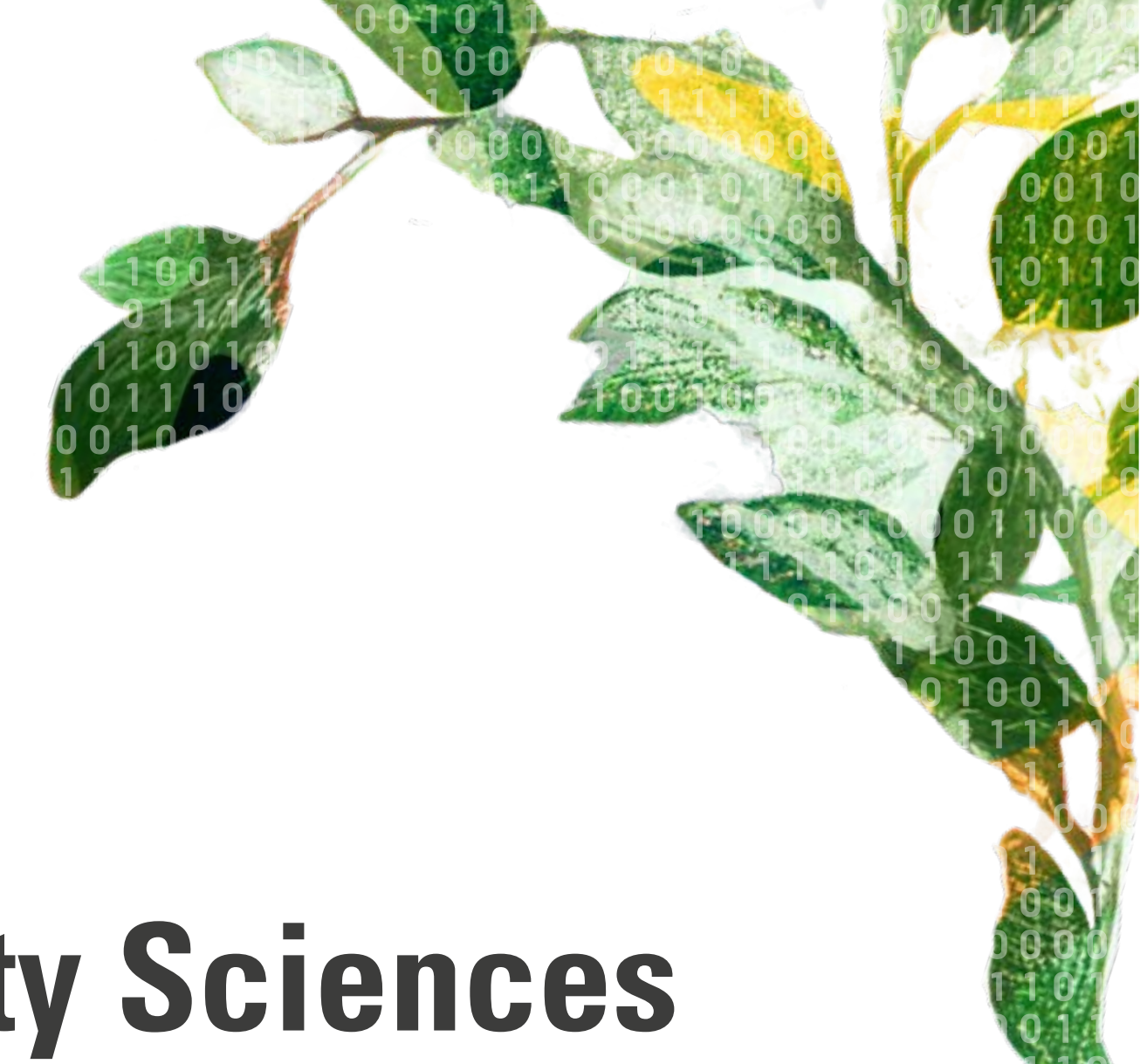

# AI for the Sustainability Sciences —A Literature Review

Authors: Erik Zhivkoplias, Maria Schewenius, Ingo Fetzer, Victor Galaz

*To what extent has AI been used for research in the eight issue areas in the last five years? Which specific AI methods dominate in which fields? These are some questions we explored in our comprehensive literature review.*

## Introduction

This chapter summarizes our AI-assisted literature overview, based on a combination of strategic searches in literature databases, expert assessments, and AI-supported analysis. The aims of this literature analysis are twofold. **First**, to generate a high-level overview of AI methods across the selected sustainability issues based on keyword frequency in abstracts. **Second**, to compile a set of relevant scientific literature to complement subsequent expert-authored sections of this report.

While far from perfect, this approach allows us to map methodological innovations, application trends, and knowledge gaps at the intersection of AI and sustainability science. The analysis builds on the methodological taxonomy introduced earlier in the report, with a special emphasis on *democratization*—defined as the process by which access to, understanding of, and application of AI methods expanded beyond specialized research communities into broader scientific practice.

## Results

We distinguish three generations of AI methods in our classification:

- **Classical machine learning (ML)**. Dominant before widespread GPU adoption for training (~pre-2010). Characterized by traditional statistical and algorithmic methods without deep neural networks, primarily trained on central processing units (CPUs). This era saw increased access through libraries like *scikit-learn*, enabling broad scientific use.
- **Focused deep learning (DL)**. Defined by the democratization/growing public access of DL starting around 2010–2012, enabled by GPU advancements and marked by superhuman performance in specific, narrow tasks (e.g., *DanNet 2011*,[1] *AlexNet 2012*[2]). While core concepts (e.g., convolutional and recurrent neural networks [CNNs and RNNs, respectively], and backpropagation) predate this, their practical feasibility and widespread adoption surged in this period.
- **New generation of AI**. Characterized by the democratization of generative capabilities and foundation models since ~2020 (e.g., *ChatGPT*, *Stable Diffusion*), making advanced AI accessible to non-experts via prompting. While generative AI (GANs, VAEs, Diffusion) and transformers are central, this era also encompasses critical advancements like multimodality (e.g., *CLIP*), knowledge-guided/physics-informed ML, explainable AI (XAI), and smart robotics.

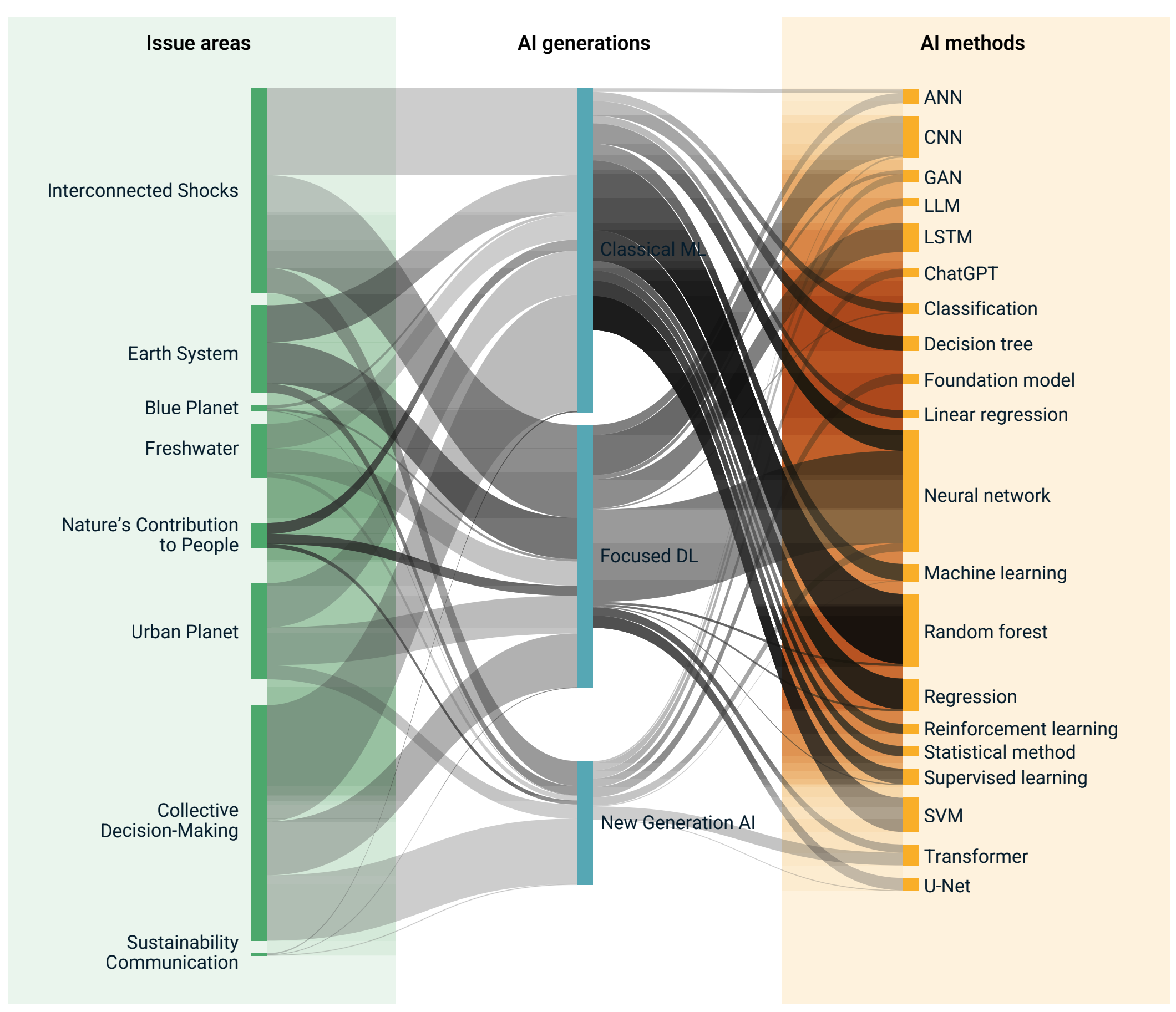

Fig. 3. Issue areas in sustainability science connected with the most frequently used methods and applications of AI, grouped by AI generation. Analysis includes 5,600 articles. See Appendix 4 for details.

The use of AI methods in sustainability research has been substantive for the five years analyzed. A total of 5,603 papers (out of the total 8,504 included in the literature review, see *Methods*) were successfully classified into the AI generations listed above. The distribution shows that classical ML is the dominant approach for all the sustainability issues selected (2,381 articles), followed by focused DL (2,163 articles), and new generation AI (1,059 articles) (Fig. 3).

There are, however, distinct methodological patterns that differ between areas. Articles belonging to the issue area *Enhancing Nature's Contributions to People*, for example, show the frequent use of AI methods like CNNs. This could be a reflection of frequent uses of AI, for example for species identification tasks. Neural networks and classical ML methods such as random forests and support vector machines (SVMs) are also widely used, for example in ecological modeling contexts. Transformers appeared infrequently, suggesting a more limited adoption (Fig. 4e).

AI methods used in the issue area *Collective Decisions for a Planet under Pressure* feature neural networks and random forests, alongside notable occurrences of transformers, LLMs, and reinforcement learning (Fig. 4h). The issue area *Understanding a Complex Earth System* shows frequent uses of recurrent CNN and long short-term memory (LSTM), alongside persistent SVM and regression methods (Fig. 4b). The area *Preparing for a Future of Interconnected Shocks* displayed high occurrences of CNNs and LSTMs, potentially linked to hazard prediction applications, followed by frequent mentions of generative adversarial networks (GANs) and SVMs (Fig. 4a).

*Stewarding Our Blue Planet* revealed frequent uses of neural network methods, alongside unsupervised techniques such as principal component analysis (PCA) and k-means clustering (Fig. 4c). *Improving Sustainability Science Communication* was characterized by natural language processing (NLP) terms, particularly transformers and LLMs (e.g., *BERT*), implying a strong focus on language-based methods (Fig. 4g).

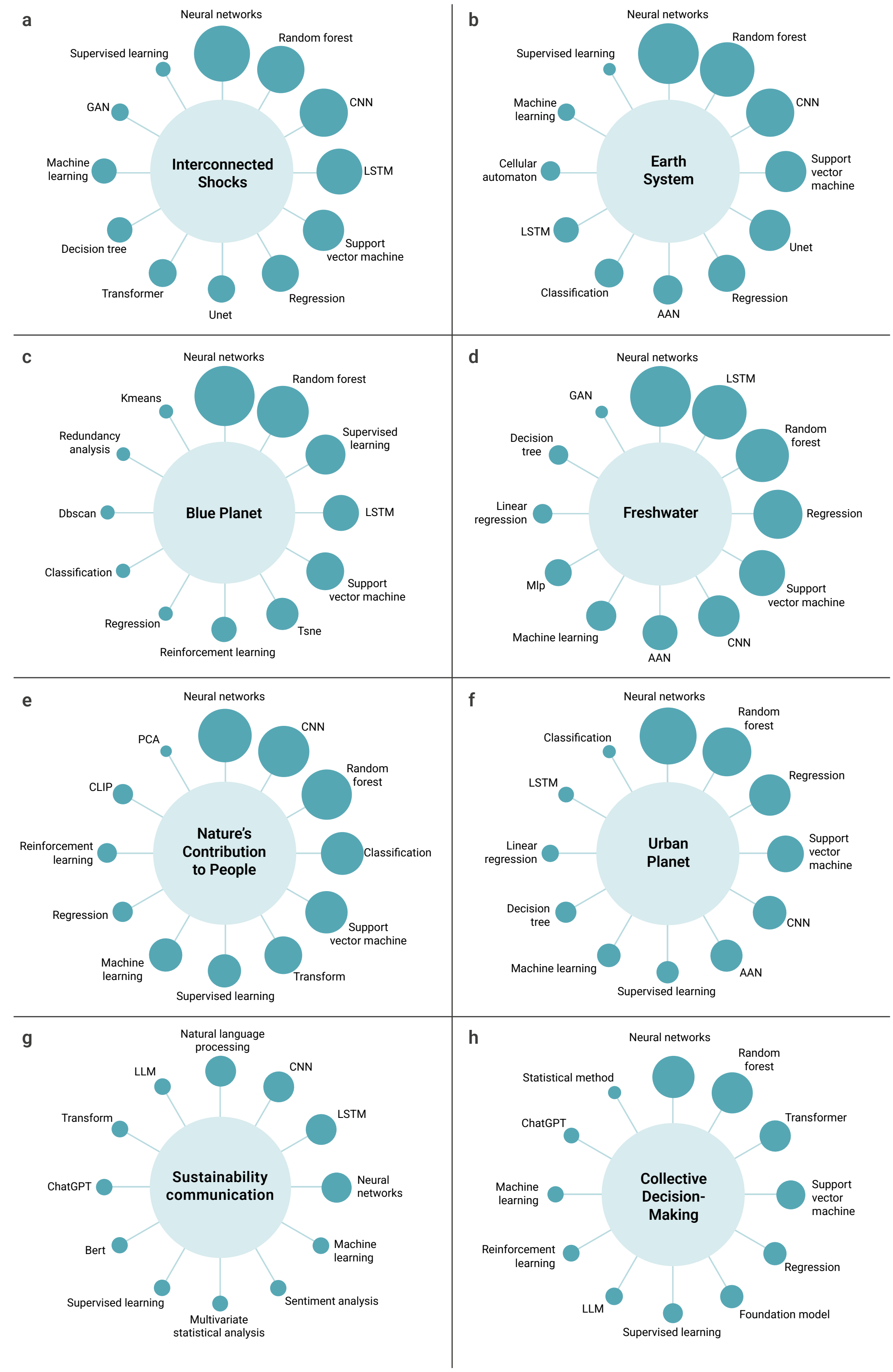

Fig. 4a-h. The most frequently mentioned AI methods in each issue area as identified in the literature review. Analysis includes 5,600 articles. See Appendix 2-5 for details.

*Prospering on an Urban Planet* displayed strong uses of CNN and GAN methods followed by frequent SVM and regression techniques (Fig. 4f). *Securing Freshwater for All* demonstrated high occurrences of LSTMs and random forests, potentially indicating their use in hydrological forecasting applications, with CNNs also frequently represented (Fig. 4d).

As a general pattern, older AI methods seem to be more commonly used in the areas of *Understanding a Complex Earth System*, *Securing Fresh-* *water for All*, and *Stewarding Our Blue Planet*. Newer AI methods can facilitate participation by non-specialist stakeholders and showed notably higher adoption in the domains of *Prospering on an Urban Planet*, *Improving Sustainability Science Communication*, and *Collective Decisions for a Planet Under Pressure*.

## Method Summary

The scoping literature search used the *OpenAlex*[3] database for peer-reviewed articles published between January 2020 and December 2024. Search queries combined AI-related keywords (e.g., "machine learning," "deep learning," "generative AI") with sustainability science concepts across the eight predefined focus areas (e.g., *Nature's Contributions to People*, *Stewarding Our Blue Planet*, *Prospering on an Urban Planet*) defined by researchers from the Stockholm Resilience Centre and Potsdam Institute for Climate Impact Research (full search strategy in Appendix 2).

After removing duplicates, 21,648 articles remained and were thus included in the first step of the analysis. We later employed a tiered screening approach and selected the 1,670 most-cited articles (citation count >10 for articles published between 2023 and 2024, and citation count >30 for articles published between 2020 and 2022). These underwent manual title/abstract review against predefined inclusion/exclusion criteria (Appendix 3). Each included article was categorized as belonging to the single most relevant sustainability science issue area. The remaining 19,978 articles were screened using *Rayyan*,[4] an AI-assisted systematic review tool, which flagged 7,679 articles as "Likely" or "Very likely" to be relevant. This yielded a final corpus of 8,504 articles combining manually selected and AI-flagged papers.

For the AI method classification of each paper, we used a locally deployed large language model (LLM) (*DeepSeek-R1-7B*). The model was used to extract AI methodology keywords from article abstracts. Extraction was guided by a taxonomy of AI and representative examples (Appendix 4). The taxonomy included keywords mapped to consolidated categories to prevent duplication (Appendix 5). Our manual validation on 100 randomly sampled articles demonstrated 93% accuracy in keyword extraction, and 88% agreement in method categorization. No formal quality assessment of individual papers was performed, as the objective was to characterize broad methodological trends and application breadth rather than assess individual study quality. Note that all methodological details of this review can be found in the appendices.

## Limitations

This literature analysis has several limitations. **First**, our use of *Rayyan*, a proprietary AI screening tool, meant we couldn't access or verify its internal algorithms, training data, or decision thresholds. The limited access could potentially introduce selection bias for lower-citation articles. **Second**, all screening relied solely on titles and abstracts rather than full texts, risking oversight of relevant studies where key details (e.g., methodology or sustainability links) appear only in the main content. **Third**, while manual validation showed strong accuracy, the LLM-based keyword extraction depended heavily on abstract clarity; ambiguous phrasing or niche methods might have been missed or misclassified.

To address these limitations, we implemented several mitigation strategies throughout the review process. To reduce potential selection bias from Rayyan's screening, we confirmed the relevance of selected studies through title keyword analysis, providing supporting evidence that the included articles aligned with the intended scope of the sustainability issue areas. Nonetheless, we acknowledge that some domains, particularly those with consistently low representation (e.g., Stewarding Our Blue Planet and Improving Sustainability Science Communication) may still be subject to undersampling.

To counter limitations of LLM-based keyword extraction, we manually validated a random sample of 100 articles, assessing accuracy in AI method categorization and finding high accuracy rates (93% accuracy in keyword extraction, and 88% agreement in AI-generated categorization of AI-methods). While reliance on abstracts and LLM-assisted extraction cannot fully replace exhaustive full-text review, these iterative validation steps served as robust checkpoints for both inclusion and classification.

Despite the limitations, we believe that this review offers valuable broad-scale insights into AI's evolving role in sustainability science.

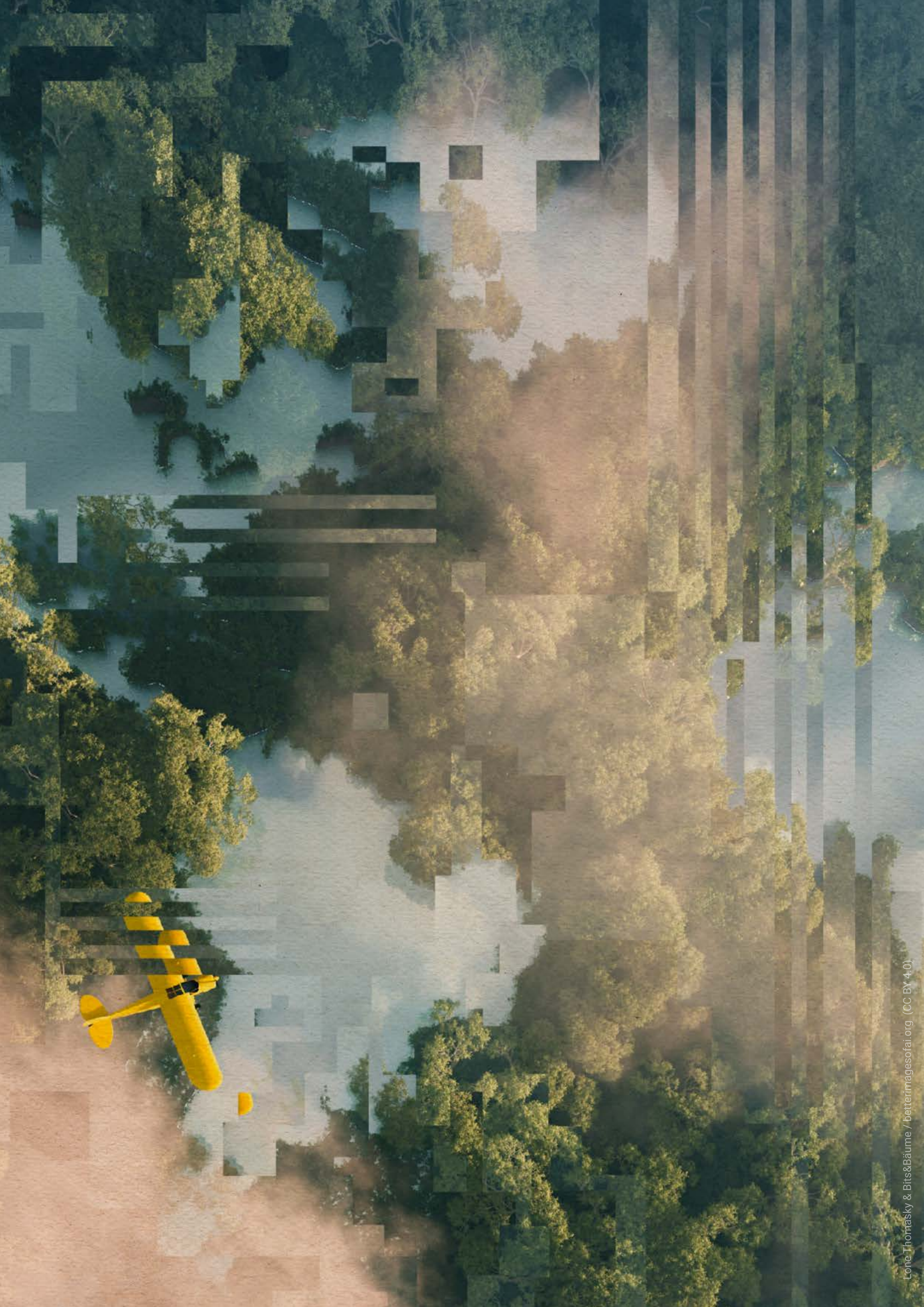

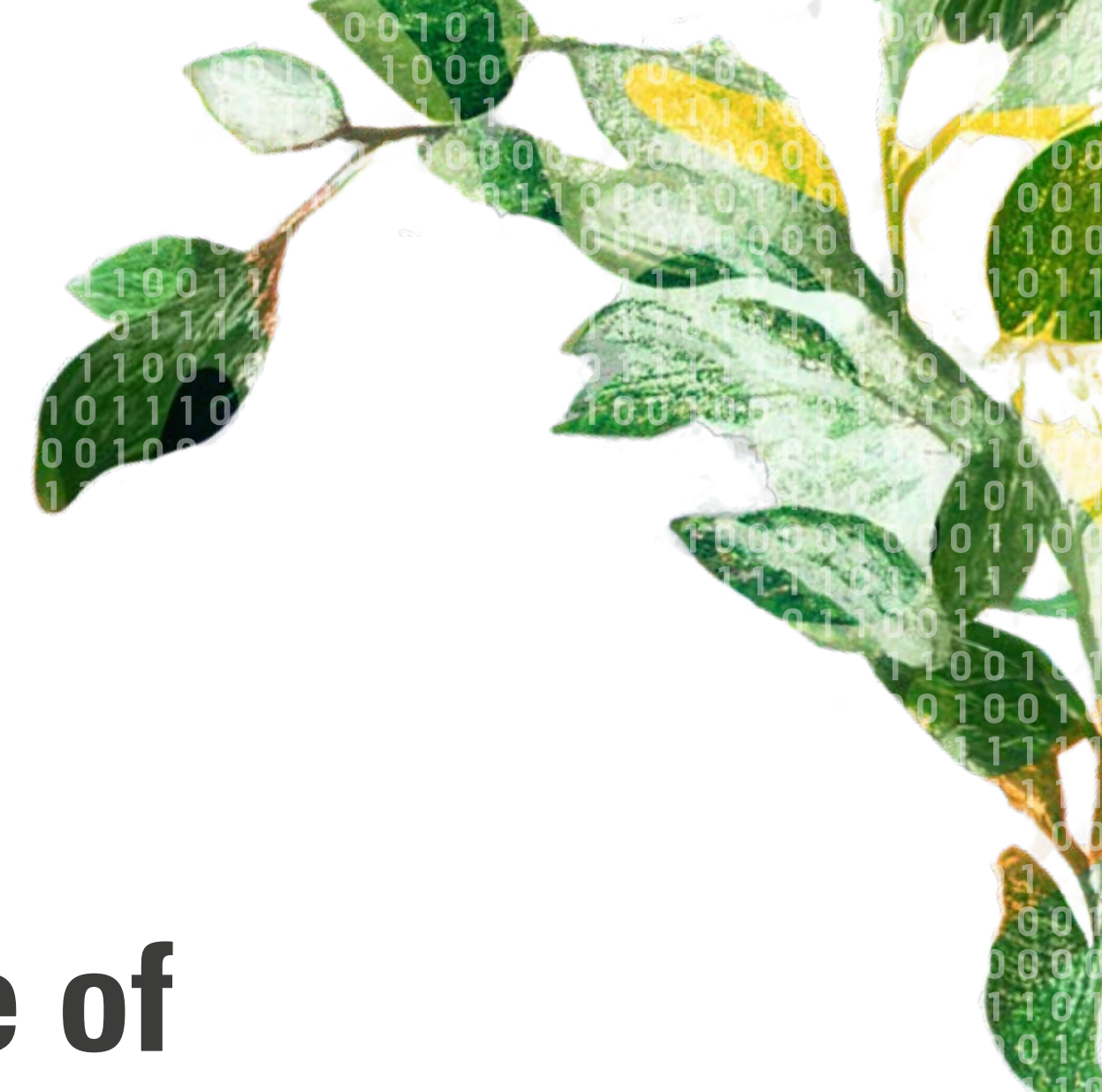

# 1 Preparing for a Future of Interconnected Shocks

Authors: Louis Delannoy, Magnus Nyström, Juan C. Rocha

*Our global future is projected to be marked by turbulence and a changing landscape of shocks like floods, fires, conflicts, and disease outbreaks. Can AI contribute to our analysis of such risks, and help identify novel solutions?*

## Introduction

The Anthropocene is characterized by a changing landscape of shocks, which are broadly defined as sudden events with noticeable impact (e.g., floods, fires, conflicts, disease outbreaks). As Delannoy et al. (2025)[1] show, not only have shocks become more numerous but also more co-occurring, especially between 1970 and 2000, meaning that multiple types of shocks increasingly struck the same country within the same year. Shocks also appear to revolve around the climate–conflicts–technology nexus, which suggests ramping up interconnection and potential lock-in effects between these sectors.[2] At the same time, shocks are difficult to unpack due to their non-linear features, resulting in changes that are difficult or even impossible to reverse (e.g., tipping points, regime shifts, trap situations) with large repercussions for people.[3]

Our knowledge about such shocks has expanded over time, as reflected in the gradual uptake of the multi-hazards and compound events literatures, which aim to describe how multiple drivers or hazards combine and contribute to societal or environmental risk.[4,5] However, multiple challenges remain for this knowledge to translate into practice. For instance, most shock databases suffer from significant limitations: non-standardized spatial and temporal classifications,[6] underreporting of events,[7] and gaps in impact estimates—particularly in low-income countries.[8] Other issues lie for instance in the lack of understanding of causal relationships, or the difficulty in modeling uncertainty, whether that applies to single shocks or shocks' interrelationships.[9] A final challenge remains in the decision-making stage, as most existing approaches to risk management are designed for single shocks and single sectors,[10] thus often failing to grapple with multiple co-occurring shocks. Artificial intelligence (AI) can in this context be a way forward to tackling some issues, but is not without limitations.

## AI for Shocks Data

AI can be used for three distinct data-related processes in the context of interconnected shocks. **First**, AI can improve the reporting of shocks. For instance, convolutional neural processes (ConvNPs) have successfully been used to suggest where to place sensors to measure air temperature anomalies while minimizing uncertainty.[11,12] AI can also be used for extracting shocks data, notably from textual content. Sodoge et al. (2023)[13] for example, applied several natural language processing (NLP) methods to automatize the detection of drought impacts from newspaper articles in Germany between 2000 and 2021. In the same vein, Li et al.[6] created a database of climate extreme impacts (*WikiImpacts*), based on information extraction from textual content from *Wikipedia*, combined with a large language model (LLM)-based text classifier (*ChatGPT 4o*) for information selection. This effort allowed for the assemblage of essential data from over 22,000 disasters from 1900 to the present, and for benchmarking it against existing databases. The results show that extraction accuracy varies for different types of disasters, which suggests that AI is useful for enabling meta-comparison of databases. However, these methods did not outperform previously existing databases and still face coverage biases (e.g., an earthquake's impacts can be under reported or even overlooked in news if it is elections week).

**Second**, AI can help break down broad or complex data into more detailed, local information, so-called downscaling. This is important because shocks like floods, heatwaves, or food shortages often have very different impacts at the local level. By using AI to zoom in on specific regions or communities, researchers can better understand who is most at risk and suggest more accurate responses. Lin et al.[14] exemplified this by downscaling high-resolution daily near-surface meteorological variables over East Asia, using convolutional neural networks (CNNs). Mobin and Kamrujjaman,[15] as another notable example, downscaled epidemiological time series data through their proposed Stochastic Bayesian Downscaling (SBD) algorithm, inspired by machine learning (ML) methods. However, while promising, AI-based downscaling techniques face challenges in ensuring accuracy and consistency across different variables and resolutions, especially in data-scarce environments.[16,17]

**Last**, AI can be used to correct data, for instance by learning systematic biases between forecasts and observations, and correcting the probability distribution. A wide spectrum of approaches coexists, ranging from support vector machine (SVM)[18] to ridge regression[19] and random forest,[20] showing the high flexibility and applicability of AI data correction. However, such methods are mostly used in Earth observation contexts, where ML-based models may introduce statistical biases. Such bias stems from the fact that they are often trained on datasets calibrated with data from resource-rich regions, where the majority of weather stations are located. Another issue is that AI techniques can struggle to capture "extreme" events or shocks (those in the tails of the distribution) because the models are not explicitly penalized for failing to capture outliers.[21] As such, a model can be good at predicting averages but perform poorly at detecting extremes.

## AI-Assisted Predictive Modeling

AI is for now mostly used to model climatic, ecological, and economic shocks, yet in different contexts. For climatic shocks, deep learning (DL) architectures such as transformers, graph neural networks (GNNs), and physics-informed neural networks have a) transformed weather forecasting, and b) enabled the modeling of complex physical systems and long-range dependencies. Prominent DL models are employed across domains such as flood forecasting, air quality prediction, and drought assessment, with GNNs and transformers demonstrating superior performance, scalability, and interpretability in spatiotemporal forecasting tasks.[17,21]

Analyses of ecological shocks also benefit from advances in ML methods for improved modeling. For instance, Ahvo et al. (2023)[22] modeled the effects of agricultural input shocks (e.g., an abrupt export ban on fertilizers that affects the importing country) using a random forest algorithm, with considerable variation between crop–climate bin combinations, but overall good or very good model performance. Modeling contributions also explored the propagation of ecological shocks, notably in disease outbreaks, using DL methodologies like long short-term memory algorithm.[23]

For economic shocks, a strong focus is given to modeling financial shocks, especially through advanced ML techniques such as random forest and extremely randomized trees.[24] However, the usefulness of such techniques depends on the underlying economic theory, as models based on neoclassical economics often overlook financial instability, whereas post-Keynesian approaches offer a more suitable framework for analyzing financial shocks, for instance.[25]

For disaster-related shocks, the literature analysis conducted for this report indicates that natural disaster management studies predominantly rely on classic ML algorithms, such as random forests, SVMs, and ensemble methods. Studies on DL techniques are increasing, including CNNs and recurrent neural models for tasks like flood forecasting, risk mapping, and event detection.

However, most hazard prediction applications remain single-hazard focused, or implicitly incorporate multi-hazard interactions through the integration of diverse static and dynamic variables. In other words, existing approaches focus on the “what,” “when,” and “where” questions, not on the “why,” “what if,” and “how confident.”[21]

## AI-Assisted Decision-Making

AI is rarely used openly in decision-making for interconnected shocks. Although such uses have shown potential to improve real-time responses to extreme weather[26] or to enhance crisis communication,[27] its use in crisis decision-making is limited by institutional and ethical challenges. Indeed, such decision-making processes often happen under high pressure, with incomplete or unreliable information and high levels of uncertainty, and involve difficult moral choices. These conditions make it hard to trust or rely on AI systems.[28] However, AI can still play a valuable supporting role in specific contexts, particularly where cognitive biases, urgency, and data overload might limit human performance. For example, while doctors may not fully trust AI in medicine, it’s used to screen large volumes of medical images to identify potential cancers[29] and prioritize which cases need human review. Similarly, AI-based forecasting models are deployed in response to pandemic outbreaks, before widespread testing becomes available, to guide early containment strategies and resource allocation.[30]

## Potential and Limitations

AI holds promise for shocks-related context, and especially for developing early warning systems (EWSs) for conflicts,[31] ecological shocks,[32,33] and their intersections.[34] EWSs such as the Violence & Impacts Early Warning System (VIEWS)[35] have demonstrated the capacity to forecast political violence up to three years in advance,[35] while AI tools for climate hazard detection now offer probabilistic assessments to support agricultural resilience.[36]

Yet, AI-enabled EWSs are only as reliable as the susceptibility layers that underpin them. Susceptibility captures the conditions that turn a hazard into harm: exposure to the shock, sensitivity of livelihoods and ecosystems, and the capacity of people and institutions to cope or adapt.[3] It is therefore affected by trade and other forms of connectivity that transmit disturbances, unequal access to economic resources and information leading to harm on marginalized groups, and by the capacities of governments to respond. Recent AI pipelines can blend satellite imagery, mobility data, and economic networks to sketch these patterns,[37,38] but the source data still lean toward well-monitored, affluent regions and may reproduce historical social bias.[39]

As a way forward, Reichstein et al. (2025)[40] propose shifting toward causal AI models and decadal early warnings to avoid misleading short-term predictions. They also call for strict adherence to the FATES principles (Fairness, Accountability, Transparency, Ethics, and Sustainability), including open access to training data and source code to ensure replicability. Still, major obstacles remain: institutional inertia, ethical dilemmas, infrastructure limitations, and the broader difficulty of embedding AI analysis into decision-making processes already under stress by urgency, uncertainty, and moral complexity.

THEME BOX 1

# Foundation Modeling in Climate and Sustainability Science

Authors: David Montero, Miguel Mahecha

## What Are Foundation Models?

Foundation models are large-scale AI systems trained on vast datasets to serve as general-purpose tools across multiple tasks. Instead of training separate models for each problem, a foundation model learns broad representations, allowing adaptation to diverse applications (Fig. 5). These models, typically transformer-based, process large volumes of data, capturing patterns that can be fine-tuned for specific tasks.[1]

## How Can Foundation Models Help in Climate and Sustainability?

Climate and sustainability sciences deal with complex, data-intensive challenges that benefit from AI-driven pattern recognition.[2] Foundation models trained on decades of ecological ground data, Earth observation (EO) data from space, and climate records over land, ocean, and in the atmosphere shall generalize across modalities, variables, and scales. This is important, given the coupled dynamics of Earth System processes,[3] for enabling consistent cross-domain AI applications. Examples of foundation model applications for sustainability include: forecasting of multiple variables of interest (e.g., weather-related variables), classification of land-surface properties (e.g., land cover, land use, their change, and objects), environmental disaster detection, and biodiversity monitoring. By leveraging pre-trained knowledge, foundation models require less data and computational effort for specialized tasks.[4]

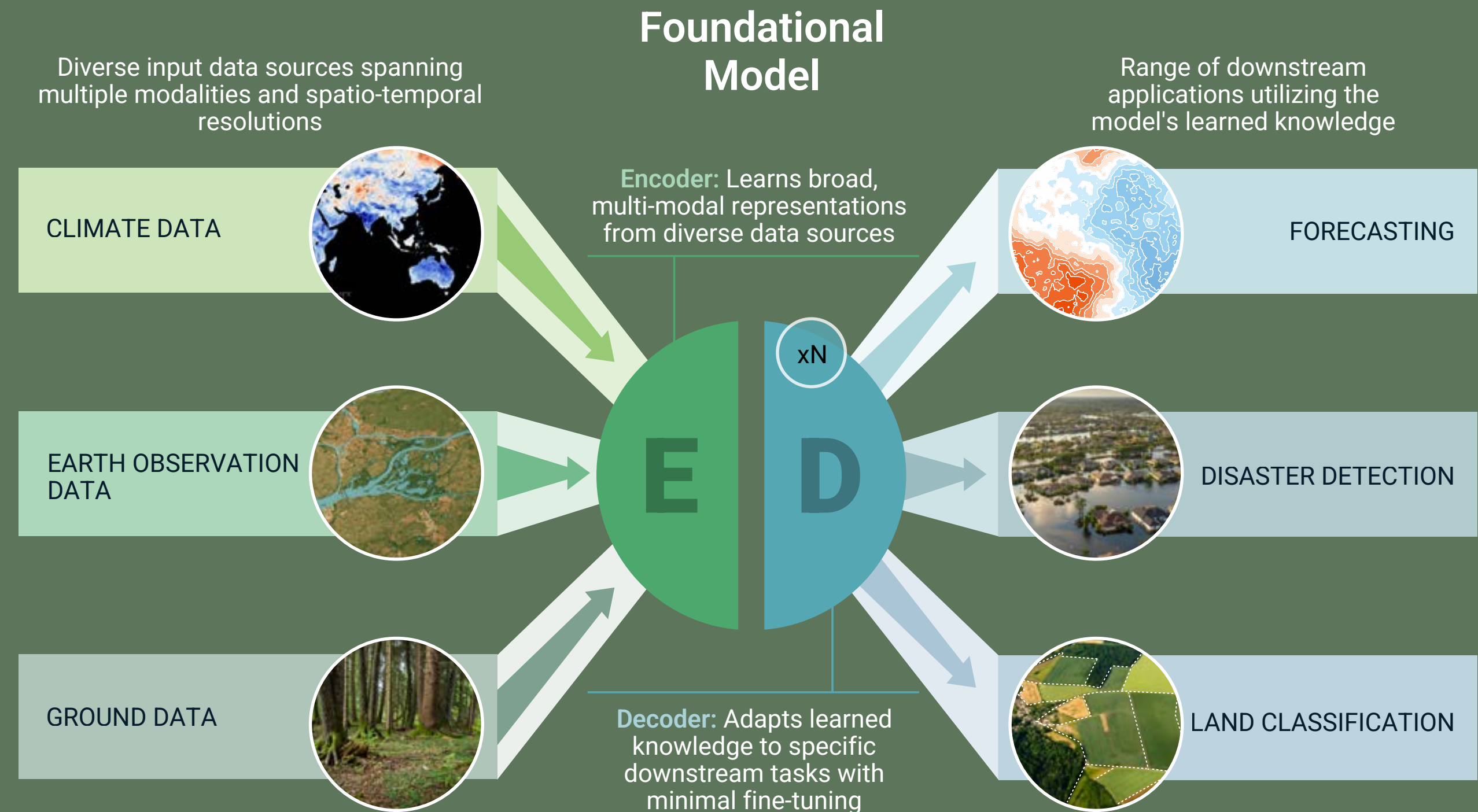

Fig. 5. An overview of foundation models data and examples of outputs.

## Existing Climate and Sustainability Foundation Models

Several foundation models have emerged to support climate and sustainability sciences.[5] In EO, models such as *Clay*,[6] *Prithvi-100M*[7] (IBM and NASA), and *SpectralGPT*[8] leverage state-of-the-art AI models trained on satellite data, including Landsat and Sentinel archives. These models enable applications like land cover classification, disaster monitoring, and environmental change detection. Beyond EO, climate-focused foundation models are also gaining traction. Models such as *ClimaX*[9] (Microsoft), *Aurora*[10] (Microsoft), and *Prithvi WxC*[11] (IBM and NASA) are trained on diverse climate datasets and reanalysis data. Furthermore, the *AlphaEarth* (Google and Google DeepMind) model was trained on a combination of multiple sources, including EO and climate datasets, as well as geophysical parameters and text embeddings.[12] These models support downscaling, climate projections, and probably extreme event forecasting, although the latter poses particular challenges.[13]

## Potential of Foundation Models

Foundation models optimize AI model development through transfer learning, requiring a minimal set of training data for novel tasks. Their ability to capture multi-scale patterns, both spatial and temporal, across Earth Systems allows the unification of traditionally separate tasks, such as climate scenario downscaling and extreme weather forecasting.[11] This capability also extends to bridging gaps between observational data and numerical simulations, allowing more comprehensive Earth System analyses. Additionally, numerical predictions are computationally expensive,[5] and foundation models offer a more efficient alternative by significantly reducing inference time while maintaining competitive accuracy.[4]

## Challenges and Limitations

Despite their promise, foundation models require vast computational resources for training, making them costly to develop. Additional challenges are the heterogeneity and distributed nature of Earth System data. Models trained on limited or unbalanced datasets may struggle to generalize, and purely data-driven approaches risk producing outputs that violate known physical laws. Ensuring physical consistency is particularly important for scientific credibility and decision-making. This can be addressed by incorporating physical knowledge into model design or combining data-driven and physics-based approaches. Data-related biases (e.g., clustered sensor networks for in-situ data or sensor degradations in satellite data) further complicate generalization, especially when models encounter conditions outside their training distribution, such as "record shattering" climate extreme events.[13,14] These risks are amplified in regions with sparse observations or limited data infrastructure, potentially leading to lower model accuracy and under representation. Interpretability remains a concern, as complex deep learning architectures function as black boxes, making it difficult to validate predictions. Bias in training data can also lead to uneven model performance, necessitating careful evaluation and governance.[1]

## Equity and Access Considerations

While foundation models offer powerful capabilities across climate and sustainability sciences, they also present challenges related to access and usability. Running or adapting these models often requires significant computational infrastructure and technical expertise, which may not be readily available to all stakeholders, including smaller research teams, local agencies, or community organizations. Moreover, effectively interpreting foundation model outputs frequently demands domain-specific and AI expertise. Making foundation models more broadly accessible will require efforts such as simplified model interfaces, improved documentation, and dedicated efforts to support users in building the skills needed to apply foundation models effectively.

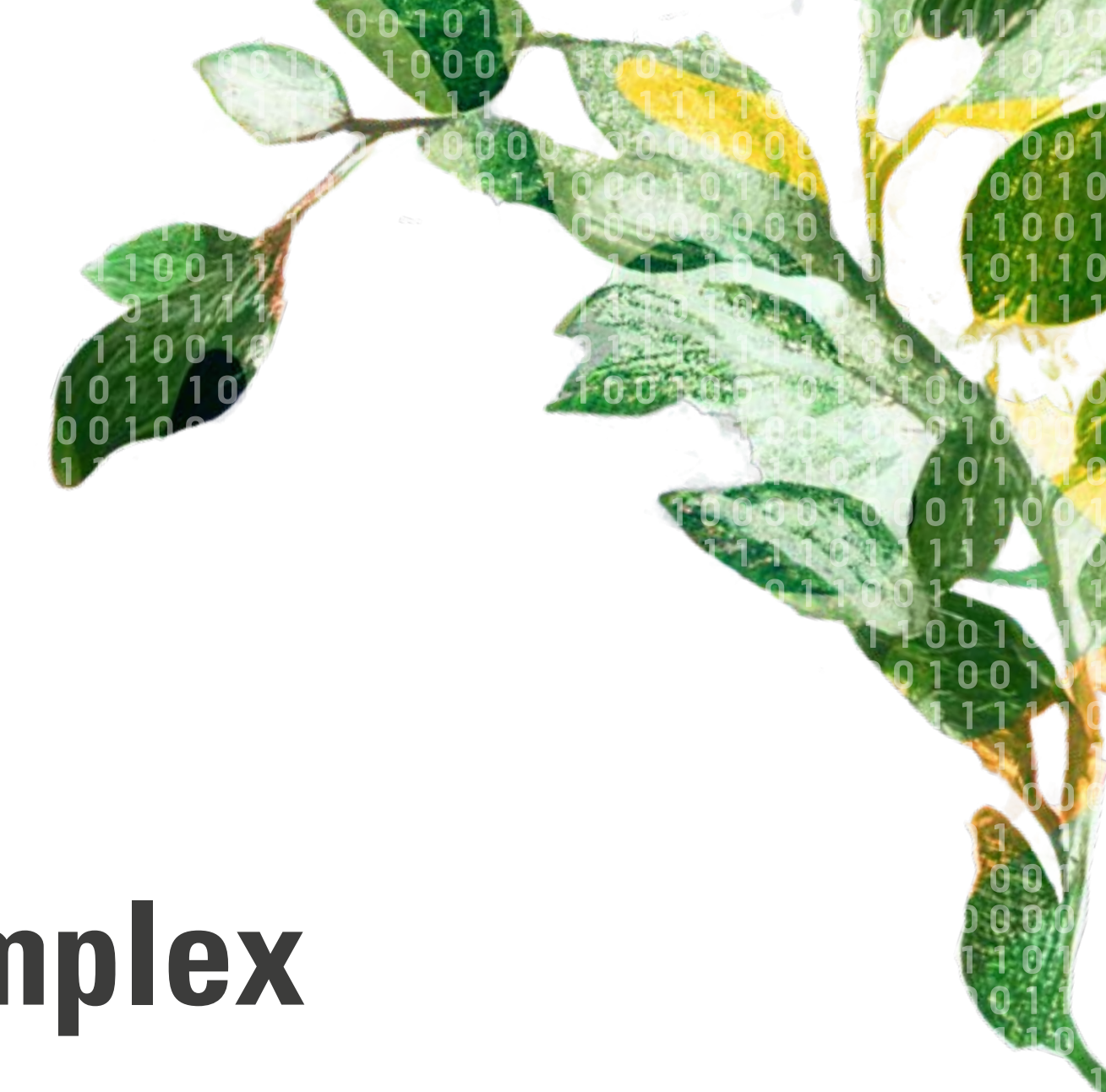

# 2 Understanding a Complex Earth System

Authors: Maximilian Gelbrecht, Ingo Fetzer

*Our planet, and the way it operates through various Earth System processes, is changing rapidly. Can AI help us understand and project these changes in ways that allow us to act in time?*

## Introduction

Earth is a complex and evolving system. It's shaped by the intricate and dynamic interdependencies between its subsystems, such as the atmosphere, biosphere, hydrosphere, and cryosphere, which are all strongly affected by human societies and their interventions.[1] The question posed here is whether artificial intelligence (AI) can play a transformative role in advancing our understanding of the Earth System.

Traditional approaches often treat different environmental and social components in separation. In contrast, AI enables the integration and analysis of a diversity of empirical data streams—ranging from integrating satellite observations and in-situ environmental measurements, to citizen science data, socioeconomic indicators, governance records, and disaster data.

Models trained on these heterogeneous and multimodal datasets can help us explore how human activities such as pollution, land-use change, resource extraction, infrastructure development, or local to global policy decisions influence key Earth System processes like the cycling of carbon, water, nutrients, and biodiversity dynamics. These processes have implications for surface-climate feedback mechanisms, and vice versa. Below, we present some potential AI contributions to address the challenges of Earth System modeling.

## AI for Earth System Data

AI offers opportunities to contribute to complex Earth Systems research in innovative ways:

**AI advances remote sensing analyses of the Earth System.** Over the last decades, satellite-based observations have revolutionized how researchers collect data monitoring Earth and its climate. Especially when combined with in-situ data, for example via machine learning (ML) or model-data integration, novel downstream data products emerge that enable a truly multivariate analysis of global Earth System dynamics.[2] Data over a few decades covers a relatively short time span, especially compared to some slow-moving key Earth System processes, such as soil, ocean, or ice sheet dynamics. Nevertheless, the data offers ample opportunities to increase our understanding of the Earth System. Training AI models on such high-dimensional data can play a crucial part in monitoring changes in data streams in time, including forecasting the complex temporal dynamics of individual Earth System components such as the atmosphere, ocean, and land,[3] or by augmenting spatial resolutions to so-called super resolutions,[4] and thus overcoming scale limitations and effectively increasing the resolution of data and models. Cloud formations are a classic example of phenomena that cannot be predicted by current numeric models at the global scale. The formation of clouds is a localized process, whereas global numeric models have a resolution that is too coarse. It creates extreme computational challenges when attempting to run a global model at such a fine resolution where cloud formations can be predicted.[5] Today, a critical frontier of science is using AI to understand and anticipate the impacts of climate extremes.[6] However, despite all challenges we still face, the increased data collection capabilities contribute to the advancement of predicting spatiotemporal Earth System dynamics, as in the examples provided earlier, and using them as references for process-based models.

**AI provides the capacity to detect new patterns.** One example where AI proved helpful is the transfer from computer vision to analyzing satellite images. These approaches lead to new and efficient solutions for classical remote sensing applications, such as monitoring land cover and land-use change,[7] and analyzing and classifying vegetation types[8] or land surface deformations, such as landslides.[9] In-painting techniques from computer vision (e.g., filling in missing or corrupted parts of an image) can be used to reconstruct missing climate data and improve our understanding of the climate of the past.[10,11]

**AI contributes to understanding interactions and non-linear relationships.** Today, big hopes are placed on interpretable and explainable AI methods (IAI and XAI, respectively) that inspect and analyze trained AI models. The idea is that if we understand what leads an AI model to predict a certain outcome, it would likewise increase our understanding of the often complex and non-linear relationships of the Earth System. Typical applications are identifying dominant driver variables or precursors of events and mechanisms that were previously not considered.[12–14] However, most methods deployed for XAI do not control for confounding factors and can be far from true causal inference. By today, the standard XAI methods remain correlative by nature but have still contributed to improved understanding of several Earth System processes. For example, XAI can help to identify the drivers and impacts of climate change on, for example, carbon dynamics,[15] vegetation,[16] urban temperature distributions,[17] dynamic regimes in ocean modeling,[18] and other domains.

## AI and Predictive Modeling of the Earth System

Earth System models (ESMs) are the primary tools used to investigate the planet's climate and how it is evolving. Predictive ESMs are numerical models that are composed of several different components, each modeling the individual parts of the Earth System (e.g., atmosphere, ocean, land) and their interactions. Traditional ESMs rely on physics-based equations and parameterizations. The parameterizations model processes that are not directly resolved by the physics-based equations, for example because they occur at smaller scales than the resolution of the model represents. However, these equations and parameterizations are computationally very expensive, and need to be improved to reduce uncertainties and biases. Such improvement is particularly important for the representation of extreme events and Earth System components with the potential of reaching tipping points, such as the polar ice sheets, the ocean current system Atlantic Meridional Overturning Circulation (AMOC), or the Amazon rainforest.[19] Potential improvements to ESMs include:

**Enhanced capacities for modeling and analyzing complex Earth System processes.** AI may help reveal previously unknown linkages, cascading effects, and early-warning signals. For instance, ML can help identify cascading interactions between Earth System tipping elements such as the AMOC and the Amazon rainforest in observational data.[20] In doing so, AI strengthens our capacity to anticipate tipping points, for example in the AMOC (Theme box 2, *Using AI to Detect Earth System Tipping Points*), assess sustainability trade-offs, and support decisions and develop strategies that align human development within planetary boundaries.[21]

**Emulating Earth System models using AI.** AI can be used to accelerate existing ESMs. While ESMs, and in particular the correct setting for each parameter within and between each sub-process ESMs simulate, are computational-ly very expensive, AI methods can be executed efficiently on state-of-the-art GPU-supported hardware once they are trained. Therefore, deep neural networks (DNNs) can be trained to be emulators of existing models or of components of existing ESMs. In this case, the DNN is trained to exactly mimic the input–output relationship of the ESM. For example, *ACE2*[22] is an atmospheric model emulator trained on *ERA5* reanalysis data and the *SHieLD* model that shows an almost hundredfold increase in computational speed and decrease in energy usage compared to the process-based model. On the other hand, analogously to super-resolution exercises in remote sensing, DNNs can be integrated into ESMs to emulate subgrid-scale processes, such as those related to cloud physics and precipitation.[23] Emulating these costly subgrid-scale processes can also massively accelerate ESMs at inference. These potential performance benefits of DNNs at inference do however come with the caveat of the expensive training of DNNs. A growing number of foundation models offer new possibilities to analyze a changing Earth System (see Theme box 1, *Foundation Modeling in Climate and Sustainability Science*).

**AI for post-processing and impact assessment.** AI methods are also particularly helpful for post-processing data from ESMs. Generative AI methods inspired by computer vision and image generation have been used for downscaling and bias correction of model output data (e.g., precipitation fields) that are needed for climate impact assessment.[24,25]

**AI in weather prediction.** Purely data-driven DNN models have also made rapid advancement in weather prediction in recent years.[26–28] These models achieve similar accuracy to physics-based weather forecasts with respect to benchmarks,[29] while being computationally more efficient at inference. Studies have also showcased the potential of DNN models for seasonal forecasts up to three months in advance.[30,31] Impactful climate phenomena like the El Niño–Southern Oscillation can even be forecasted up to two years in advance with AI methods.[32,33]

## Limitations

Some key challenges in evaluating complex Earth System research with AI include:

**Challenges for long-term AI climate projections.** While the AI weather prediction systems are remarkable successes, the application of purely data-driven AI models to long-term climate projections is limited by the lack of observational data. Another problem is the intrinsic lack of internal physical consistency of AI-based predictions. High-quality and high-resolution data have only been available for a few decades,[34] which is short compared to the time scales of many natural variability modes of the Earth System. Naturally, there are also no observations of future climate changes from which AI methods could learn. The current generation of AI-based weather models has demonstrated surprisingly strong capabilities in generalizing to previously unseen, warmer climate scenarios. However, these models are also considerably biased toward colder climate conditions in such scenarios. However, they are also considerably biased toward a colder climate in these scenarios.[35] Analogous models for other critical components in the Earth System such as vegetation dynamics and their feedback mechanisms have not yet been developed.

**Hybrid models: merging physics and ML.** Hybrid approaches that combine process-based ESMs with DNNs seem particularly promising for the use of AI in long-term climate projections. The model elements that are based on empirical relations, such as many biological feedbacks, or are computationally extremely expensive, are learned by a deep learning model. Meanwhile, the elements whose physics are already understood by science, such as the dynamical cores of fluid dynamics in atmosphere and ocean models, are inherited from the ESM.[36] Differentiable programming that leads to models that are both objectively calibratable on observational data and in which DNNs can be more easily integrated are a potential perspective for such models.[37] One such differentiable, hybrid model is *NeuralGCM*.[38] It is an atmospheric model that combines physics-based fluid dynamics with an AI parameterization. *NeuralGCM* is competitive to purely physics-based models for weather prediction and can be used for decadal simulations. Hybrid models have both improved stability and short-term predictability compared to purely data-driven DNN approaches.[39] Further work has also shown the potential of hybrid models for the global hydrological cycle[40,41]; for the modeling of turbulence, convection, and radiation in the atmosphere, and of fire dynamics in land models[42]; and for bias correction of sea ice modeling.[43] While a new generation of automatic differentiation tools such as *JAX* and *Enzyme*[44,45] makes writing such hybrid models easier, they do usually need a complete rewriting of the physics-based code, which is a considerable hurdle in their adaptation.

## Potential

Finally, there are many promising sustainability science areas in which AI methods can **augment our understanding of the Earth System**. Some examples include modeling and scenario-building of high-risk phenomena, such as climate tipping dynamics and Earth resilience loss, enhanced Earth monitoring, and incorporating the increasing amount of data that we collect into our models. Physics-based and classical statistical models will remain instrumental in Earth System modeling, especially when making climate projections into future scenarios. AI can augment those models efficiently and leverage available data.

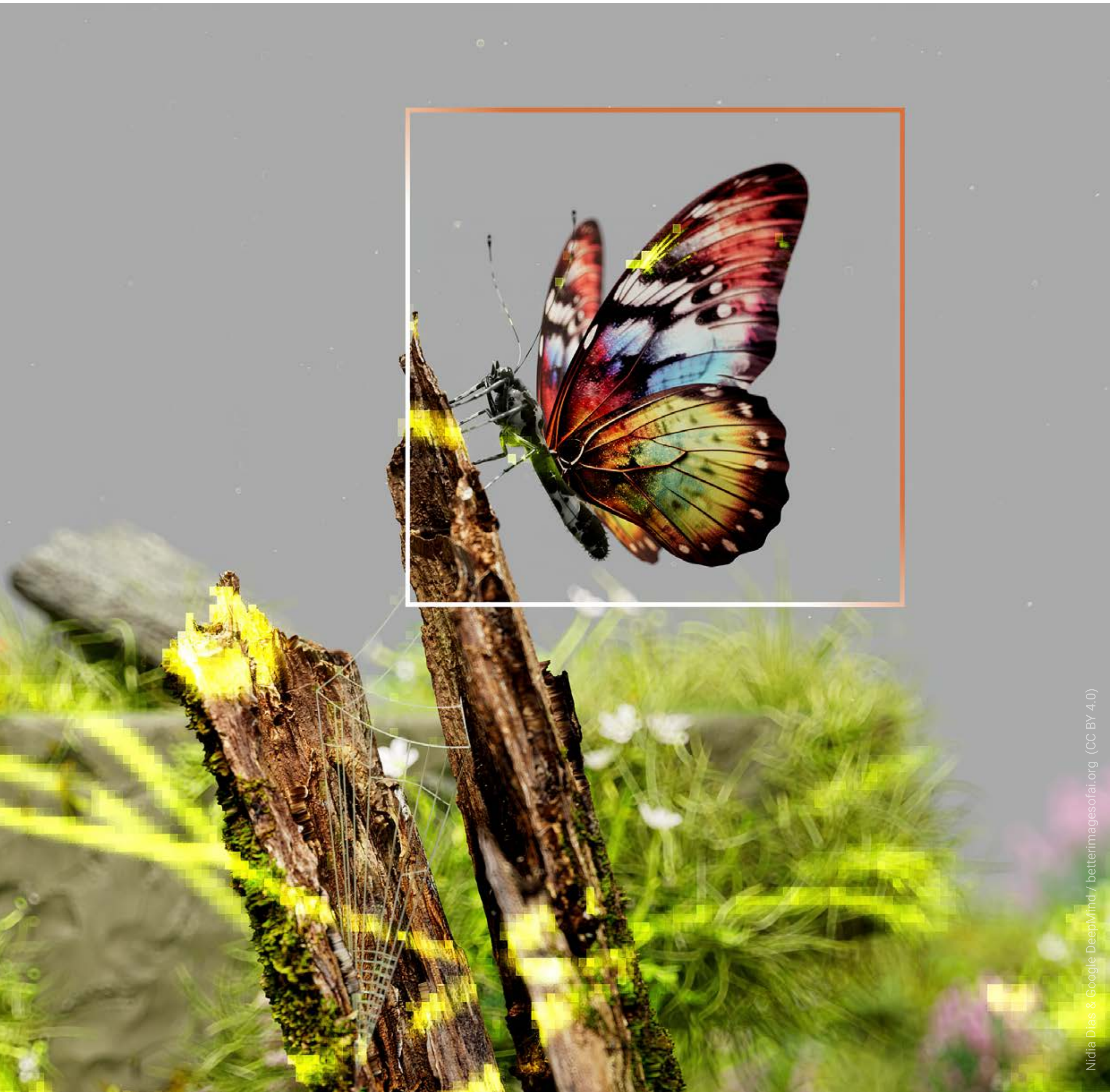

THEME BOX 2

# Using AI to Detect Earth System Tipping Points

Authors: Juan C. Rocha, Jonathan F. Donges, Ingo Fetzer, Maximilian Gelbrecht

Tipping points are defined as critical thresholds in the parameters of a system, transgressions of which can trigger a substantial qualitative change in the behavior of its long-term dynamics, often driven by amplifying feedback mechanisms.[1] Key large-scale examples in the Earth System include the irreversible collapse of major ice sheets, the dieback of the Amazon rainforest into a savannah, and the shutdown of major ocean current systems like the Atlantic Meridional Overturning Circulation (AMOC). In the field of mathematics, these phenomena are known as bifurcations, referring to the characteristics of nonlinear dynamics.[2] Hysteresis, defined as the degree of the reversibility of these shifts, is a sufficient but not necessary condition for tipping points. This limitation is demonstrated by the existence of continuous tipping phenomena, such as pandemics, the disappearance of Arctic sea ice, and certain social diffusion processes. Artificial intelligence (AI) and machine learning (ML) methods are well suited to learn complicated nonlinear functional forms from data. Hence, these methods have been proposed as potential promising future avenues to detect, predict, and anticipate tipping points–through early warning signals–in a variety of systems relevant to sustainability and Earth System science. These include, for example, climatic, biological, social, and social-ecological systems.[3–5] The AMOC is an important part of the Earth's climate system. Mounting scientific evidence shows that this system of ocean currents possesses potential tipping points.[6] These potential tipping points have, in turn, been scrutinized using deep learning (DL),[7] reservoir computing,[8] or ML-based rare event detection techniques.[9]

Initial experiments in detecting and predicting tipping points with ML have mainly used deep neural networks, support vector machines, or random decision forest approaches. The ML techniques have been applied so far mostly on synthetic data from computer simulations, but only scarcely on real-world data.[3–5,10–12] Due to the enormous learning skills of ML, tipping points in artificial data can be successfully detected. However, it should be noted that being based on computer simulations, these approaches are constrained by certain assumptions of an upcoming and anticipated bifurcation, and most significantly about their well-defined and often low dimensionality. These assumptions impose limitations on the applicability of these approaches in real-world systems as it is not known if a bifurcation actually exists and knowledge on their dimensionality is hypothetical.[13]

Recent efforts have attempted to incorporate novel approaches such as deep neural networks for higher-dimensional bifurcations.[14] Efforts have also been made to include a more diverse array of pathways, more likely leading toward bifurcation, such as rate-induced or noise-induced tipping.[15] Even more recent efforts have focused on the prediction of real observed abrupt transitions from observational data with promising outcomes for future research.[16] One key challenge in implementing and scaling up these ML methods for the prediction of tipping points is the scarcity of observed annotated datasets from real systems on which the models can be trained and evaluated. Consequently, until such datasets become available, the evaluation of precision and reduction of uncertainty will remain very much constrained.

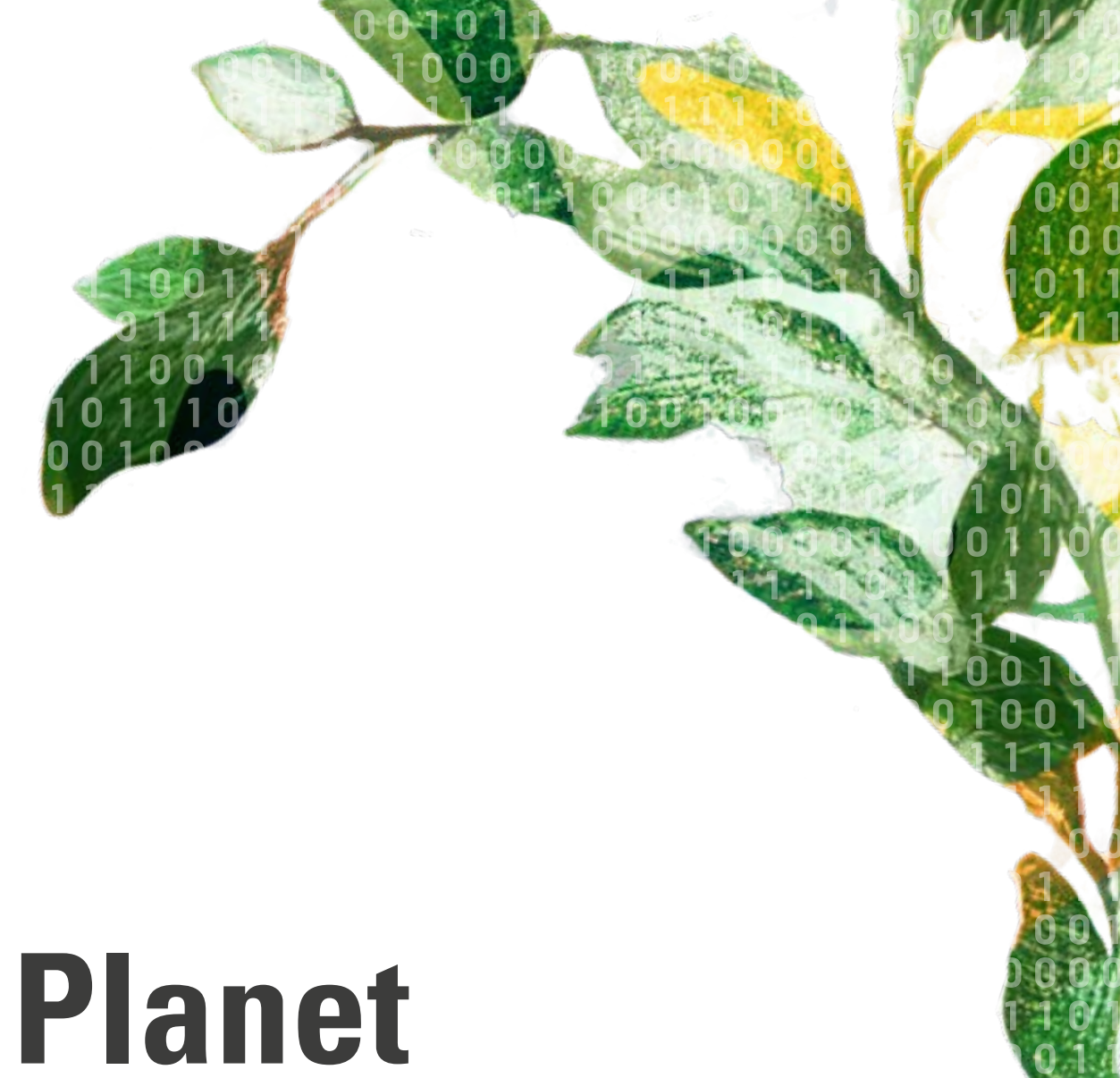

# 3 Stewarding Our Blue Planet

Authors: Erik Zhivkoplias, Alice Vadrot, Jonas Hentati Sundberg

*The ocean transcends national boundaries as a shared global common, where collective responsibility is challenged by fragmented jurisdiction. It requires unprecedented international cooperation for its stewardship. Can AI help navigate the complex challenges of the ocean?*

## Introduction

The ocean, a vast interconnected system fundamental to planetary habitability, regulates climate by absorbing anthropogenic $CO_2$ and over 90% of excess heat.[1] It sustains crucial biodiversity and food webs underpinning global ecosystems and resources. However, escalating anthropogenic pressures, overexploitation, pollution, habitat degradation, and climate change impacts demand effective governance through robust sustainable and equitable management frameworks.

Sustainable ocean governance is complicated by a fragmented legal landscape and diverse international agreements and management efforts. Key measures include mitigating climate change impacts (e.g., by improving energy efficiency, promoting sustainable transport, and protecting and restoring blue carbon ecosystems), establishing marine protected areas, implementing science-based fisheries management, and tackling pollution.[2] However, persistent gaps in scientific research and monitoring, and inconsistent data, hinder evidence-based decision-making. Traditional monitoring methods are often resource-intensive and limited in scope, impeding accurate ecosystem assessment and adaptive policy development.[3]

Artificial intelligence (AI), encompassing machine learning (ML), deep learning (DL), and advanced analytics, emerges as a transformative tool to bridge these gaps. It can facilitate large-scale, real-time surveillance of marine activities, automate the identification of marine species and threats, improve predictive modeling of oceanographic changes, and advance data processing efficiency.[4] This technological advancement can provide critical insights for informed, adaptive ocean governance, supporting the implementation of international agreements. Nevertheless, AI integration poses challenges, including data representativeness, algorithmic biases that may exacerbate existing inequalities, and ocean justice concerns.

This chapter illustrates AI applications in ocean governance, conservation, and area-based management for understanding dynamics and predicting change. Several applications that focus on optimizing commercial operations, infrastructure, and engineering efficiency (e.g., port operations and logistics, engineering, and offshore infrastructure) fall beyond this chapter's scope.

## AI for Data Collection, Monitoring, and Surveillance

Addressing the ocean's persistent monitoring deficit requires novel, AI-driven approaches that can complement traditional resource-in-tensive methods. Autonomous sensor-equipped platforms can use onboard analytics power to optimize survey routes and extract features from data in real time, potentially informing real-time ecosystem management.[5,6] Such platforms have fundamentally expanded our capacity to collect and analyze data by integrating a broad variety of sensors for detecting biological and non-biological scatterers in the water, and above-water and underwater imagery including hyperspectral imaging for water quality assessment. Many of these data streams are suitable for developing automated analysis pipelines around, using ML and DL approaches for real-time feature identification:

**Marine acoustics is a research area with large potential** for automation and real-time feature extractions. It includes both passive acoustics (with autonomous recorder units capturing impulsive underwater sound signals) and active acoustics (devices emitting sound waves and recording the returned echoes, i.e., echo sounders). These complex acoustic datasets can be processed with various ML methods to, for example, distinguish seabed types and structures,[7] identify fish schools,[8] detect hydrothermal emissions, and monitor small whales based on their emitted contact sounds.[9] This use of AI in analyzing marine soundscapes not only advances mapping and classification, but can also help detect ecological changes and anthropogenic impacts that might not be visible using optical techniques.

**Big data is starting to play a major role in imagery and video data.** Underwater imagery,[10] as well as above-water imagery of marine organisms such as seabirds[11] and marine mammals,[12] is increasingly collected using automated systems, and analyzed using DL frameworks enabling large-scale, automated identification of marine species. Platforms such as *FathomNet*[10] leverage DL to analyze millions of underwater images and recordings, dramatically accelerating species classification and population tracking. Advanced ML applications are also starting to expand beyond simple classification tasks toward identifying behavior, growth, and individual performance, which is often highly informative for revealing population status and threats to marine animal populations.[13,14]

**Such automation of data collection and feature extraction can aid biodiversity monitoring.** Emerging threats to marine biodiversity, especially in remote or under-observed regions, are increasingly studied using a combination of remote sensing and sequencing data.[15] Remote sensing technologies allow near real-time tracking of changes in large, inaccessible marine areas,[16] while tagging and in-situ fieldwork support a robust picture of ecosystem health.[17] Meanwhile, environmental DNA analyses allow ML models to detect rare or endangered species from trace genetic material in ocean samples,[18] uncovering microbial diversity and revealing how different organisms contribute to nutrient cycling and ecosystem resilience. This integration of remote and autonomous sensing, acoustic monitoring, and advanced analytics is fundamentally redefining our ability to assess ecosystem changes by delivering comprehensive, near-real-time insights that support both scientific discovery and adaptive, evidence-based ocean stewardship.

## AI-Assisted Predictive Modeling

Predictive modeling plays an increasingly vital role in helping scientists and policymakers anticipate oceanic responses to accelerating anthropogenic pressures. However, the complexity of marine systems, marked by feedback loops, threshold-driven events, and persistent data scarcity, often limits the capacity of traditional modeling frameworks. Advances in AI are beginning to address some of these challenges by enabling integration of multimodal data streams and supporting higher-resolution forecasts.

**DL methods are advancing the detection and classification of marine pollutants.** For example, *U-Net* neural networks trained on Sentinel-2 satellite imagery demonstrate great potential for large-scale identification of floating marine plastics, providing a scalable monitoring framework aimed toward automated global plastic tracking systems.[19] Similarly, integrating hyperspectral sensors with ensemble algorithms enhances the monitoring of oil spills and other contaminants. When combined with ocean current modeling, these approaches help anticipate pollutant drift and accumulation, enabling more targeted cleanup efforts. Emerging evidence shows that AI algorithms (e.g., ML, DL), along with big data analytics, Internet of Things (IoT)-enabled sensors, and smart sensor technologies, can improve environmental monitoring, disease and feed management, production optimization, and traceability in fisheries and aquaculture. The combination of technologies can thus support sustainability, waste reduction, and increased supply chain transparency.[20]

**The application of AI in modeling global marine changes is advancing**. Convolutional neural networks (CNNs) help correct systematic biases in sea ice projections—crucial indicators of climate-driven ocean changes—while physics-informed neural networks trained on long-term hydrographic and turbulence observations more accurately represent small-scale ocean processes such as vertical mixing.[21,22] These models, which integrate data-driven insights with physical constraints, not only outperform traditional parameterizations, yielding more realistic predictions of ocean temperature and heat fluxes in both stand-alone and coupled climate models, but also improve projections of phenomena like marine heatwaves and ocean acidification.

AI-driven analysis is also **expanding our understanding of the interplay between human activities and marine environments**. Notably, ML models trained on global datasets of daily ship movements have revealed fundamental patterns in maritime trade evolution, enabling accurate forecasts of shipping routes and trade flows.[23] These predictive models might improve our ability to anticipate the impact of maritime logistics on ecological systems and climate change, potentially supporting more informed sustainability planning. Another example is the global surveillance of fishing activities.[24]

AI is also being used to **model changes in marine biodiversity**, for instance by predicting coral bleaching through analysis of thermal stress, nutrient levels, and ocean chemistry. In blue carbon ecosystems, ML algorithms like random forest and *XGBoost* are used to synthesize light detection and ranging (LiDAR), remote sensing-, and field data, resulting in more accurate estimates of carbon stocks across mangroves and seagrasses.[25] The use of cloud computing and high-performance hardware further enables the processing of these increasingly complex, multi-source datasets.

## AI-Assisted Decision-Making

AI is increasingly supporting decision-making in ocean governance, particularly in fields such as marine protected area (MPA) management, maritime spatial planning (MSP), and combating illegal, unreported, and unregulated (IUU) fishing. These applications are grounded in the integration of diverse, spatially explicit ecological and operational datasets, with the goal of informing management that balances conservation needs and human activities. Several important application areas and contributions can be identified:

For MSP, **coupling Bayesian networks with GIS maps enables multi-scenario assessments** of cumulative impacts under various climate and management conditions. The coupling also allows the identification of key pressure-driving impacts and potential protective measures to reduce environmental vulnerability.[26] In the European context, a case study was recently conducted on the Italian region and emerging AI hub of Emilia-Romagna. The case study demonstrates how integrated AI-supported MSP platforms, which combine multiple geospatial decision support tools with extensive ecological and socioeconomic datasets, can enable transboundary, ecosystem-based MSP. The MSP platforms do so by providing rapid, spatially explicit outputs for scenario analysis.[27] Globally, AI can help identify high-priority areas for MPA establishment by analyzing integrated global datasets on species diversity, habitat heterogeneity, benthic features, productivity, and human activities such as fishing effort. AI can also support the implementation of the UN's Biodiversity Beyond National Jurisdiction Agreement (BBNJ)—the key multilateral framework for high seas conservation.[28]

**AI contributes to the detection and analysis of IUU fishing**. CNNs, applied to satellite imagery (including synthetic aperture radar, vessel monitoring systems, and automatic identification system data), have enabled more accurate and persistent monitoring of vessel activity on the open ocean. In particular, these methods can help to identify vessels operating without tracking signals ("dark vessels"), recognize atypical fishing patterns, and estimate unreported fishing effort.[29,30] Identifying such deviations from normal vessel behavior can also support enforcement efforts by authorities.[31] Combining fishing activity data with satellite altimetry, this approach can also reveal fine-scale fishing activity patterns linked to ecological drivers, facilitating sustainable management and the protection of marine biodiversity.[32]

## Limitations and Key Challenges

Oceans remain critically understudied and "under-governed," with 64% of waters beyond national jurisdiction and deep-sea ecosystems largely unexplored.[6,33] This reality intensifies three interconnected challenges that constrain AI's transformative potential and risk perpetuating existing inequities:

**Persistent data gaps** form the first challenge. AI's performance depends on abundant, high-quality data, yet ocean observations remain sparse and fragmented. The Intergovernmental Oceanographic Commission of UNESCO (IOC-UNESCO) estimates that more than 90% of marine ecosystems lack sufficient data for effective management,[34] which hinders progress toward global targets such as the Convention on Biological Diversity's goal to protect 30% of the ocean by 2030—a target aimed at conserving marine biodiversity and ensuring the sustainable use of ocean resources.[35]

These issues stem from deep-seated **inequalities in scientific capacity between the Global North and the Majority World**, which manifest in disparities in funding, access to research infrastructures, thematic priorities, and influence over global ocean governance discourses.[36,37] These data coverage and scientific capacity gaps converge on issues of ocean justice. Without dedicated capacity building aimed at more inclusive, diverse, and equitable exploration and exploitation practices,[38] AI models risk producing inaccurate or unjust outcomes due to bias in training data. Algorithmic biases emerging from unrepresentative training data can skew conservation priorities away from vulnerable regions or communities. Making private data available to coastal states that cannot afford to buy or collect it themselves can represent a first step in that direction.[39]

Developing adaptive, multi-scalar governance regimes will be essential to harness AI's potential for addressing the "wicked problems" of the global ocean and the challenge of scaling up the potential of AI within ocean governance frameworks.[4] AI makes it possible to monitor ecosystems faster than ever before, but it also brings new challenges. The timing and scale of ecological data don't always match up with what's practical or realistic for social, economic, or political decisions. Bridging that gap will be key if policymakers are going to make real use of AI.

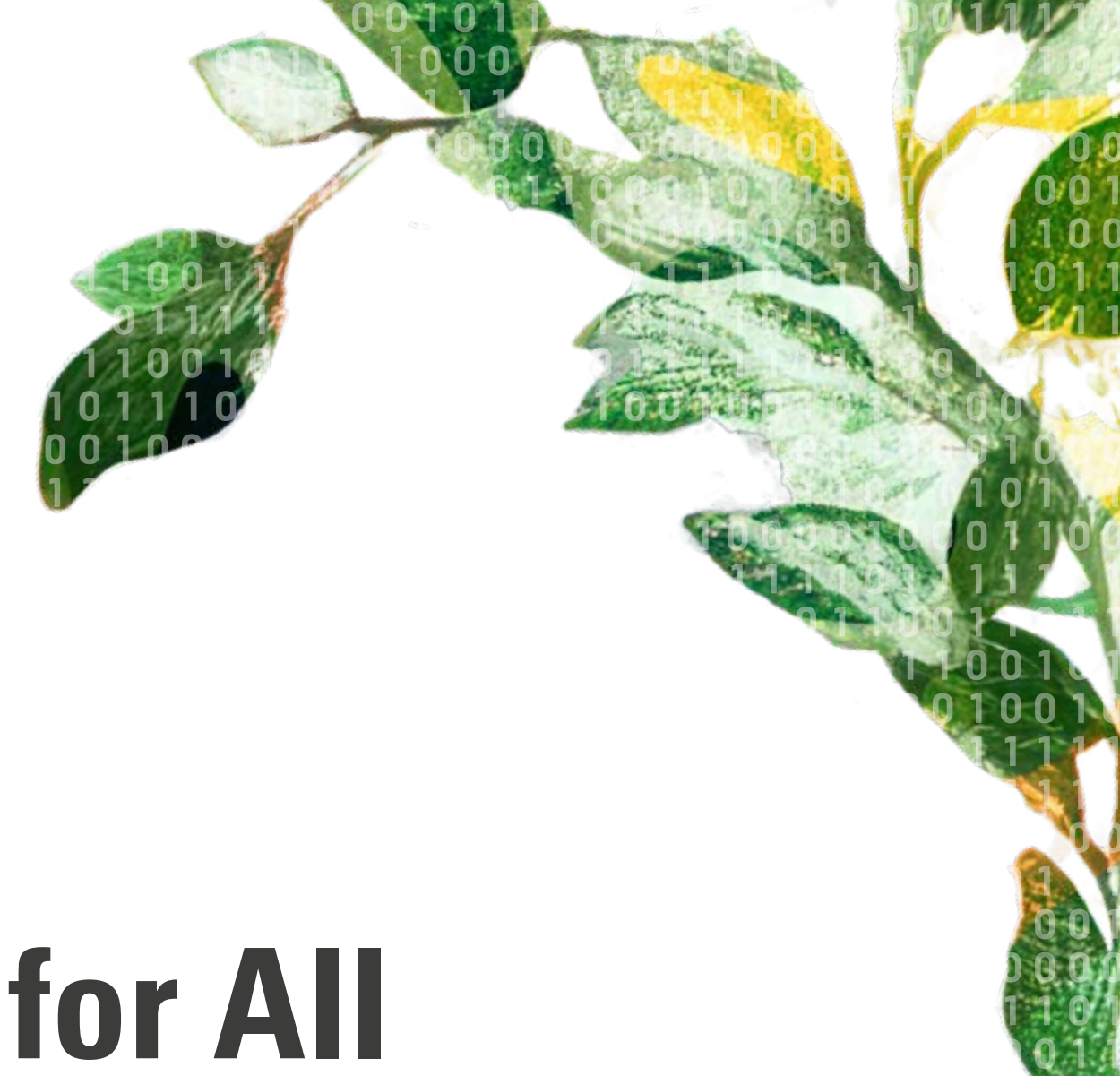

# 4 Securing Freshwater for All

Authors: Lan Wang-Erlandsson, Nielja Knecht, Romi Lotcheris, Ingo Fetzer

*Fresh water, the bloodstream of the biosphere, is fundamental for all life on land and all societal functions. Can AI be used to map, analyze, and protect freshwater flows on a changing planet?*

## Introduction

Freshwater—"the bloodstream of the biosphere"—is fundamental for all life on land and all societal functions. The availability and dynamics of freshwater play a critical role in the regulation of Earth's climate system and, at the same time, are among the most visible manifestations of climate change. Droughts and extreme precipitation events are becoming more severe, intense, prolonged, and abrupt, resulting in costly damages to agriculture, infrastructure, and livelihoods, with implications for migration patterns, conflicts, and security. Shifting hydroclimate predisposes ecosystems to regime shifts, thereby increasing the risk of disease outbreaks. Four billion people are estimated to face severe water scarcity conditions at least one month per year.[1] Freshwater flows in the landscape and across the atmosphere do not respect administrative borders, complicating decision-making.

Hence, securing freshwater for all demands a comprehensive understanding of complex system dynamics spanning diverse scales and sectors, which necessitate high-quality data and modeling, as well as just and effective decision-making processes. Conventional approaches to supporting freshwater sustainability are hampered by data scarcity in many regions, slow model development, and the inherent complexity of decision processes.[2,3] Artificial intelligence (AI) can accelerate progress by bridging data limitations, enhancing model accuracy and speed, and facilitating informed decision-making. While AI can play a transformative role in strengthening freshwater's role for resilience and sustainability, proactive consideration of emerging ethical and governance challenges that arise with its application is crucial for ensuring equitable outcomes.

In this chapter, we illustrate some applications of AI in agriculture, including disaster reduction, water resources management for forecasting, understanding water cycle dynamics, and supporting flood and drought prediction. In all these areas, the application of AI has demonstrated usefulness and continues to develop rapidly. The ethical and governance challenges that arise from the use of AI in freshwater management will also be briefly discussed.

Several key aspects of water sustainability, beyond the scope of this chapter, include AI applications for ensuring access to clean water and sanitation, addressing water quality and aquatic ecosystem health, water in industries and energy production, and understanding water-induced collapse and tipping points.

## AI for Freshwater Data

Informed decisions and management of water resources require accurate water-relevant data at the proper resolution. As several studies have demonstrated, AI can have a transformative impact on water sustainability in multiple ways, including:

**Counteracting geographical bias.** Ground-based data collection of freshwater-related variables is unevenly distributed across the world, which has detrimental implications for security in some of the world's most socioeconomically vulnerable regions. Many regions in sub-Saharan Africa, parts of Asia, and South America lack comprehensive monitoring networks for basic hydrological and hydroclimatic data, including precipitation, evaporation, and runoff. For example, a lack of warning systems and preparedness contributed to making flood events in Africa four times more deadly than in Europe or the US in recent years (2020–2022).[4] AI methods, particularly supervised and semi-supervised machine learning (ML) approaches, offer diverse opportunities to contribute to water-related data generation and collection, and can help mitigate global sampling biases.[5] Because differences in data formats and standards can make it challenging to combine datasets from different regions, AI-based approaches can complement traditional statistical techniques (e.g., kriging, a method for spatial interpolation) to improve data integration.[6] Verification methods that combine satellite observations with local ground measurements can also enhance data quality for AI-based hydrological models, especially in regions with limited data coverage.[7,8]

**Increasing the quality and speed of data analyses.** A lack of freshwater-related data (i.e., accurate, continuous, and with high resolution) hinders the monitoring, assessment, and prevention of unsustainable water use and other adverse human impacts. Reanalysis datasets of hydroclimatic variables combine observations with model outputs to create consistent, globally gridded time-series data. They are useful for regional and global applications but often lack sufficient granularity for local applications.[9] Remote sensing provides high-resolution observational data but remains limited by uncertainty. This is especially true for variables such as evaporation, which are often indirectly derived and rely on further assumptions. Recent efforts have shown promise in addressing these limitations with AI techniques. For example, deep neural networks have been used to improve the accuracy and reliability of precipitation data from reanalysis products,[10] and to map and classify surface water bodies such as lakes, wetlands, and reservoirs from remote sensing data.[11,12] Around a third of remote sensing-based studies on surface water detection and delineation used AI.[13] However, many hydrologically relevant variables, such as soil moisture and evapotranspiration, are not readily measurable from optical satellite data and require calibration with ground measurements.[5] Combining networks of ground measurement stations with remote sensing data and supervised ML techniques allows for the generation and extension of high-resolution, spatially and temporally continuous datasets for these variables.[14] Further, AI can also be essential in the timely delivery of information for emergency responses, for example through fully automatic detection of flood extents from satellite imagery.[15,16]

**Enabling the collection of new types of data.** Social media and citizen science measurement networks provide additional, novel data sources, particularly for real-time assessments and early warnings of risks such as floods.[17,18] Some recent research has focused on developing AI-based early warnings of flood impacts using text posts from social networks,[19–22] but given the challenges in accessibility of mostly privately held data and often a lack of georeferencing, such applications have not been widely employed yet. Supervised ML models such as deep neural networks in combination with remote sensing data also allow for detecting and mapping hydrologically relevant data that are otherwise not easily established, such as different irrigation schemes on agricultural lands.[23]

## AI-Assisted Predictive Modeling

Modeling is crucial for securing freshwater sustainability. Models (i.e., conceptual models, mechanistic models, process-based general circulation models) can help test hypotheses and support a better understanding of hydrological dynamics and processes of varying complexity. Operational models can, as an additional example, be used for forecasting and early warning of water-related disasters (such as drought and flood prediction). AI can help to improve predictions and understanding gained from freshwater modeling efforts by enhancing model components (e.g., parameterization, equations) or simulation outputs (post-processing of model outputs), allowing for alternative modes of modeling that are computationally less expensive and/or more accurate (e.g., data-driven, surrogate, and hybrid models). Some of the important contributions of predictive modeling for securing freshwater provisioning are presented in more detail here:

**Enhancing model parameters, equations, and outputs.** Key applications of AI in this context include identifying optimal parameter values for specific subprocesses in process-based models and developing context-specific bias correction algorithms for predicting hydroclimatic variables.[24] For advancing basic scientific inquiry, deep learning (DL) may be leveraged to enable data-driven equation discovery.[25] AI approaches have been utilized for the spatiotemporal downscaling of coarse outputs from process-based models,[26] for example using neural networks to increase the spatial resolution of precipitation predictions by a factor of up to 25.[27] This generates higher-resolution and regionally accurate predictions of the key climatic and hydrological processes, which are necessary for local management plans.

**Diversifying alternative modeling approaches.** In data-driven modeling, AI is used to directly derive relationships based on empirical data, thereby better accounting for complexity not captured in models,[28] or to enhance model ensemble data when there are limited observations, thereby saving computational time.[29] AI-based surrogate models are trained on the input and output parameters of process-based models, learning to provide faster and less computationally expensive predictions of expected climatic or hydrological conditions under specific scenarios. They are particularly useful for assessing model sensitivity, as well as the risks associated with a wide range of possible scenarios.[30] In hybrid modeling approaches, AI algorithms have been integrated and combined with process-based models in various ways to enhance their performance, interpretability, and applicability for decision-making. One common strategy is to use AI-based post-processing of hydrological model output to better capture local ground truth conditions. Generally, the process-based submodels are useful for well-understood physical processes, while AI methods are employed for fine-tuning and representing complex and poorly described subprocesses.[31] Such approaches have proven valuable in enhancing flood prediction,[32] groundwater simulations,[33] and drought forecasting.[34]

## AI-Assisted Decision-Making

AI can be used throughout the entire decision-making chain, from data retrieval and sense-making to facilitating discussions on risks.[35] High speed and accuracy in forecasting are critical for saving lives in the context of flood and drought disasters. AI methods, such as the coupling of artificial neural networks, the Bayesian framework, and genetic algorithms, can also directly assist decision-making for water sustainability by improving the use and allocation of water resources, forecasting the consumption and demand of water resources,[36] facilitating communication and dialogues, and enhancing the understanding of risks and uncertainty. Among others, AI approaches such as ML, artificial neuro-genetic networks, and AI-powered drones and satellites have been used to forecast, monitor, and optimize irrigation in agriculture, reservoirs and dams management, and industrial and household water use.[37–39]

Moreover, communication and dialogue are essential in facilitating decision-making in multi-stakeholder contexts, which water decisions inevitably are. Large language models (LLMs) can, for example, be used as moderators in meeting contexts to help navigate complex issues and/or provide information that can be perceived more neutrally in contexts with diverse views and needs, such as urban planning.[40] Furthermore, AI can also aid in understanding risks and uncertainty by simulating multiple scenarios, analyzing larger sets of data, and generating visualizations. For example, *WaterGPT* is an LLM designed to act as a hydrology expert by processing and analyzing images and text, and answering questions in natural language.[41] General-purpose multimodal LLMs such as *GPT-4 Vision*, *Gemini*, etc., showed potential in various hydrological applications.[42]

## Limitations and Key Challenges

The full transformative potential of AI for securing freshwater for both people and the planet remains to be explored, and key limitations and challenges need to be addressed. Critically, AI needs to be more understandable by users, and its continued application should pay close attention to ethical implications, continued investments in ground measurements, and the considerable water requirements of the data centers used to power AI. Some of the more pressing concerns include:

**Explainability and transparency.** AI models, especially DL algorithms, are highly complex and opaque. DL models often act as "black boxes," making it difficult to understand the causal mechanisms behind the results and build trust among decision-makers, especially since the academic experts developing these models may not be able to explain outcomes. Understanding, transparency, and trust are particularly relevant in water management, where decisions impact resource allocation, flood prevention, and climate response strategies. Developing models with greater transparency, such as decision trees[43] or *SHapley Additive exPlanations* (SHAP),[44] can provide clear insights that help water managers understand key factors driving hydrological variability.

**Access, fairness, and other ethical concerns.** The computational demands of AI models also present a challenge. Training and optimizing complex models requires substantial computing power, particularly for large-scale simulations and high-resolution predictions.[44] This can pose challenges for the use and development of AI in resource-limited settings. Ensuring that AI technologies are accessible to all regions, developed in collaboration with local agencies, and tailored to local use cases, particularly in low-income countries, is crucial to prevent widening disparities in water resource management.[45,46] Stakeholders and policymakers also face challenges when adopting AI tools in water research. A lack of technical expertise can hinder understanding and adoption of AI models, particularly for non-specialist decision-makers. Furthermore, AI models may produce results that conflict with traditional knowledge systems, creating skepticism and reducing trust. Ensuring that AI tools follow principles of trustworthy AI,[47] and are accompanied by clear documentation on how they can be used, simple user interfaces, and precise documentation of trained resources can improve stakeholder engagement and confidence.[48,49]

**Fundamental lack of observations.** Fundamentally, AI relies on data inputs, and a key challenge is the lack of observations. Existing biases in data collection are particularly problematic for addressing water issues, as water processes and dynamics are highly heterogeneous and context-dependent.[38] A key challenge lies in the availability of evenly distributed, high-quality, and unbiased data. For most regions, hydrological data are often incomplete in both space and time, can be inconsistent in their measurement, and are lacking in regions with limited monitoring capabilities. Such gaps in data reduce the development and reliability of AI-based models.[50]

**Negative impacts of AI usage on water sustainability.** While AI can be used for good, the explosive growth in AI use by a variety of users has also necessitated a significant increase in data centers that require water for cooling. Training the *GPT-3* language model has been estimated to lead to the evaporation of 700,000 liters of clean freshwater, and globally, AI is projected to lead to 4.2–6.6 billion $m^3$ of water withdrawal by 2027.[51] Such estimates are highly uncertain due to a lack of transparency from the tech companies that train AI, and this uncertainty can hopefully be lessened with improvements in computationally efficient algorithms, increased public awareness, and legislation.[52]

Addressing these challenges will require **collaboration across various disciplines**, from academic developers to applying stakeholders, as well as improved openness to data sharing from both sides. This will also involve the development of transparent AI models that align with the complexities of hydrological systems.[49]

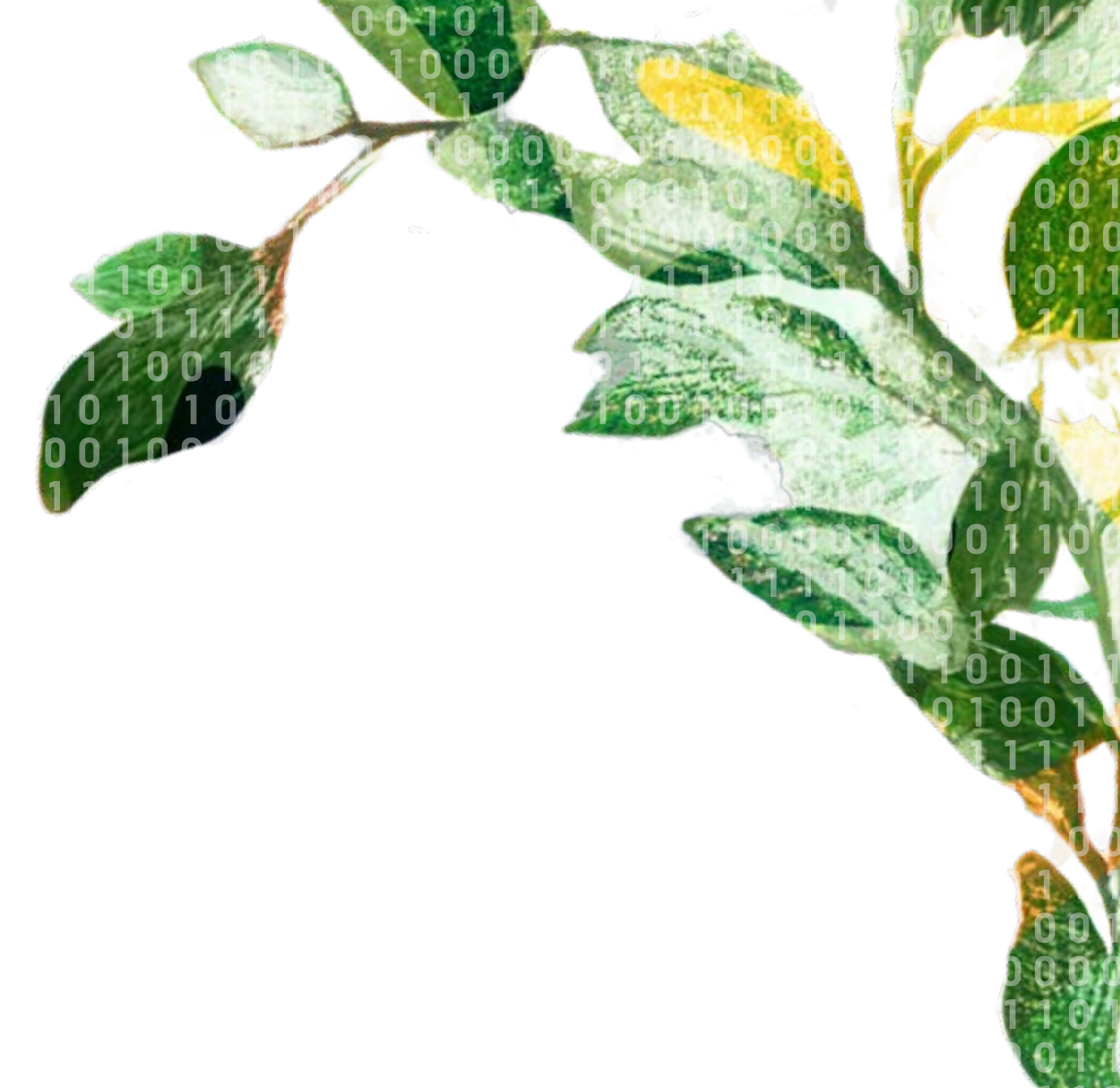

# 5 Enhancing Nature's Contributions to People

Authors: Maria Schewenius, Ingo Fetzer, Miguel Mahecha, Erik Zhivkoplias, Drew Purves

*Nature and the functions it performs are essential for planetary stability and prosperity, including human health and well-being—indeed, even our survival. Here, we explore how artificial intelligence can support the provisioning of benefits humans derive from ecosystems and biodiversity.*

## Introduction

The well-being and indeed survival of humans depend on the functions of the biosphere and the benefits it provides, i.e., ecosystem services.[1] The biosphere provides a diverse array of ecosystem services, including regulatory (for example, temperature regulation, clean drinking water, or pollination), provisioning (for example, food and fiber), cultural (for example, recreational, aesthetic, and spiritual values), and supporting (for example, photosynthesis, carbon sequestration, or soil renewal).[2]

The newer concept of nature's contributions to people (NCP) emphasizes the role of different cultures and local knowledge sources (such as local indigenous knowledge) for influencing the perception of ecosystems and contributing to shape the provisioning of services, or NCP.[3,4] However, the provisioning of ecosystem services around the world has decreased since the 1970s.[2] A continued loss, and potential widespread ecosystem collapse, is estimated to amount to a devastating annual cost of 2.3% of global GDP in monetary value alone.[5] Low- and lower-middle-income countries are particularly at risk, with an estimated potential 10% loss of annual national GDP by 2030.[5]

Biodiversity is fundamental for shaping the functions of an ecosystem and its capabilities of providing NCP. Biodiversity is a broad concept referring to the variability of, for example, genes, species, functions performed by groups of species (such as coral grazing or soil renewal), and interactions (such as pollination, seed dispersal, and pest control) in ecosystems.[2,6–8] However, biodiversity is decreasing on a global scale, at a rate and extent that has been referred to as the sixth mass extinction.[9,10] This decline extends beyond a loss of species and functions to also include a pervasive erosion of genetic diversity within species.[11–13]

The human realm is intricately intertwined with the natural world. Biodiversity status and NCP provisioning, or losses, are part of complex, linked social-ecological-technological systems[14] that also include economic dimensions, operating from the local to global scale.[15] These dynamic and interconnected systems exhibit varied responses to change, yielding both social and ecological repercussions.[16] For example, mining is a form of resource exploitation and a contributor to land-use change and biodiversity decline. However, the AI hardware industry, for example, relies on mining for provisioning of critical minerals such as silicon (which forms the basis of microchips) and boron (a rare earth mineral used for example in logic and memory circuits).[17]

Biodiversity and sustainable NCP provisioning are vital for building capacity to mitigate and adapt to social-ecological challenges (such as climate change), support human health and well-being, and support long-term economic development and Earth System stability. AI indicates promising potential to complement traditional research methods and tools for understanding and supporting NCP provisioning. However, use of AI also comes with trade-offs, risks, and uncertainties.

## AI for Data on Nature's Contributions to People

AI methods such as machine learning (ML), deep learning (DL), natural language processing, computer vision techniques,[18] and combinations thereof can contribute to NCP research in a number of ways. Specifically, to assessing, identifying, monitoring, and predicting the presence, composition and health, and people's uses and perceptions, of ecosystems.[19–21] Audio-based (i.e., bioacoustics, ecoacoustics, and soundscape) research is one area that has benefited from AI.[22] A convolutional neural network (CNN) was used for analyzing soundscape data from three urban forests in China, linking soundscapes to land use types, and assessing the impact of human activities on biodiversity.[23] After large language models, multimodal language models have emerged that can analyze combined types of data. In a recent study, an audio-language model demonstrated unprecedented capacities for recognizing animal group sounds (such as birds and frogs) on par with supervised learning—but with only basic, general training.[22]

Main advantages of AI for NCP include the capability of **interpreting vast amounts and different types of data** into training and validation of AI models.[24] In London, AI was used to support automated analyses of big data surveys on the prevalence and characteristics of green roofs, to increase climate-resilient green roof designs. An ML algorithm was used to segment aerial imagery at single-building level, covering 1558 km$^2$ of Greater London.[25] Combinations of spatial remote sensing and social data (e.g., data from surveys or personal monitors, such as fitness trackers or social media applications such as Flickr) can help with insights on connections between socio-economic status, land use, biodiversity and ecological habitat qualities, and their spatial distribution.[26–28]

Diverse data and improved models can also provide a **holistic view of the human pressures on ecosystem**s and provide comprehensive information to guide conservation priorities. This has been shown in a study on species classified as Data Deficient in the IUCN Red List, in which a global multitaxon ML classifier predicted the probability of species facing extinction. The study re-evaluated previous threat level estimates and identified various threat probabilities, with potential implications for conservation policy.[29] Loss of genetic diversity and extinction risk across taxa are quantified using genetic data from environmental DNA (eDNA) and whole-genome sequencing (i.e., determining an organism's entire DNA sequence).[30] Here, AI could make important contributions such as classifying the often highly dimensional genomics data using predictive biomonitoring models.[31]

AI methods can complement traditional tools such as photo traps and manned observations to identify new patterns in data. For example, genetic data analysis enables the **identification of cryptic biodiversity and indicators of ecosystem health** that is not identifiable by conventional photo imagery alone.[32] Sampling and analyzing eDNA from water or soil further supports the **detection of elusive species** including those considered extinct in specific regions.[33] More frequent sampling also enhances the ability to **detect early warning signals** of potential ecosystem shifts.[32] Furthermore, DL approaches can now directly estimate alpha (local), beta (between-community), and gamma (regional) diversity from plot data, **bypassing traditional species-distribution modeling**. For example, neural networks trained on vegetation plots across Australia predicted plant diversity patterns at high resolution by learning species–area relationships from climatic, geographic, and human-impact variables, offering a scalable framework for automated biodiversity mapping in data-poor regions.[34]

Moreover, **AI can support conservation efforts** and **empower local communities**. In the Brazilian Amazon, for example, one of the world's most biodiverse regions, Indigenous groups have used drone data and AI-based monitoring tools to protect their territory against invasions of illegal miners.[35] Digital AI applications based on citizen science, in which people contribute with data and identifications, such as *iNaturalist* and *Flora Incognita*, can help identifying different individual plant species and contribute to our understanding of plant trait patterns on a global scale.[36]

## AI-Assisted Predictive Modeling of Nature's Contributions to People

Modeling of NCP and biodiversity is primarily conducted using ML, DL, and computer vision techniques. Potential outputs include predictions, pattern and trend detections, and impact assessments across spatiotemporal scales.[36]

AI's ability to **analyze complex remote sensing data** and **identify key indicators** increases the availability of data for modeling scenarios. AI can extract and evaluate indicators (such as changes in vegetation cover, or air and water quality) from remote sensing images and create climate adaptation scenarios that benefit both humans and ecosystems.[25] Such capabilities can make significant and potentially even lifesaving contributions. For example, by mitigating the impacts of natural hazards, or contributing to NCP-enhancing design suggestions for natural (green-blue) infrastructure in cities.[25]

The predictive capabilities of AI can inform the analysis of complex data such as the **combined effects** of multiple interacting stressors. For instance, AI has been employed to evaluate the cumulative impact of climate and land-use change on NCP provisioning in Yunnan Province, China.[37]

Earlier ecosystem models have demonstrated a capacity to identify thresholds (such as removing 80% of biomass in an ecosystem[38]), for sudden and often irreversible changes with cascading effects, known as "tipping points."[39] Recent advancements in AI also enables integration of species abundance, genetic diversity, and functional traits within phylogenetic frameworks, enabling **multidimensional predictions** that capture complex eco-evolutionary dynamics beyond traditional threshold models[40,41] (see also Theme box 2, *Using AI to Detect Earth System Tipping Points*).

AI methods can augment human capabilities and assist discerning patterns in complex social-ecological data.[42] For instance, CCNs in species distribution models (SDMs), such as those relating to pollinator diversity, can extend beyond site-specific predictions to also estimate the **potential influence of surrounding landscape structures**. This capacity is particularly useful in the study of highly heterogeneous landscapes such as urban and peri-urban areas.[43] Other predictive studies have used ML to assess the suitability of plant species based on their adaptability to projected future conditions.[44]

Studies on land use and land-use change have benefited from AI, in part due to ML's ability to **analyze large and high-resolution datasets**, such as satellite data. For example, recent AI-supported analysis at a 1 km resolution indicates that 63% of urban expansion will likely occur on current cropland, which could decrease the global crop produc- tion by 1–4% (equivalent to the annual food requirements of 122–1389 million people).[45] Large-scale biodiversity data analyses supported by AI methods have enabled new forms of modeling of global microbial soil systems (crucial for soil renewal and thus supporting NCP). The approach has enabled projections of ecological functional and diversity **responses to perturbations**, such as climate change, invasive species, and pollution.[46]

## AI-Assisted Decision-Making for Nature's Contributions to People

Participation in land management and decision-making can foster a sense of meaning and responsibility toward the natural environments, i.e., stewardship —an important factor for ecology conservation and long-term support for NCP.[47] AI can support **decision-making concerning NCP at various levels**—for example, mobile applications that can advise small-scale farmers or citizen gardeners on everything from sowing to harvesting[48–50]—while larger-scale efforts include guiding conservation policy,[51] biodiversity protection, and land management.[52] One example of AI research providing actionable advice to **mitigate ecological and social harm** is a study of the wetlands of Kolkata, India. The AI analysis of combined biogeophysical and social data revealed that 60% of the wetland's NCP provisioning was threatened.[53] The information can support planners in assessing preferred sites for, and characteristics of urban development, and for conservation areas.

AI can inform decision-making processes by **identifying nuances** in social-ecological data. Examples include mapping how people use ecolog-ical resources such as gardens,[54] assessing NCP, such as the stormwater absorption capacity of green infrastructure,[55] and understanding different perceptions of ecology and NCP.[56,57]

AI-driven analyses of crowdsourced and citizen data, such as photos of natural areas or fitness tracker data, can inform decision-makers about how people utilize and value natural spaces.[28,58]

## Potential

AI has the potential to contribute to the analysis and management of NCP in various ways, particularly in areas where research has previously struggled. Promising application areas include: **a) integrating different data** such as specimen location, habitat preferences, ecological data, and spatial datasets on human impact, to advance social-ecological predictions and maps,[59] **b) evaluate ecological and environmental policy and practice** by combining different types of AI, such as natural language processing and computer vision, to enhance the understanding of how natural spaces and resources are used and monitor changes[60] **c) improving remote sensing analyses**, e.g., for cover classification[61] and threat detection,[62] and **d) accelerating analyses** of images and recordings.[63]

Exciting opportunities also lie in advancing AI-driven scenario modeling of NCP-based strategies. Specifically, for strengthening the adaptive and mitigative capacities of both ecosystems and human societies in response to, for example, climate change and biodiversity loss.

## Limitations

The **geographical bias** identified in our literature review indicates that a known issue in ecological research[64] persists in AI applications for NCP. By predominantly focusing on upper-middle- and high-income countries, AI-generated NCP outputs, such as recommendations on ecosystem management, risk misaligning with the perspectives of people and the ecological, climatic, and socio-economic contexts in other regions, i.e., the Majority World.

Several of the less studied countries, such as Brazil, Colombia, Indonesia, and in sub-Saharan Africa, often have limited AI capacities. They are also exceptionally rich in biodiversity, with a high number of threatened species,[65] and with the highest number of people who depend directly on ecosystems for their livelihoods. For instance, it is estimated that over 50% of Africa's workforce is employed in agriculture.[66] However, even in Africa's burgeoning AI sector, NCP and biodiversity risk being overlooked when other areas are prioritized.[67]

Predictions of NCP, such as the decomposition of organic matter, pest control, crop pollination, and seed dispersal for agriculture and habitat maintenance, require modeling of both biophysical variables and of biodiversity.[68] However, these two areas have largely been **studied separately**, and biodiversity information is often lacking in NCP predictions.[69]

AI-generated insights on NCP alone do not guarantee **sound decision-making**.[70] Preferences and resource management histories vary across groups,[71,72] yet underrepresented groups–such as women, the elderly, Indigenous peoples, and low-income communities–often lack visibility in both data and decision-making.[73,74] Non-traditional sources like social media risk reinforcing inequalities by reflecting the behaviors and values of user groups with access to specific digital applications. Access to, and use of digital applications vary by gender, geography, age, income level and so on, and not everyone uses digital platforms or has reliable internet access.[75,76]

The dynamic nature of NCPs, as they arise from interconnections between humans and the biosphere, combined with data gaps, can lead to differences between AI predictions and actual outcomes. This phenomenon is known as "**predictive dissonance**".[77] Such dissonance could undermine trust in AI NCP recommendations, which is one reason why it's important to clearly communicate uncertainty factors in AI modeling outputs.

Only around half of the models in NCP predictions have provided **quantifications of uncertainty**,[68] and NCP research suffers several blind spots.[78] Analytical uncertainty in NCP modelling can occur in several steps of the research process: in data quality and biases,[78] data preprocessing and supervised classification,[79] the selection of predictor variables,[80] the behavior of model algorithms, and decisions concerning baseline data, to name a few.[81]

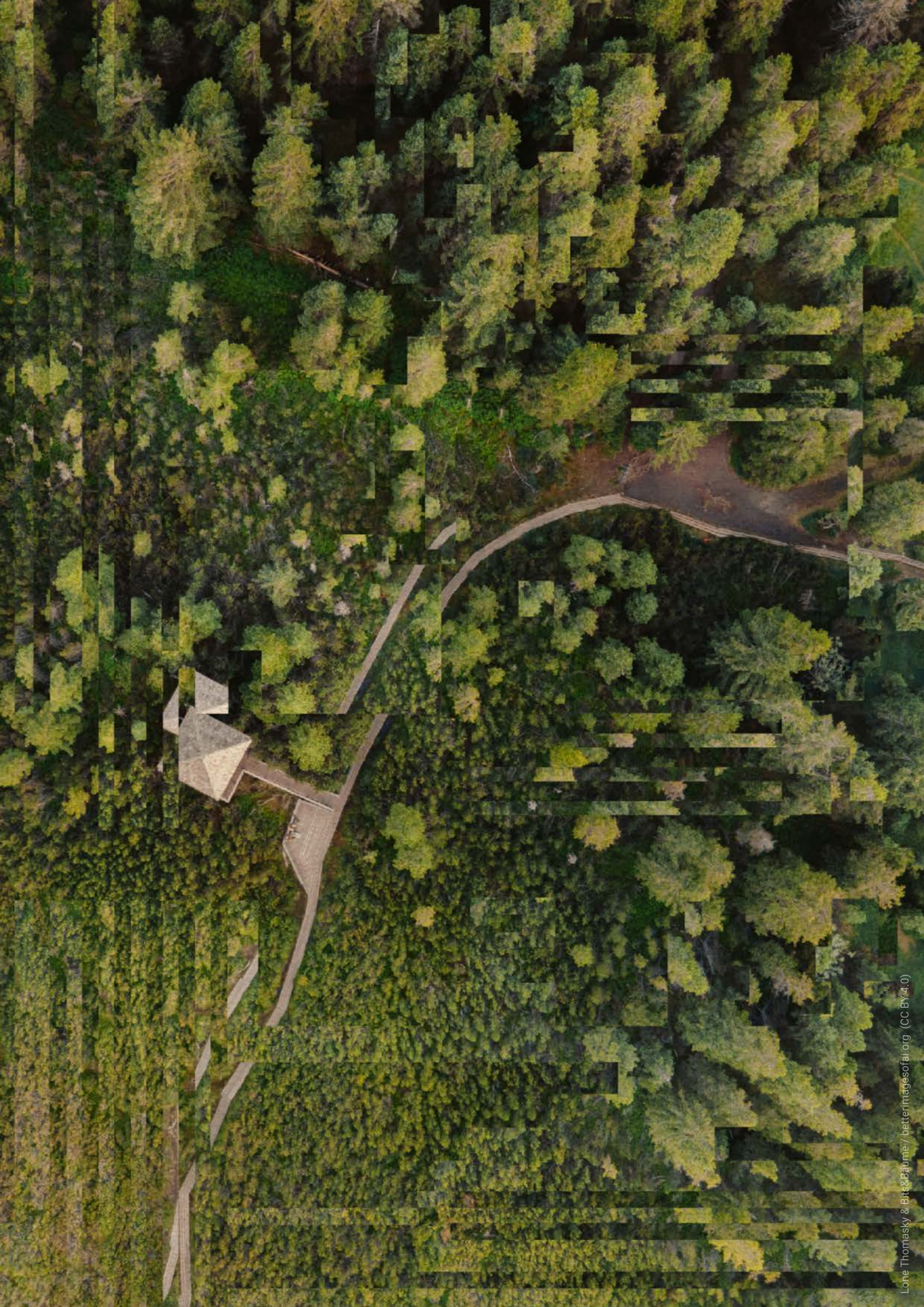

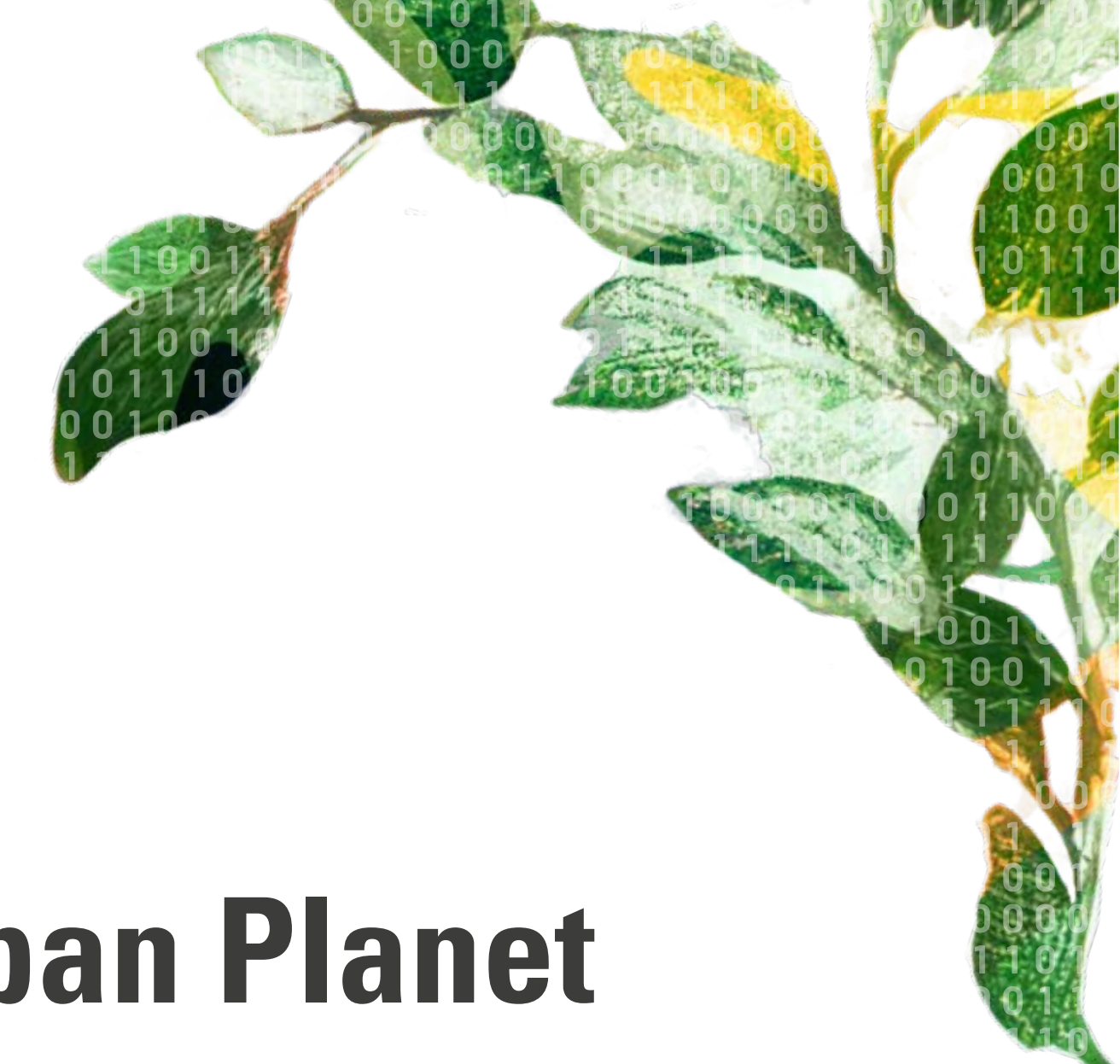

# 6 Prospering on an Urban Planet

Authors: Maria Schewenius, Timon McPhearson, Ahmed Mustafa, Elizabeth Tellman

*Urban areas and artificial intelligence (AI) are increasingly interconnected, with urban AI offering significant opportunities to advance both urban and planetary sustainability. We provide an overview of the essential field of urban AI, and present some key considerations for its research, development, and practical application.*

## Introduction

From London, UK (DeepMind), to Menlo Park, USA (Meta), San Francisco, USA (OpenAI), Beijing, China (Baidu), and numerous others, cities are hubs for artificial intelligence (AI) development and practical applications. Investments in AI in cities (urban AI) are exponentially increasing. For example, some sources estimate that AI in sustainable urban planning will increase by 20% annually, from $12 billion in 2022 to over $54 billion by 2030.[1]

The efficiency and capacities of AI, such as the rapid cycle from data collection to model output, influence the very core of urban research, policy, and practice. AI can support urban planning for example by summarizing, analyzing, and providing feedback on plans, helping to incorporate diverse perspectives on urban challenges and AI applications. It also shows promise in enabling two-way information exchange between planning authorities and the public, thereby supporting participatory planning.[2] Multi-stakeholder partnerships and platforms emerge that can jointly contribute to further urban AI research, develop industry solutions, and support policy. Examples include the UN and government-led, Africa-focused initiative *AI Hub for Sustainable Development,*[3] think tanks such as the Paris-based *Urban AI*[4], and transdisciplinary academic institutions such as *MILA*[5] in Quebec, Canada.

However, research on AI and the digitalization of urban areas, the so-called “smart city” discourse, has been fragmented, mainly focused on the built and engineered environment, and mainly highlighted positive contributions of technologies and successful projects.[6]

Urban AI research, development, and deployment need to recognize the complex interactions between society, ecology, and technology in urban areas. Cities function as interacting social-ecological-technological systems (SETS).[9,10] They are shaped by patterns of architectural design, people’s habits such as movements and activities, the norms and rules of the governing institutions, and the resources locally and from afar that sustain the urban fabric. As data-generating, pattern-generated systems, cities can both inform and benefit from AI. For example, in Singapore, the advanced digital twin *Virtual Singapore*[7] integrates data from both above- and below-ground infrastructure. It is one of the application areas highlighted in the *National AI Strategy*,[8] providing managers with diverse datasets and supporting a wide range of uses. These include educating the public about urban nature, managing traffic in real time, and enabling long-term planning through scenario modeling, such as evaluating strategies for responding to disease outbreaks.

Cities are both contributors to and recipients of global trends such as climate change, and are projected to experience increasingly turbulent futures. Changes such as increasing temperatures, increasing frequency and intensity of floodings, increasingly expansive and dense built-up areas, and biodiversity decline often co-occur[9] and create compound effects that planning has traditionally struggled to manage. For example, urban greenspaces are widely recognized for providing essential services, such as shade and rainwater absorption. However, their average share of urban landscapes decreased from 19.5% in 1990 to 13.9% in 2020.[10] Meanwhile, it's estimated that by 2040, only 1% of the urban populations could escape temperature increases.[10] Urban populations are projected to continue to increase. From 2018 to 2050, 2.5 billion more people are projected to be living in cities, primarily in low- to lower-middle-income countries in Asia and sub-Saharan Africa.[11]

It is crucial to harness the potential in the burgeoning emergence of AI to support urban sustainability and ensure planetary stability. On an urban planet, urban resilience is central to planetary sustainability. The often superhuman capacities of urban AI could provide unprecedented potential to support development toward desirable urban and planetary futures. However, responsible use of AI,[11] and recognition of the diversity and complexity of urban SETS in research and development, is critical to ensuring urban development is equitable, resilient, and sustainable.

## AI for Urban Data

With AI, urban governance is becoming increasingly autonomous and data-centric.[12] The influence of AI on urban societies, infrastructure, and institutions, and how cities are monitored, controlled, and planned, is rapidly growing. In China, for example, the (Alibaba) *City Brain* is an AI-powered, computer vision system intended to optimize traffic flows and support traffic planning.[13] Cities and urban regions **generate massive amounts of data** through sensors, cameras, and monitoring systems, requiring increasing use of AI to manage data, make decisions, and integrate data for interoperability. It is not only cities that generate data; people do as well. Digital footprints from social media platforms, mobile apps, and other user-generated content are increasingly leveraged as data sources for AI models applied in urban analytics and planning.[14] At the same time, cities are also **massive consumers of data** to guide urban planning, development, policy, management, and more, thus serving as key systems for the continued development of AI.[15]

AI has shown unprecedented capacities to **analyze different and large quantities of data**, often in real time, and **dealing with non-linear data,** such as the impact of urban features on surface temperatures.[16] Urban AI data has made valuable contributions, for example, to mapping,[17] identifying,[18,19] assessing,[20] and monitoring[21,22] urban green-blue and built infrastructure, and understanding user perceptions.[19,23,24] Increasingly high-resolution remote sensing (satellite) data is improving the precision of AI models and supports more informed planning. Lesser-known areas can be better understood, such as informal urban areas (often called slums or favelas when mainly inhabited by the financially poor), which often have limited population data and can be difficult to access.[25] AI models can produce urban plans and greenspace recommendations with superhuman efficiency, for example suggesting designs that increase people's proximity to greenspaces[26].

**AI can support combinations of data**, such as local sensors and remote sensing, for mapping urban areas, for example night lights, vegetation cover, and socioeconomic factors such as population density.[27] As was shown in the Three Gorges Reservoir Area in China, AI has the capacity to identify complex social-ecological relationships in urbanizing regions, for example assessing the presence and behavior of ecosystem services bundles and their drivers.[28]

## AI-Assisted Urban Predictive Modeling

AI is increasingly used to **predict and simulate complex urban systems**, enabling more accurate forecasting of trends such as land-use/cover changes and land surface temperature,[29] traffic demand and flow,[30] and energy and water demand.[31–33] Increasingly spatiotemporally precise urban models contribute to, for example, forecasting weather, hazards, and their impact. The capability stems from AI's ability to **a)** process vast amounts of diverse data, **b)** combine geospatial, weather, human, infrastructure, and critical facilities data,[34] and **c)** identify complex patterns. Advances in explainable AI can identify, for example, which urban infrastructure features or inequalities lead to hazard predictions (e.g., air pollution or urban heat[33]). The advances allow at least a partial view or some light into the AI "black box" of algorithmic processing.

**AI can generate forecasts and nowcasts,**[35] and increasingly realistic models of future hazard and flood damage patterns.[36] As AI modeling moves beyond simple random-forest models, and incorporates the spatial and temporal structure of urban environmental cities through graph networks with spatial and temporal convolutions (or transformers), urban digital twins begin to better approximate real urban environments.[34] For example, flood depths and extent are predicted not just from physically based models or weather data and remote sensing, but are updated with human telemetry (e.g., movement or cell phone use) and urban resident reports of traffic to cities or on social media.[35] AI can combine traffic, flood depth, power grid, human movement, and critical facilities data to assess if evacuation orders were effective, where people moved, and which hospitals or food facilities are compromised or stressed (e.g., in Hurricane Beryl, see *Resilitix*[37]).

**The generative capabilities of urban AI** extend beyond mere predictions to actively shape the future of our built environments based on predictive insights. For instance, Mustafa et al. (2020)[38] demonstrated this by using an AI model to generate urban layout designs specifically engineered to be more resistant to urban flooding. This model achieved flood mitigation by optimizing variables such as the redistribution of green areas and the redesign of road networks to reduce water depth. This showcases how urban AI moves beyond simply predicting problems to also predict, design, and optimize potential solutions. Research is still limited regarding impact forecasting, communication of information (such as early warning signals), and the potential of support for rapid decision-making.[39]

## AI-Assisted Urban Decision-Making

Urban AI offers capabilities that can **deepen the understanding of urban areas** by merging different data for sensing, imaging, and mapping, often in real time,[40] and presenting complex and detailed data analyses and predictions to support decision-making[41] (see Theme box 3, *ClimateIQ: Empowering Communities to Prepare for Climate Threats with Hyperlocal data*). Its support spans crucial decision-making phases—for example, from problem identification, where AI-powered analyses of real-time traffic sensor data can swiftly pinpoint congestion patterns[42]; option generation and evaluation, where AI models can simulate the impact of new zoning regulations on housing affordability[43]; resource allocation, where AI optimizes the deployment of limited urban resources based on data-driven insights[44]; to performance monitoring and adaptive management, such as analyzing bus passenger ridership data to optimize demand.[45]

Beyond these direct decision-making applications, urban AI **contributes to a deeper understanding** of urban built-up structures,[46] processes such as land-use change,[47,48] and proposals for innovations, such as new and resource-efficient approaches to urban agriculture.[49] Foundation models using self-supervised learning can facilitate the development of other digital applications[50] and could potentially revolutionize decision-making by offering area-specific machine capacities.[39] After large language models (LLMs), which typically deal with text, large multimodal models (LMMs) are emerging that can combine text, images, and audio.[51] The improved observation, imaging, and analysis capabilities of urban AI provide potential to increase the understanding of urban challenges and identify (multifunctional) solutions appropriate for the local context.

## Potential

We present some key areas where urban AI can support prosperity on an urban planet:

**Spatially precise, detailed, and comprehensive SETS modeling.** The enhanced forecast capacities and detailed understanding of potential urban scenarios provided by some AI systems can support solutions designed for the local area and that take into account the interactions among multiple urban systems domains.[40] Graph networks are presenting new opportunities, allowing modeling of the spatial temporal and connected structure of urban elements across diverse data, such as traffic flow simulations. Importantly, graph networks allow using urban data for more realistic urban digital twins and improved urban structures.

**Powerful scenarios and predictions,** from spatiotemporally precise hazard prediction to impact forecast, communication, and decision support.[39] AI acting inside digital twins offers unparalleled potential to examine how climate change and other urban stressors can be mitigated and adapted to.

**Improving social sensing capabilities.** AI capacities to preprocess and incorporate social sensing and human telemetry data, such as Flickr photos, are improving. The capacities can mitigate some of the known biases related to, for example, social media, and contribute to more nuanced understandings of how urban spaces are used and function.

## Limitations

**Urban AI research incorporates different biases.** Urban AI studies tend to focus on a few specific social media apps and their users, large cities, and cities in upper-middle- to high-income countries. Urban areas tend to be highly heterogeneous, which can increase the risk of so-called transfer bias,[11] in which models trained on data from one region risk producing incorrect and potentially damaging recommendations when applied to another region. For example, sudden and intense precipitation might have little impact in one urban area, whereas in another it could cause flooding or landslides.[52] AI has also been found to reinforce existing socioeconomic differences, in part due to unequal representation in training data.[53–55] Such bias risks limiting the potential of AI models to identify sustainable urban solutions.

**Studies have mostly focused on the built-up and engineered environment.** Green-blue infrastructure and ecosystem services are still relatively under-explored. There remains untapped opportunity to use AI to better understand and address biodiversity loss and ecological resilience, which is critical to social and infrastructure resilience.

**Limited preparedness.** Studies indicate a low awareness among urban planners of how to utilize AI.[56] This lack of knowledge and preparedness often stems from curriculum gaps in higher education, a scarcity of accessible and tailored training for professionals, limited exposure to compelling real-world case studies, and resource constraints within planning departments. Furthermore, for many, AI can still be perceived as a “black box,” contributing to skepticism.[57] The lack of knowledge and preparedness could become a problem not necessarily through low uptake, but through damaging uptake, for example of AI systems with limited capacities to deal with given tasks or specific SETS contexts.

AI’s ability to **manage interconnected shocks in urban SETS** remains underexplored. While AI typically learns from routine patterns, like traffic flows,[58] urban futures are expected to be increasingly turbulent and diverge from historical trends (see chapter 1, *Preparing for a Future of Interconnected Shocks*).[59,60] Research has also primarily focused on isolated elements of SETS, such as the effect of expanding urban areas on the provisioning of ecosystem services. Predicting the behavior and impact of shocks is especially challenging due to the complexity and variability of urban SETS, which differ across socio-economic, climatic, ecological, and infrastructural conditions, even within short distances or timeframes. Urban AI modeling holds promise for identifying responses to change, tipping points (when change becomes irreversible), feedback loops (self-reinforcing desired or undesired behaviors), cascading effects across SETS–and strategies for adapting to, and mitigating turbulence. Yet despite its potential to support sustainable and livable urban futures, this research area remains largely untapped.

THEME BOX 3

# *ClimateIQ*: Empowering Communities to Prepare for Climate Threats with Hyperlocal Data

Authors: Christopher Kennedy, Timon McPhearson

The AI-powered online platform *ClimateIQ* aims to democratize access to high-resolution, neighborhood-scale climate hazard information, particularly for under-resourced and vulnerable communities worldwide. The goal is to improve urban resilience, emergency preparedness, and climate adaptation planning. The tool is designed to support city planners, community organizations, and decision-makers by identifying localized risks from extreme heat and urban flooding (fig. 6).

Rapid urbanization, rising temperatures, and the increasing frequency of extreme weather are escalating risks to cities worldwide. However, many local governments lack the resources, data, or technical capacity to generate meaningful climate hazard assessments. *ClimateIQ* addresses this gap by providing a free, open-access tool and map-based dashboard that delivers high-resolution and scientifically robust climate risk information, leveraging both traditional physical models and modern machine learning (ML) approaches.

*ClimateIQ* uses ML models trained on the outputs of physics-based simulations—specifically, the *Weather Research and Forecasting* (WRF) model for urban heat (Fig. 7a) and hydrodynamic models like *CityCAT* and *HEC-RAS* for flood hazards (Fig. 7b). The ML models replicate the behavior of traditional simulations but with significantly reduced computational demands, enabling rapid, scalable forecasting across diverse urban areas (Fig. 7c).

In an optimal use scenario, *ClimateIQ* takes in global and local datasets—such as topography, land use, weather patterns, and urban infrastructure—and applies trained ML models to simulate climate hazard exposure at fine spatial resolution. Through an intuitive digital dashboard and APIs, planners and communities access current and projected hazard maps and data layers to inform emergency response, adaptation planning, and investment prioritization (Fig. 6).

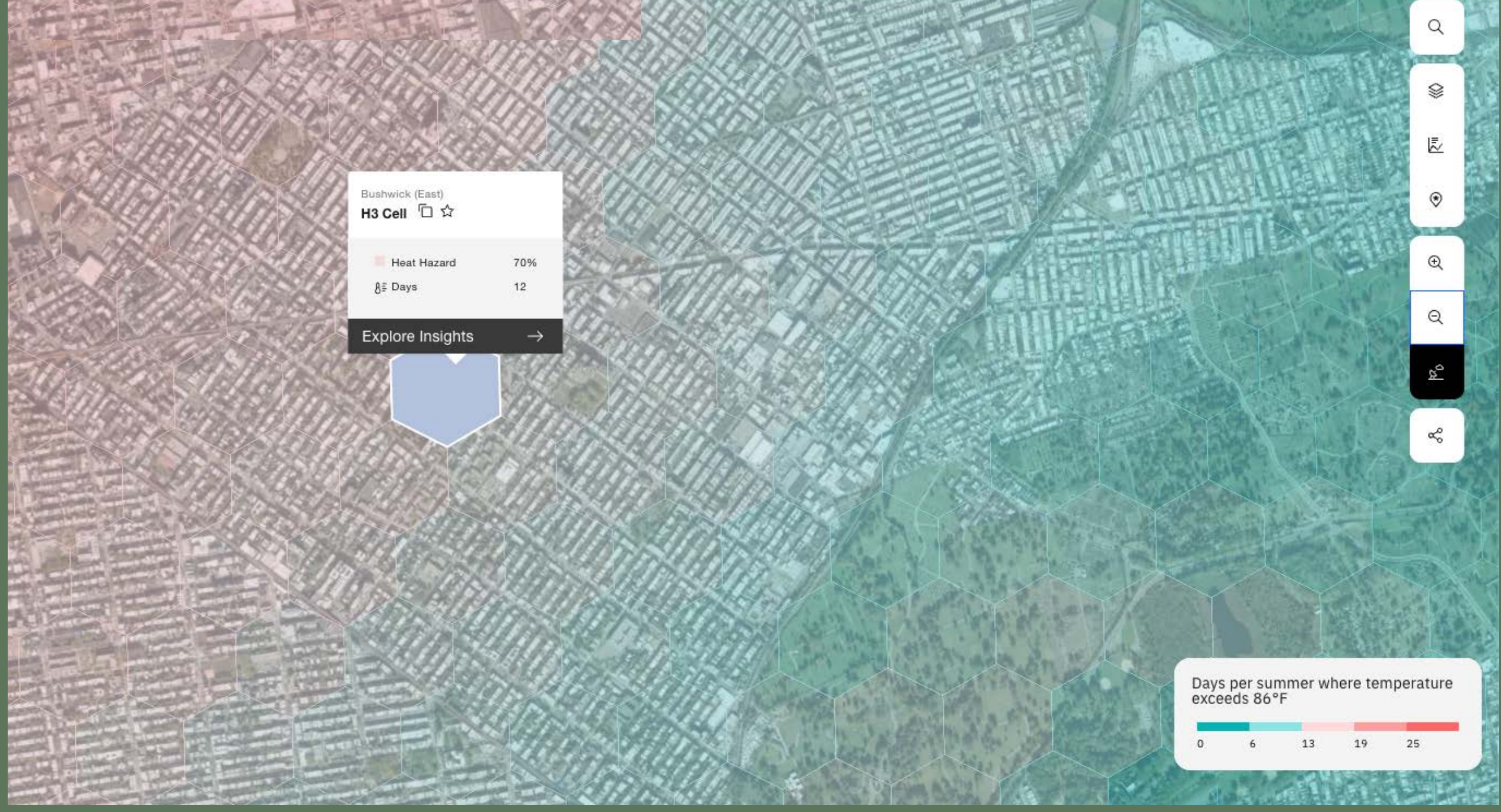

Fig. 6. *ClimateIQ* dashboard interface prototype, with hazard scenario selector on right panel. Dashboard developed in collaboration with ClimaSens.

While powerful, *ClimateIQ's* accuracy depends on the quality and availability of underlying data. In regions with sparse datasets, model performance may be reduced. Additionally, like many AI models, predictions can be less reliable in entirely novel geographies unless the model is retrained or supported by additional ground truth data. Further, while the tool removes many barriers to access, sustained engagement with end users is needed to ensure integration into local planning processes. For more information, visit: https://climateiq.org

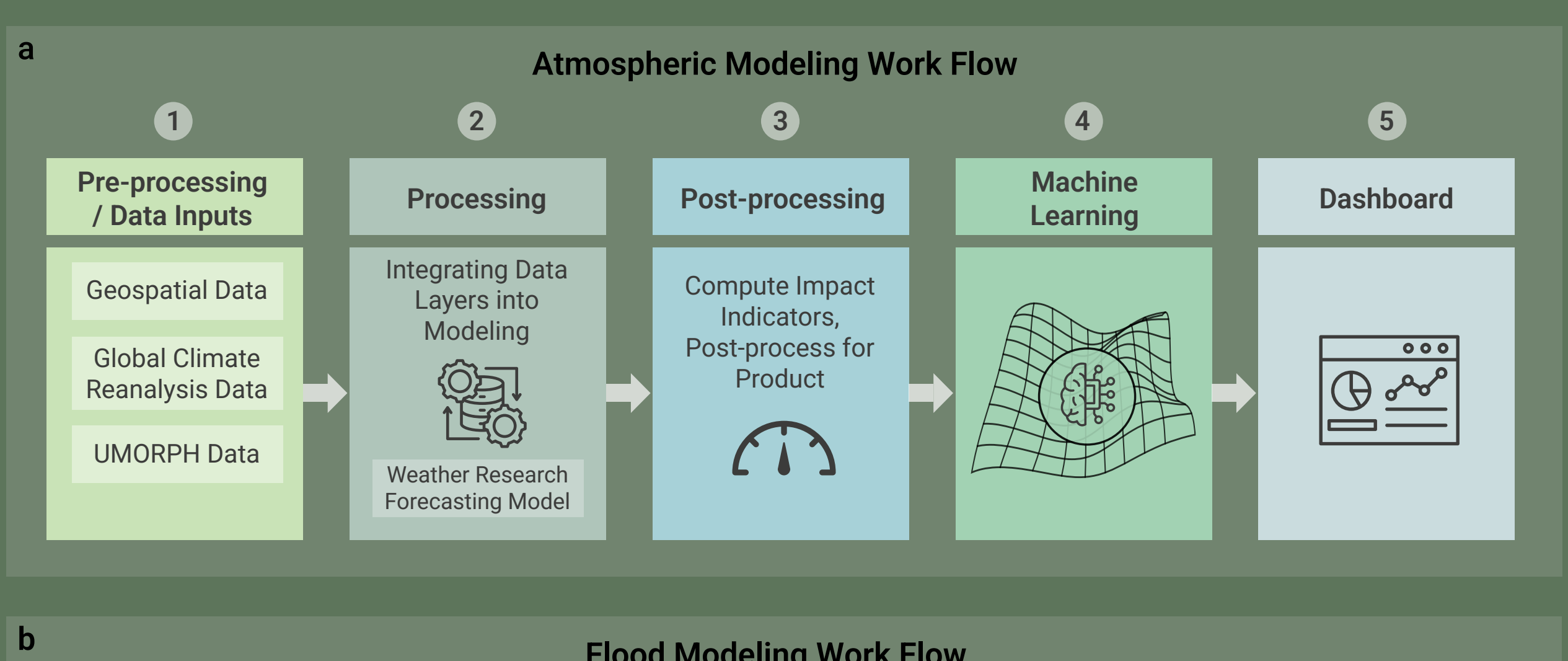

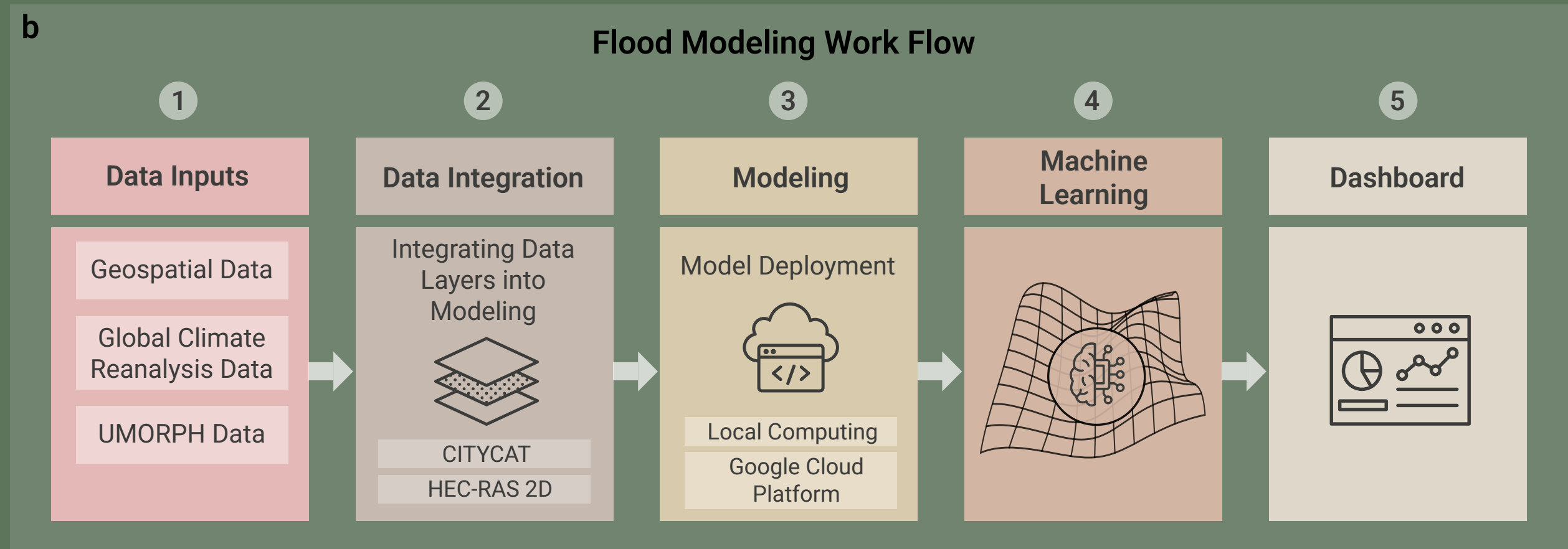

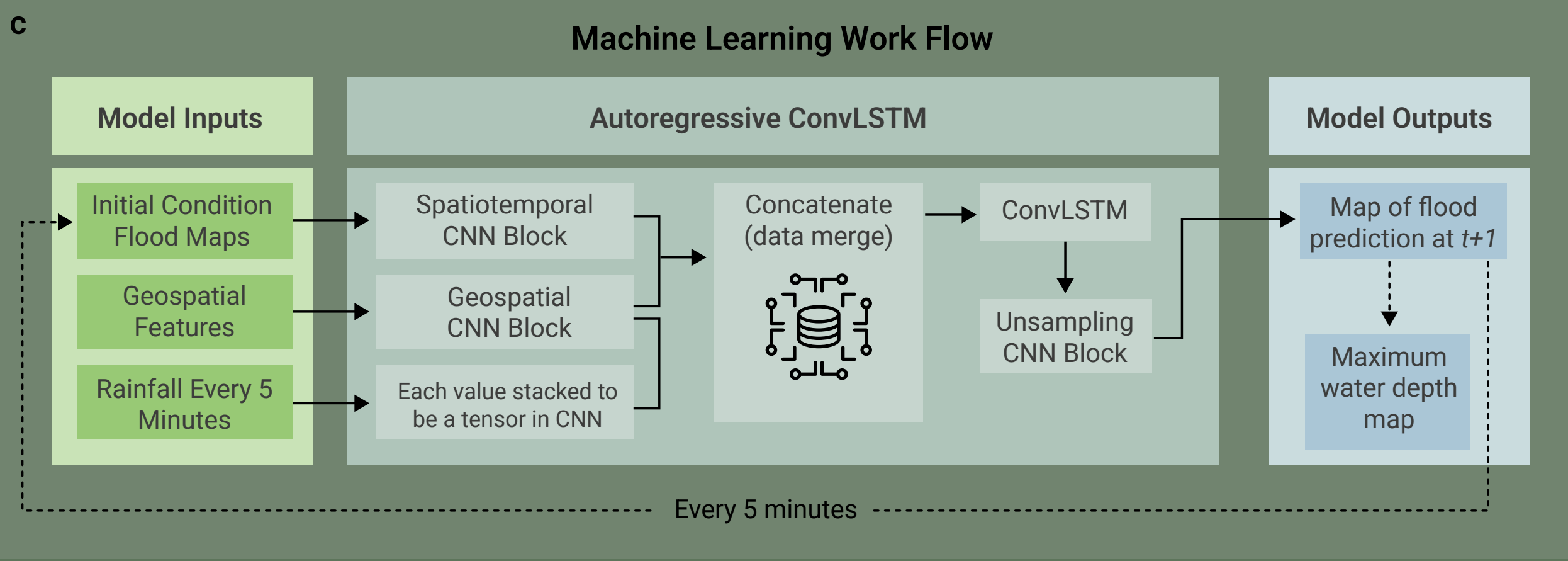

Fig. 7 a-c. The AI processes behind the flooding and atmospheric modeling in *ClimateIQ*: a – atmospheric modeling workflow diagram illustrating the primary components of the extreme heat physics-based model, b – hydrological modeling workflow diagram illustrating the primary components of the inland (pluvial) flooding physics-based model, and c – machine learning workflow diagram illustrating the core components of the modeling pipeline.

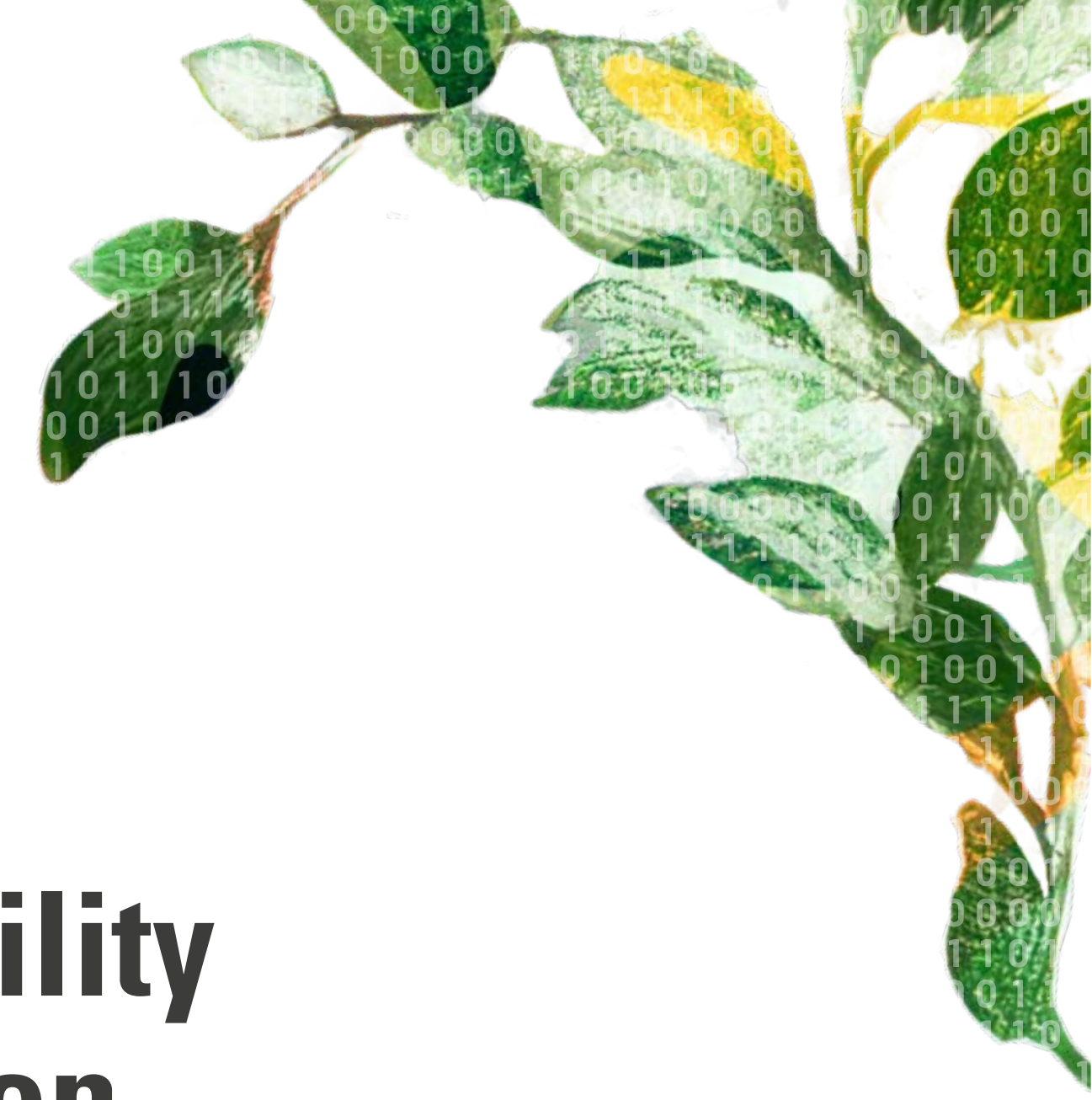

# 7 Improving Sustainability Science Communication

Authors: Masahiro Ryo, Victor Galaz, Andrew Merrie

*Communicating climate and sustainability sciences has proven challenging, especially in strongly polarized contexts. Can AI support new ways of science communication, storytelling, and scenario visioning?*

## Introduction

Our collective abilities to act on climate change and sustainability challenges require up-to-date information. For example, governments trying to set targets and design conservation policies need to base such decisions on the best available evidence. Local farming communities preparing for a changing climate need to be able to assess future risks to their crops and livestock, receive evidence-based guidance, and collaborate with their members. Engaging with young people on climate issues requires styles of communication that are perceived as both attractive and actionable.

Climate and sustainability-related mis- and disinformation pose serious threats to information integrity all over the world. Such risks may very well become amplified by the increased use of generative artificial intelligence (generative AI) through, for example, personalized false messages, an increasing volume of false news online, or the proliferation of chatbots that distort scientific facts.[1] There is also increased evidence that the combination of expanding digital social networks and the infusion of AI in digital media platforms is influencing human emotions at scale (see Theme box 4, *Climate Emotions and AI*).

Uses of AI can, however, also help support science communication in new ways.[2] Generative AI, such as *ChatGPT* and multimodal language models (multimodal LLMs)—that is, models that combine text, images, data, and even sound—offers a novel opportunity to enhance science communication for sustainability. Traditional science communication methods often struggle to reach diverse audiences in a timely and engaging manner. In contrast, AI-powered tools enable personalized, interactive, and multilingual communication that adapts to users' needs, local contexts, and decision-making scenarios in real time.[3] By transforming complex sustainability science into accessible narratives, visualizations, and dialogues, AI can, at best, help empower both citizens and policymakers.

## What Makes Science Communication Effective?

Truly effective science communication is not only about sharing facts from scientists to non-scientists, but also about making scientific knowledge reachable, understandable, reliable, and actionable. This suggests the importance of recognizing the diversity of the audience, what they care about, what they already know, and how they prefer to receive information (see Moser et al., 2010[4] for more details).

The effectiveness of science communication differs largely from different stakeholders, educational backgrounds, ages, and beliefs. Using clear language, relatable examples, and practical stories helps people engage with complex issues such as climate change and biodiversity loss. Ideally, science communication should also allow scientists to listen and respond to the audience. People are more likely to trust and act on scientific information when they feel respected and included. Unfortunately, scientists are in general not trained to develop such skills, and often lack the organizational incentives to engage in such forms of communication (see Markowitz and Guckian, 2018[5] for details).

## Engaging with Scientific Knowledge Through AI

One of the observed benefits of large language models (LLMs) is their ability to rephrase text in ways that suit the profile of the user. For example, an LLM can explain complex things in plain language (ELI5: Explain Like I'm Five[6]). It can translate academic jargon and explanations into more accessible, easy-to-understand explanations.[3,7] Small and purpose-built language models can also be developed in ways to support grassroot communication and education, as proven by several African-led initiatives.[8]

A number of studies show the potential of using LLMs to expand education opportunities in areas of the world where teaching resources are limited, potentially helping bridge educational inequalities,[9] although they may also foster educational inequality and digital divide unless such risks are mitigated proactively.[10,11] Multimodal LLMs (or multimodal AI) can help explain issues in engaging and understandable ways. Google's *NotebookLM*, for example, creates a podcast based on uploaded documents. Uses of multimodal AI can act as helpful "co-communicators," providing personalized answers, visuals, and summaries, all while adapting to local context and real-time information. Multimodal AI can, at least in principle, be developed to show how climate change might affect a specific town. It can also help citizens understand scientific reports through interactive maps and stories, thus allowing people to understand (and maybe even enjoy) complex scientific materials.

Table 1. Examples of how LLMs and generative AI can enhance science communication.

| Communication Aspect | Traditional Science Communication | With Human–AI Interaction (LLMs, Multimodal LLMs, Generative AI) | Climate and Sustainability Communication Example |
|---|---|---|---|
| Audience Centeredness | One-size-fits-all messaging; static content tailored to broad audiences | Personalized communication adapting in real time to individual users' knowledge, language, and preferences[3,7,14] | Enhancing climate awareness and managing climate anxiety for teenagers[15] |
| Medium and Mode | Pre-defined information mode | Seamless transition across text, image, audio, and video (e.g., any-to-any multimodal model[16]); LLM predicts which modality the user wants to access[17] | A multimodal LLM can enhance information accessibility for people with blindness and low vision[18] |
| Trust and Credibility | Credibility tied to human experts and institutions; trust earned over time | AI agent provides the relevant information source (cf. hallucination; retrieval augment generation)[19] | A chatbot explains scientific facts based on the IPCC report (e.g., *ChatClimate*[20]) |
| Anytime Two-way Conversation | Public talks, Q&A, citizen panels | Conversational interfaces enabling continuous dialogue, real-time feedback, and exploratory questioning by users[7] | An AI podcast can summarize multiple, complex scientific information sources in a narrative way, and the listener can interrupt and ask questions anytime (e.g., Google *NotebookLM*) |
| Localized Information Delivery | Scientific evidence is generalized and the answer to a context is "it depends" | AI answers a general question while accounting for personalized context[14] | A farmer asks how to cultivate wheat under climate change in his locality[21]; disaster prediction with personalized early warning system[22] |
| Language Barrier | Translator required for low-resource languages | Language models can be locally adapted for indigenous languages; information dissemination at massive scale and speed[7] | The local disaster news in its local language can be accessed in various languages without bypassing public/social media (e.g., by using Meta AI *NLLB-200*, which can translate across 200 different languages) |

LLMs can also be used in multi-agent system architectures that mimic scientific discussions, such as a roundtable or panel, with several agents with different perspectives communicating iteratively.[12] In the context of sustainability and climate change issues, "virtual stakeholders"—each of them based on one specific type of text corpus, say IPCC reports or FAO reports—can bring different stakes and perspectives to answer questions. A moderator can then summarize the different viewpoints and identify commonalities and differences in expert opinions.[13]

Table 1 lists a number of idealized possible applications of LLMs and generative AI on various science communication aspects.

## AI for Storytelling and Futures

Much of the science that is being published today indicates a rapidly changing world and the complex developments that lead to many potential futures. The fields of futures studies, anticipatory governance and scenario planning, and the applied fields of strategic foresight and speculative design use a variety of methods and tools to try to make sense of what might be emerging.[23–25] More importantly, these fields also connect what might happen in the future to decisions that are being made today.

These more formal methods and fields of study often incorporate aspects of art and literary forms with a specific focus on the future, such as science fiction, which at its heart is concerned with reflecting on the human condition and our relationship with new forms of technology. In recent years, there has been an increasing focus on applying futures methods and developing scenarios specifically in the context of sustainability.[26–29]

Many of these future scenarios, especially when applying narrative futuring methods or working at the intersection of art and science, can themselves be powerful science communication tools that can bring a number of different types of scientific findings and driving forces together and present them in compelling ways.[30] Science has many stories to tell, and scientists may need extra support to find the story in the data.

## How Can AI Tools Support Foresight and Futures Work Connected to Sustainability

There are several different applications of AI to support horizon scanning, foresight, futures, and science-based storytelling work in the context of improving science communication.

AI tools can help with signal detection and horizon scanning in the early phases of finding the anchor points for scenarios. For example, in Lübker et al.,[31] the authors used a specific type of ML algorithm to scan thousands of scientific papers on ocean governance in the high seas in order to find thematic clusters, which could then be the different scientific anchor points for the eventual scenarios. The scenarios were as a last step written up as short science-fiction stories. A number of quite surprising or unexpected connections were made, which could then be incorporated into the stories and help the authors to find new interesting angles[31] (see also Carvalho, 2024[32]).

As always with any application of AI, it is necessary to actively engage with the outputs of the work, test the output, and validate the results against the sources. Specifically in the context of science, recent work by Messeri and Crockett[33] makes the case that different AI solutions as applied to science communications can exploit cognitive shortcomings, thereby making us vulnerable to "illusions of understanding" whereby we believe we understand more about the world than we actually do. AI then has the potential to supercharge the Dunning-Kruger effect—when lack of knowledge and skill in a certain area causes people to overestimate their own competence.

## Limitations and Key Challenges

LLMs have a well-explored capacity to quickly summarize, tailor, and communicate climate and environmental information.[3,34] However, their potential to hallucinate—that is, create inaccurate or misleading information, for instance false facts or citations—remains a concern.[35] LLMs can refer to or cite nonexistent material, like news and scientific articles; produce factually incorrect information; introduce subtle inaccuracies; and oversimplify responses. As discussed earlier, hallucinations are not only the result of the technical features of LLMs, but also evolve as the result of how humans interact with the LLMs through prompting and feedback.[36] The absence of industry standards, along with a lack of agreed-upon and regulated best practices, presents severe challenges in mitigating hallucination risks.[37] These limitations should be taken seriously and serve as reminders that uses of AI require careful, and time-demanding, curation and oversight.

As for any other area discussed in this report, AI cannot be viewed as a panacea to deeper science communication issues—many of which are rooted in broader social issues such as unequal access to education, feelings of exclusion, polarization, identity politics, and more.[38–40]

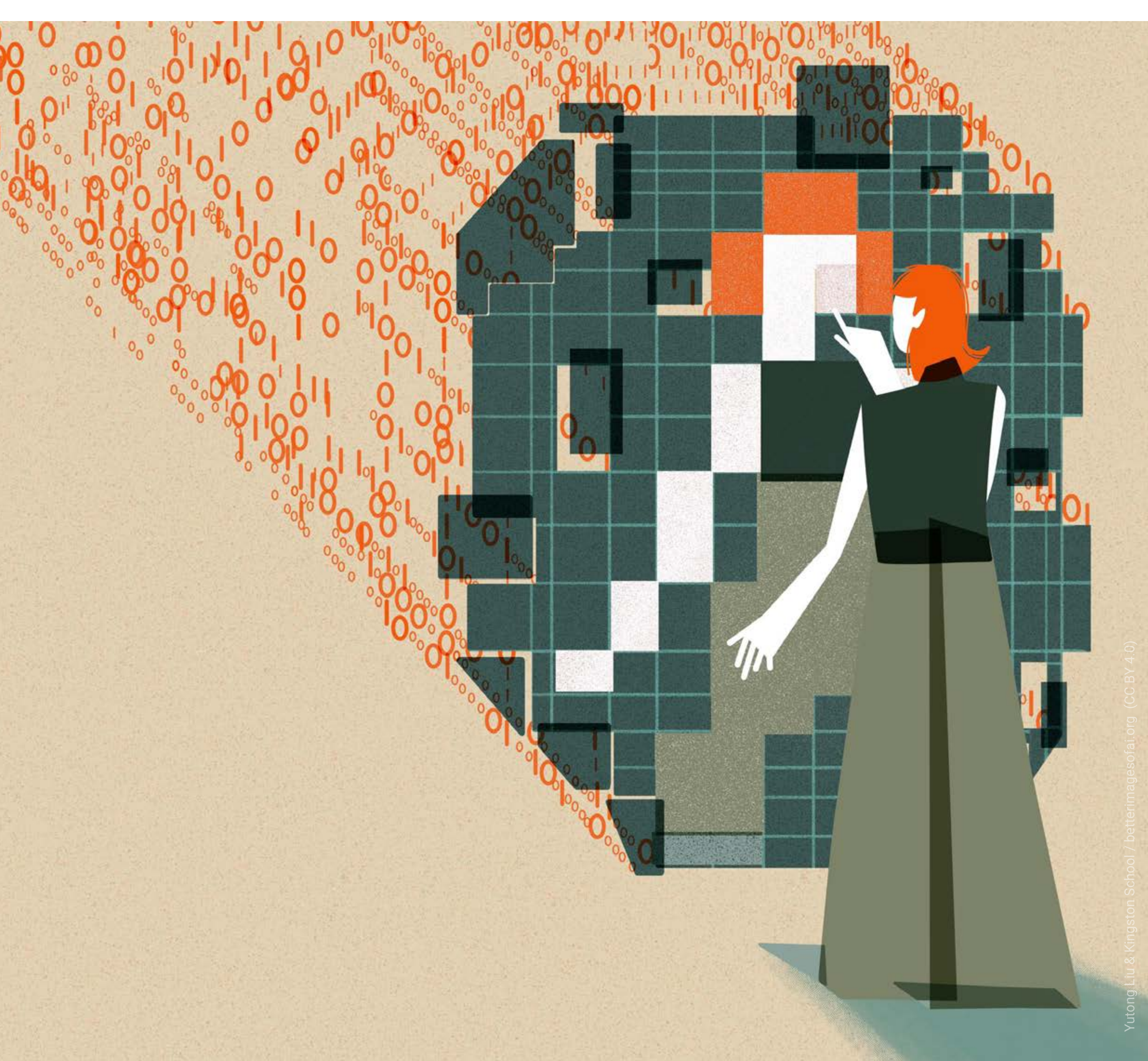

THEME BOX 4

# Climate Emotions and AI

Author: Victor Galaz

*Artificial intelligence (AI) is often viewed as a method to help support rational decision-making across a whole range of sustainability and climate issues. However, various uses of AI also shape our emotions—another key aspect of human decision-making.*

Emotions such as fear, joy, anger, and empathy play a crucial role in shaping human perceptions and responses to crises, including climate change. Fear of extreme weather events, anger at political inaction, and hope for a greener future all influence how individuals and societies respond to environmental challenges. AI and associated technologies may shape climate emotions in several ways, thus influencing human behavior at scale. Social media platforms, for example, powered by AI-driven recommender systems, curate content based on emotional engagement, often prioritizing emotionally charged posts—whether hopeful, outraged, or fearful. Generative AI can be used to craft highly persuasive digital content, further influencing public opinion and activism in complex ways.

The individual and social impact of the spread of climate emotions is far from linear and predictable. Instead, it's the result of interactions between technology, social relations, and psychological mechanisms. On the one hand, AI-powered social networks can help mobilize climate action. Online-supported movements like *Fridays for Future* gained traction through emotionally resonant social media messaging, helping millions of people rally around climate advocacy. On the other hand, digital networks can contribute to climate anxiety, spread misinformation, further disconnect people from Nature, and lead to affective polarization. AI-generated content can reinforce such processes and thus deepen societal divisions on climate issues.

In an era in which AI contributes to shaping how we feel about climate change, the question is not whether technology influences our emotions—but how we understand its role in shaping our emotional connection to the planet, to future generations, and to one another.

This text builds on the article by Galaz, V., *et al.* (2025).[1]

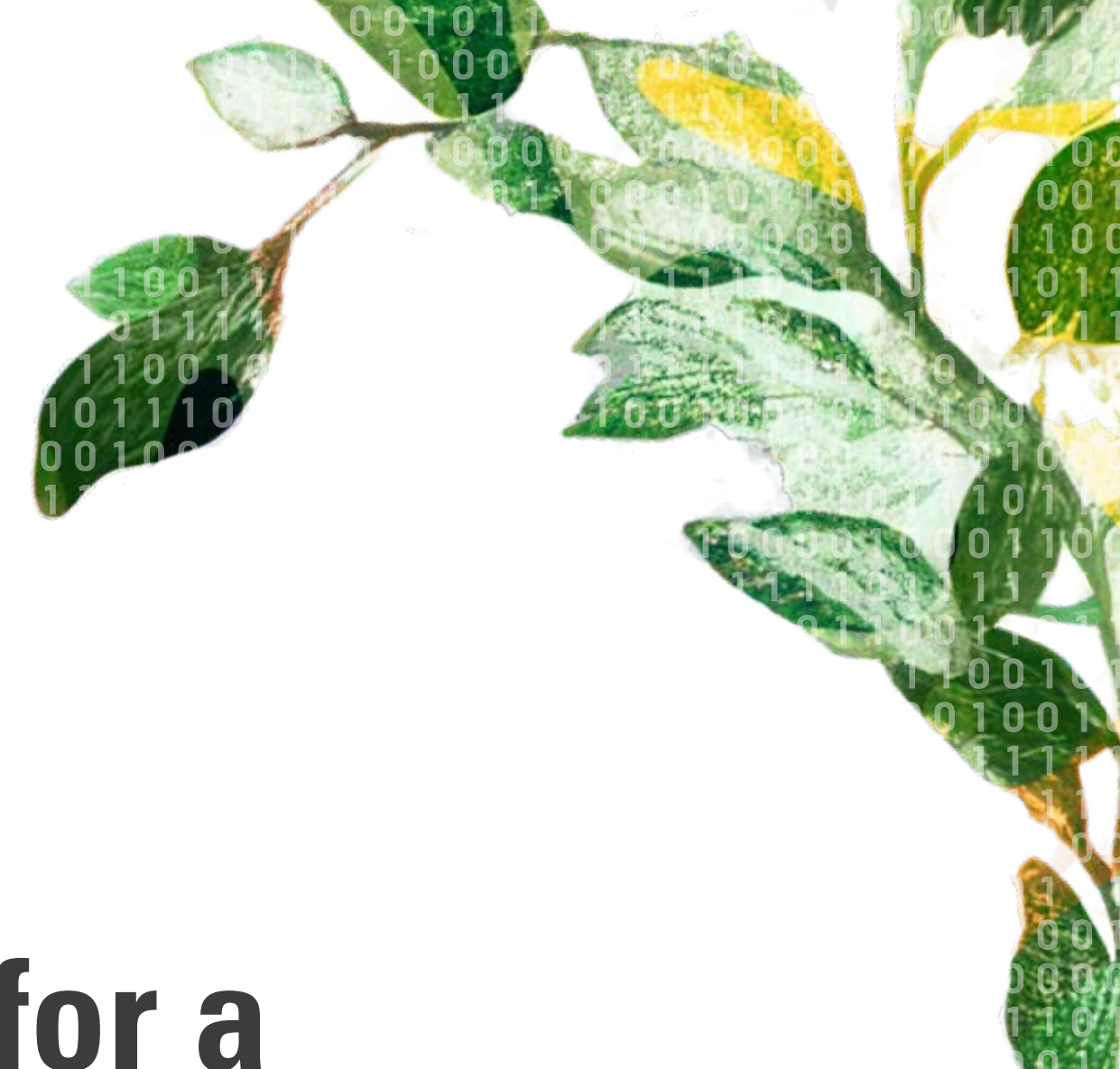

# 8 Collective Decisions for a Planet Under Pressure

Authors: Wolfram Barfuss, Victor Galaz, Anna B. Stephenson,
Erik Zhivkoplias, Jobst Heitzig

*Securing a just and safe future for all requires cooperation and collective decisions at scale. Can AI facilitate concerted and informed decisions for a planet under pressure?*

## Introduction

Cooperation and collective decisions are critical for achieving a sustainable future for all. Such cooperation is vital for managing both environmental and social commons, such as fisheries and public infrastructure. Cooperation involves a group's ability to agree on and work together toward common goals, even when individual selfishness is tempting. The importance of collective action is often seen in social dilemmas where collective benefit is at stake. The study of cooperation spans various disciplines, including biology and social sciences, and identifies mechanisms like external authorities (e.g., governments taxing uncooperative behavior) and social reciprocity to promote cooperative behavior. However, several challenges hinder the emergence of cooperation[1]:

**Large and heterogeneous collectives.** Cooperation becomes difficult in large groups due to the lack of effective enforcement mechanisms and the anonymity of participants, making it hard to stabilize reciprocity. Also, heterogeneous groups with differences in preferences and resources are often a challenge for collective action.

**Complex human behavior.** Understanding the diverse motivations and behaviors of individuals is crucial for fostering cooperation, especially in the context of sustainability and policy changes.

**Environmental complexity.** The dynamic nature of environments, including feedback loops, thresholds, and uncertainties, complicates the stabilization of cooperative efforts.

**Transient dynamics.** There is a lack of focus on the temporary phases that lead to cooperation, including the timing and stability of cooperative arrangements, which are essential for sustainability transitions.

In what ways could uses of AI help us explore and better understand the dynamics of human collaboration? Below we discuss a couple of interesting, related areas, including the potential and limitations of AI.

## AI for Data Generation and Collection

One key aspect of cooperation and collective decisions is related to the ability of those making decisions—such as state and non-state actors—to act on the best available information. While this might sound straightforward, numerous studies show the difficulties decision-makers face in trying to match collective decisions and institutions to the rate of change in the climate system and ecosystems.[2,3]

A number of intriguing applications of large language models (LLMs) illustrate the potential to support the synthesis capacities of international institutions,[4,5] including the Intergovernmental Panel on Climate Change (IPCC)[6]. LLMs can also help structure and summarize increasingly information-dense policy areas, such as national climate pledges.[5]

Scientific syntheses help inform political agendas and public debate, and the use of domain-specific LLMs could thus potentially help speed up the collection and analysis of a growing body of scientific studies and national reports in the climate and sustainability domain.

## AI-Assisted Predictive Modeling

There are several ways in which predictive modeling can contribute to collective decision-making, which are presented here:

**Mathematical models can be of help to understand and support cooperation.** Process-based, mechanistic models facilitate theory development and simulations when traditional experiments are impractical. Approaches that build on complex systems science have enhanced our understanding of how simple agents interact to form larger structures, evident in fields like swarm intelligence and evolutionary games.[7,8] However, the individual behaviors in biological, social, and artificial systems are often complex and not easily captured by simplistic models. Here, AI models can help in multiple ways.

Agent-based modeling and artificial life allow for more **nuanced insights about individual decision-making and diversity among agents.** Multiagent reinforcement learning (MARL) offers a way for agents to learn their behaviors for themselves in dynamic environments. Recent examples of deep MARL applied to collective decision-making and cooperation include the project *AI Economist*[9] and the *AI for Global Climate Cooperation* initiative.[10] These projects implement an inner-loop outer-loop reinforcement learning setting. The outer-loop deep reinforcement learning agent aims to obtain a taxation and subsidies policy and negotiation protocol, respectively, while the inner-loop agents continuously adapt, aiming to maximize their individual welfare. The learned *AI Economist* policy outperforms several established taxation and subsidies policies in terms of the total welfare achieved and inequality avoided.

**Bridging complex systems science and MARL together also offers new possibilities.** MARL operationalizes complex cognition processes, while complex systems approaches provide a structured understanding of how cooperation emerges. For example, Barfuss et al. (2020)[11] use dynamical systems theory in combination with reinforcement learning agents to analyze how agents learn sustainable collective action under a variety of conditions. Such integration can enhance our understanding of collective decision-making and cooperation, and inform the design of cooperative algorithms and strategies for achieving sustainability. For example, which features of the agents' perception, internal representation, and decision-making systems are particularly apt to promote cooperative behavior in the face of environmental disaster.

Another way forward is to **combine the complex systems frameworks of agent-based models (ABMs) with LLMs**. Recent advances in LLMs have enabled a new class of ABMs that combine complex, adaptive decision-making with large-scale simulation capacity. These new, generative agent-based models (GABMs) allow agents to reason, communicate, and act in contextually appropriate ways using natural language.[12] By capturing both rich individual behavior and social interaction, this approach offers a framework for linking local actions to emergent collective phenomena.

Tools like *Concordia,* a generative agent-based modeling library, provide a modular framework where agents interact in physical, social, or digital spaces. The agents' actions are mediated by a Game Master that interprets their language-based intentions and enforces environmental rules.[13] Foundation models have been adjusted to capture and predict human cognition across a variety of behavioral experiments.[14] Scalable approaches such as LLM archetypes allow for simulating millions of agents with distinct behavioral patterns by grouping similar individuals and sampling representative decisions, as demonstrated in simulations of pandemic response in New York City.[15]

## AI-Assisted Decision-Making

There is a growing discussion about the **need to increase the availability of AI tools to non-experts.**[16] A number of initiatives and projects have been developed in recent years to explore how human-centric, explainable AI approaches can support collective decisions in various ways. LLMs have, as one example, been successfully studied as mediators in deliberative dialogues.[17] In urban planning, the *Potential allocation of urban development areas* framework integrates geographic information systems (GIS) with participatory scenario planning, enabling non-experts to visualize urban development–ecological preservation trade-offs across case studies.[18] Participants reported greater confidence in AI-assisted decisions due to clear scenario visualization and collaborative refinement of adaptation pathways, highlighting how these tools helped reconcile competing priorities through accessible interfaces.

Similarly, Utrecht University's project and framework *Human-Centered AI for Climate Adaptation* combines interactive visualization and explainable AI to co-design Dutch flood resilience strategies, in which stakeholders dynamically evaluate AI-generated approaches balancing agricultural, urban, and shipping needs.[19] User-friendly AI applications that use Bayesian networks,[20] as another example, can empower non-experts and multidisciplinary teams to analyze complex sustainability datasets and explore *in-silico* policy scenarios.

Some of these uses can be **informed by predictive analysis** that combines AI-augmented modeling with scenario processes. For example, the digital platform *ClimateIQ* combines physical models and CNNs to help project urban climate risks in ways that allow for local engagements and collaborative climate adaptation planning (see Theme box 3, *ClimateIQ: Empowering Communities to Prepare for Climate Threats with Hyperlocal Data*). The project *ARtificial Intelligence for Environment & Sustainability* (ARIES) uses a semantically connected network of AI models that allows users to ask ecosystem service-related questions or build scenarios to support engagements and planning.[21,22]

## Limitations and Key Challenges

While their potential is intriguing, each use of AI listed earlier is associated with known limitations. Uses of deep MARL are associated with **high computational costs** and **difficulties in interpreting learned behaviors,** due to the numerous parameters of these simulations. These models can also be difficult to analyze and may suffer from issues like overparameterization, leading to unreliable outputs. As we discussed in the previous chapter *Improving Sustainability Science Communication*, hallucinations from LLMs also remain a concern.

Both LLM archetypes and GABMs offer new ways to simulate collective decision-making, but each has important limitations. Archetypes **simplify individuals into a few representative types**, which can obscure meaningful variation within groups. Generative models like *Concordia*[13] produce more detailed behavior by prompting LLMs directly, but these behaviors **reflect bias or patterns in training data.** This can lead to biased or flattened portrayals, especially of marginalized groups.[23] It becomes a problem when the model is wrongly applied to represent a population where the patterns are different, which is most likely in marginalized groups. The meta problem is that we don't know or cannot say, even, when the patterns are sufficiently different for the application to become a significant problem.

These risks are serious if the models are used to inform real-world decisions, and illustrate the need for AI model transparency and interpretability. However, even though increasing transparency and interpretability in AI model design is important, it does **not automatically lead to greater user trust or better decision-making.**[24,25]

It is worth noting, however, that traditional agent-based models also simplify human behavior. The use of novel AI methods may introduce different risks of abstraction, but can, if conducted with nuance and skepticism, help simulate unprecedented complexity and scales of collective behavior.

Lastly, while uses of AI have the potential to help model and inform collective decisions, their **proper use is not a guarantee for a successful outcome**. Climate and sustainability questions entail a suite of challenging value judgements, distributional issues, and inequalities, which lead to decision-making and political gridlock.

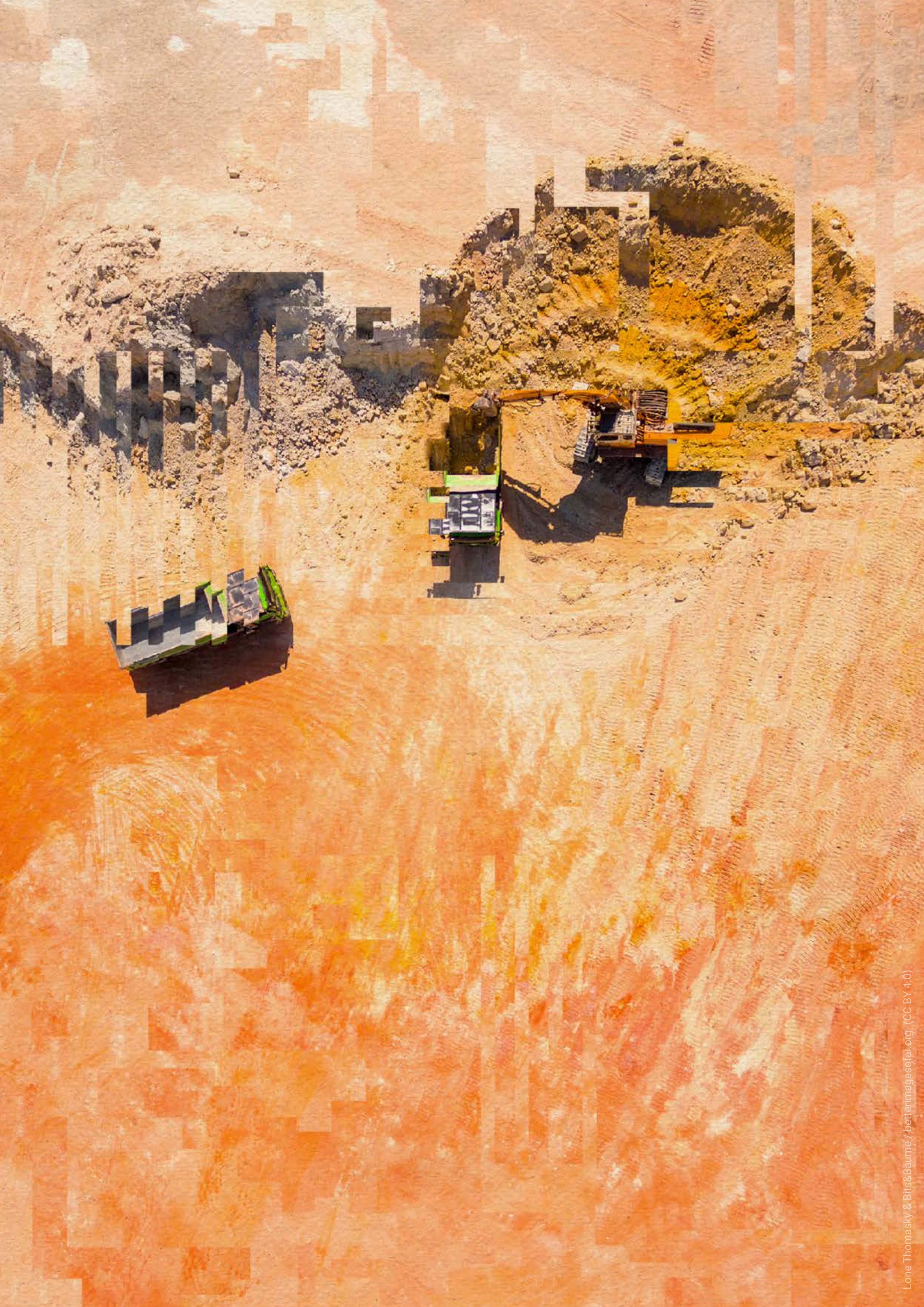

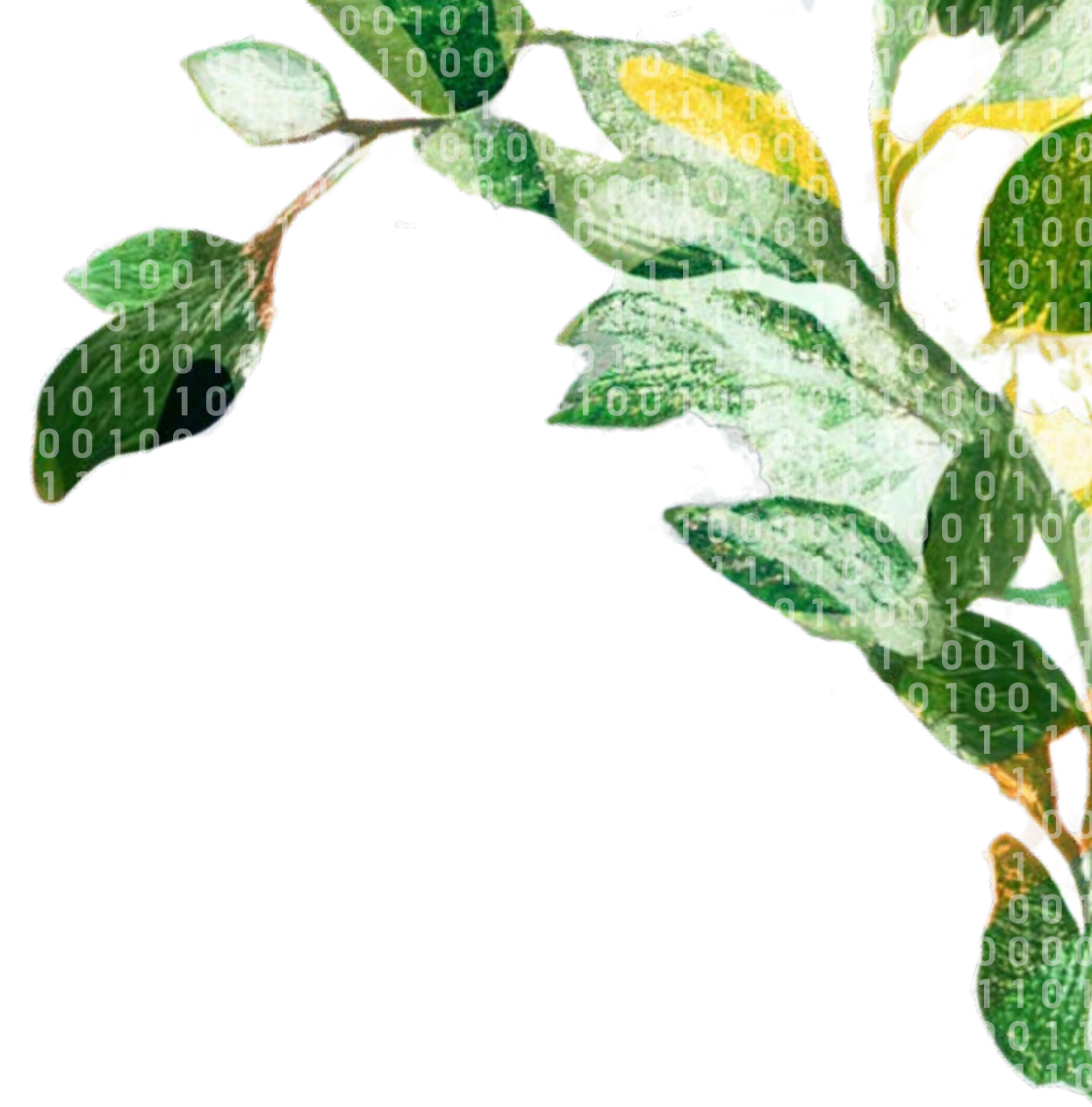

# Conclusions and Recommendations

Authors: Victor Galaz, Timon McPhearson, Samuel T. Segun, Jonathan F. Donges, Ingo Fetzer, Elizabeth Tellman, Magnus Nyström, Lan Wang-Erlandsson

The planet on which we live is undergoing rapid changes. The conversion of land to agricultural and urban areas, the extraction of natural resources, the loss of biodiversity, and rising carbon emissions are altering planetary processes on land and in the oceans in ways that increase risks to people all over the world. Novel digital technologies and advancements in artificial intelligence (AI) have profoundly reshaped social interactions, resulting in a more connected global population, but simultaneously creating a diversity of novel risks.

Science plays a key role in this rapidly changing context. It provides evidence of how planetary changes could affect people's lives, offers analyses for making informed decisions to mitigate risks and reduce social vulnerabilities, and spurs the innovation needed to drive social transformations. At its best, science can guide decisions that help societies move toward a safe future for all. Exploring such pathways is far from straightforward, however. Analyzing the multiple and interacting complexities of human behavior, ecosystems, a changing climate, and technological change remains a considerable research challenge.

While AI is already driving breakthroughs in fields from the life sciences to mathematics, its role in sustainability research is currently less established. This raises a critical question: could AI be effectively applied to the complex, interconnected challenges of sustainability?

This concluding chapter is structured in three parts. **The first part** summarizes an internal survey where we assess and rank the different issue areas according to their present level of AI maturity (i.e., access to data, and application of AI methods), as well as their wider accessibility. **The second part** summarizes high-level insights from our analysis. **The third and last part** ends with a list of recommendations to colleagues in academia, to governments, the public sector, entrepreneurs, and to philanthropic funders.

## Comparing the Issue Areas

The use of AI methods differs considerably between the eight chosen issue areas, as shown by our literature overview and deeper analyses of each area (fig. 4a-h). In general, classical machine learning methods are more commonly used in areas exploring complex interactions in climate and nature (e.g., *Understanding a Complex Earth System, Securing Freshwater for All, and Stewarding Our Blue Planet*), while newer AI methods are more frequent in areas where stakeholder participation and interactions are more prevalent (e.g., *Prospering on an Urban Planet, Improving Sustainability Science Communication*, and *Collective Decisions for a Planet Under Pressure*).

How do these issue areas compare in terms of their AI capabilities, and how accessible are insights derived from such analyses to actors outside of academia? Which areas seem to have made the fastest progress? And which ones are in need of a stronger push to be able to fulfill their potential? These questions all relate to comparisons between the issue areas, thus offering insights into different levels of maturity and availability of interest to both researchers and funders.

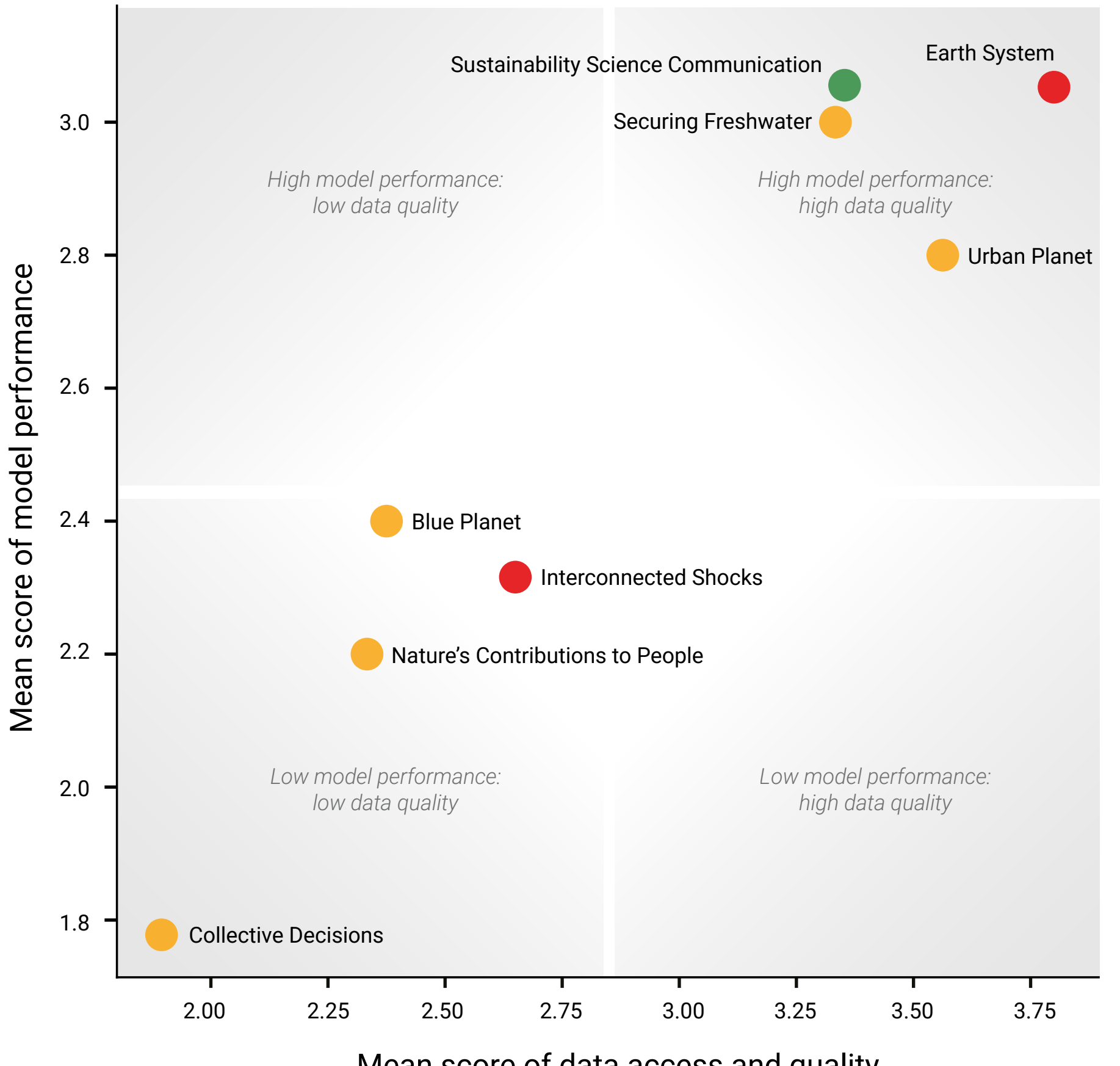

Fig. 8. Overall evaluation across issue areas. Each participating author evaluated the areas along three key dimensions: (1) data access and quality, (2) model performance, and (3) accessibility for non-experts, i.e., the extent to which different stakeholders (such as NGOs, civil society groups, and local communities) can use or develop AI tools to address the issue. Authors rated each dimension on a scale from 1-5, or opted out of rating when appropriate. The complete distribution of scores is available in Appendix 6. Produced by Erik Zhivkoplias.

We bring together data from internal expert assessments of these three dimensions—data availability and quality; AI model performance; and the accessibility of AI models and insights to non-expert communities (e.g., NGOs, civil society, local communities). The assessment builds on an online survey sent out to each contributing author of this report in the final part of the writing. The literature overview and review of each issue area thus informed the individual assessments. Between 15 and 23 experts contributed depending on the issue area, forming the joint assessment (from here on referred to as "the assessment") seen in Figure 8.[*]

The assessment shows several notable findings. Some issue areas stand out as more mature, characterized by relatively high scores along the dimensions of AI-model performance, and data availability, and quality. This grouping includes the areas *Understanding a Complex Earth System, Sustainability Science Communication, Securing Freshwater*, and *Urban Planet*. In contrast, there are issue areas where data availability is comparatively sparse, and where AI-methods development is viewed to be less advanced. This includes the area *Collective Decisions for a Planet under Pressure* in particular, which scored lowest on both dimensions (see Figure 8). These areas are likely to require a step-change increase in

* The interpretation of the survey results should be approached with caution. The assessment is based solely on responses from members of the author team. Moreover, the grading reveals a tangible spread across the assessed dimensions (see Appendix 6 for details on the survey methodology and results). The spread, as seen in the distribution of values, indicates several potential systemic issues. These include: limited cross-disciplinary insight among experts; the rapid pace of technological advancements outstripping the collective ability to integrate these advances into ongoing research; and insufficient coherence and integration across research communities in the different issue areas. The diversity of responses may also imply a lack of awareness of developments in the other issue areas. To strengthen systemic analyses of this kind and complement the area deep dives, future assessments should involve a larger and more geographically diverse group of experts. Such assessments should be followed up on a regular basis to monitor the development and maturity of research areas over time, and ensure representation from researchers from low- and middle-income countries in Asia, Africa, Latin America, and Oceania.

investment for research and applications (including data infrastructure, compute and expertise) to enable innovation with AI-methods at scale.

Advancements in Large Language Models (LLMs), along with their increasing accessibility to the public may, we believe, have increased the availability of AI for science communication. This trend aligns with the relatively high accessibility grade given to the *Sustainability Science Communication* issue area. It is noteworthy, however, that two crucial issue areas *Preparing for a Future of Interconnected Shocks* and *Understanding a Complex Earth System* were graded to provide low accessibility to users outside of academia, such as NGOs, civil society, and local communities. These latter areas thus offer an opportunity for funders to invest in ways that help bridge the gap between ongoing AI research for sustainability, and its uses by society. Increased collaborations across domains and targeted research initiatives may drive scientific advances and innovative uses of AI across, and between, issue areas. In summary, there is a need to bring more collaboration into the community, including knowledge sharing, to ensure that investment is flowing to the areas of biggest impact, especially where critical issues often lie at the intersection of different domains.

---

## Key Insights

1. **AI offers vast potential across the sustainability sciences**

   Recent advancements in AI, combined with increasingly powerful computing capabilities and methodologies, are enabling the discovery of patterns and relationships within vast datasets from various fields of science, leading to intriguing scientific discoveries. This capacity for data integration, generation, and analysis across global and local sustainability domains allows for insights that were previously too complex, or too interdisciplinary, to tackle. This capability extends to diverse data sources enabling automated feature extraction in realms ranging from marine acoustics and species identification, to water resources monitoring and urban climate risk mapping (see Issue Areas). Hybrid simulation-AI models, which combine AI algorithms with process-based models, are proving particularly powerful, with reduced computational cost compared to traditional methods (p. 16, 19, 34, Theme box 3). There are, however, large differences in capabilities between the different analyzed issue areas (p. 65-67).

2. **AI can inform decisions and help communicate complex sustainability challenges**

   AI can bolster decision-making and science communication for sustainability. Examples include AI-augmented analysis supporting marine protected area (MPA) management (p. 39), water resource planning and management (p. 41-43), and the development of early warning systems for hazards and conflicts (p. 27-29). Large language models (LLMs) and multimodal language models (MLLMs) can help personalize, localize, and make complex scientific information more accessible, engaging, interactive, and actionable for diverse audiences (p. 57).

3. **Environmental footprint, biases, hallucinations, and unequal access to AI pose sustainability risks**

   While AI as a research method for sustainability holds important potential, the increasing use of AI in society in general, particularly for generative AI, demands energy and freshwater across the whole supply chain, leading to increased $CO_2$ emissions and e-waste, and often impacting already vulnerable communities. This growing footprint, including toxic substances from e-waste, and water withdrawals for energy production and/or cooling in water-scarce regions, represents a sustainability risk and climate justice issue on its own, thus requiring serious consideration and mitigation (p. 10-11). Realizing AI's potential for sustainability will require overcoming issues of data scarcity, quality, and geographical imbalance, which are particularly acute in low-income countries (p. 12). Underrepresentation of low- and middle-income countries in AI research, design, and use may perpetuate existing inequalities. Algorithmic biases and "hallucinations" introduce substantial ethical and accuracy challenges, and can lead to allocative harms that disproportionately affect vulnerable groups (p. 11-12). Interpretability and vast computational demands are major hurdles for AI adoption and trust. Complex deep learning architectures until now often function as "black boxes," making it difficult to understand the causal mechanisms behind AI predictions and acting as a barrier to trust for decision-makers and users (p. 11). Training and optimizing these models, especially foundation models, require vast computational resources, creating access and usability challenges, but also opportunities once these models are made accessible (see Theme Box 1).

4. **Responsible uses of AI for sustainability research are possible**

   As all individual chapters illustrate, addressing risks associated with uses of AI for sustainability research is possible. Responsible AI development for sustainability requires fostering interdisciplinary collaboration; augmenting rather than replacing human expertise; supporting academic freedom, peer review, and open science; and establishing robust governance mechanisms addressing AI's limitations proactively. Several pioneering private, science, and civil society-led initiatives have emerged in this domain. This includes, for example, tools and methods for the development of responsible foundation models; consensus checklists to help advance valid, reproducible, and generalizable AI-based science; and Indigenous Data Sovereignty principles. The co-production of knowledge and benefit-sharing with local communities, decision-makers, and others offer ways to empower and augment human expertise, requiring careful selection of problems and rigorous evaluation.

5. **Breakthroughs in sustainability research driven by AI are within our reach and essential for our collective future**

   Current applications of AI for sustainability-related research are extensive and diverse, and show that an "AI for Sustainability Science" community has begun to emerge. Its diversity in focus and expertise (see Issue Area chapters), as well as in different levels of AI maturity (p. 65-66), can be viewed as a source of innovation. The potential is substantial, and could transform existing areas of sustainability science, and open up new research areas and practical applications. The growth in public and private investments in AI hardware, AI research, and "AI for Science" in general, however, will not drive scientific breakthroughs of relevance for sustainability at the pace and scale needed. Lack of technical expertise, bottlenecks in access to AI compute, and limited collaborations between AI scientists, developers, and sustainability researchers create obstacles that limit the current potential of AI. Supporting cross-domains collaborations; scaling up AI funding for integrated biodiversity, climate, and social-ecological research; developing clear guidelines for responsible AI development and use for sustainability; and building ambitious international research programs across domains could lead to important advances that benefit people and the planet.

## Key Recommendations for Different Stakeholders

**Researchers**

- **Explore a diversity of AI methods across various sustainability issue areas.** Build on existing work in related areas to lay a foundation for an agenda for AI for the Sustainability Sciences. Foster inter- and transdisciplinary collaboration among ecologists, social scientists, Earth System scientists, and computer scientists.
- **Address AI limitations and risks explicitly.** Pioneer work that helps reduce the environmental footprint of AI (energy, water, e-waste) and develop methods to mitigate these impacts, including rebound effects. Investigate algorithmic biases and develop methods for mitigating "hallucinations" in LLMs, especially their impact on vulnerable groups and accuracy in sustainability contexts. Develop more interpretable and explainable AI (XAI) models to address the "black box" problem. If outsourced, secure data transparency and ensure that data labelling is done by service providers that offer fair and just working conditions.
- **Counter data gaps, and geographical and social imbalances.** Develop methods for few-shot learning (enabling accurate model predictions with limited data), self-supervised approaches, and data integration techniques. Prioritize collaborations in low- and middle-income countries (LMICs) and under-represented regions, and with under-represented communities, to counteract imbalances. Ensure equitable representation in AI model development and use.
- **Champion responsible AI principles.** Emphasize that AI should augment, not replace, domain expertise, including local knowledge. Focus research on AI tools that empower scientists, decision-makers, and stakeholders. Advocate for open access to training data and source code, and rigorous evaluation of AI models, to foster transparency, replicability, and trust-building.

**Private sector**

- **Invest in research and development for AI solutions that bridge science and real-world action.** Expand the scope of work by private actors to include aspects of planetary change beyond energy and carbon emissions, for example, biodiversity loss, deforestation, and ocean acidification. Include integrated approaches that connect biosphere changes and human responses. Prioritize the development of interpretable AI systems, particularly for decision-support tools, to build trust among users by making AI models understandable.
- **Mitigate the environmental footprint of AI and other risks.** Prioritize the development of energy-efficient AI algorithms and hardware, particularly for large models and data centers. Implement sustainable practices for managing e-waste and limiting water use.

- **Identify and mitigate algorithmic biases and "hallucinations".** Implement rigorous testing and validation protocols in AI systems, particularly those used for sustainability applications. Invest in developing universal industry standards and best sustainability practices for AI safety and ethics.
- **Invest in AI that matter to people and the planet.** Develop investment strategies that consider the material and carbon footprint of AI. Support the next generation of AI companies that explicitly focus on addressing climate and sustainability challenges.

**Public agencies**

- **Invest in opportunities for AI for the Sustainability Sciences.** Establish interdisciplinary centers of excellence that bring together AI scientists, AI entrepreneurs, and sustainability researchers. Include programs that provide AI expertise and compute to researchers in the sustainability science domain. Support innovation initiatives that allow local communities and civil society to access independent AI expertise..
- **Secure robust AI governance for sustainability.** Establish and enforce governance frameworks that include ethical guidelines for AI in the sustainability domain, grounded in the principles of fairness, accountability, transparency, ethics, and sustainability (FATES).
- **Implement policies for environmental sustainability.** Governments and public regulatory agencies must ensure that the growth of AI and data centers does not compromise environmental sustainability. Mandate and enforce transparency requirements, and extend the producer responsibility for hardware to manage e-waste. Set water efficiency standards for cooling systems to conserve local water resources. Require companies to source energy exclusively from renewable sources and to conduct environmental impact assessments (EIA). Protect vulnerable communities from allocative harm.
- **Enhance data sharing, collaboration, and capacity to use AI.** Engage in open collaboration with academia and public agencies, sharing data and code where appropriate, to advance responsible AI development. Utilize AI tools to enhance public engagement and communication on climate and sustainability, adapting messages to diverse audiences and local contexts. Invest in collecting and curating high-quality, representative datasets, especially from data-scarce regions. Promote open access to training data and source code to ensure replicability and reduce bias. Invest in shared infrastructure and promote access to high-performance computing for sustainability applications.

**Philanthropies**

- **Support inter- and transdisciplinary collaborations for AI for sustainability.** Invest in research and initiatives that explore the potential of AI across diverse sustainability issue domains, particularly those that bridge scientific knowledge with real-world action. Direct investments toward underrepresented domains, for example, AI in the intersection of climate, health, and biodiversity; and AI that advances areas where the immediate return on investment is not obvious, but addresses fundamental planetary challenges.
- **Support data infrastructure and insights for action.** Support projects focused on development and implementation of AI for enhanced monitoring. Allow all sectors of society to tackle the repercussions of rapid climate and environmental change.
- **Invest in research into the sustainability implications of AI.** Focus on algorithmic bias, hallucinations, fairness, equitability, and the prevention of allocative harms. Support independent auditing and oversight mechanisms for AI systems in the sustainability domain.
- **Mitigate social and geographical bias.** Support efforts to make computational resources and robust AI tools broadly accessible to researchers, organizations, and agencies in resource-limited settings. Ensure that underserved communities benefit from advancements in AI for sustainability.

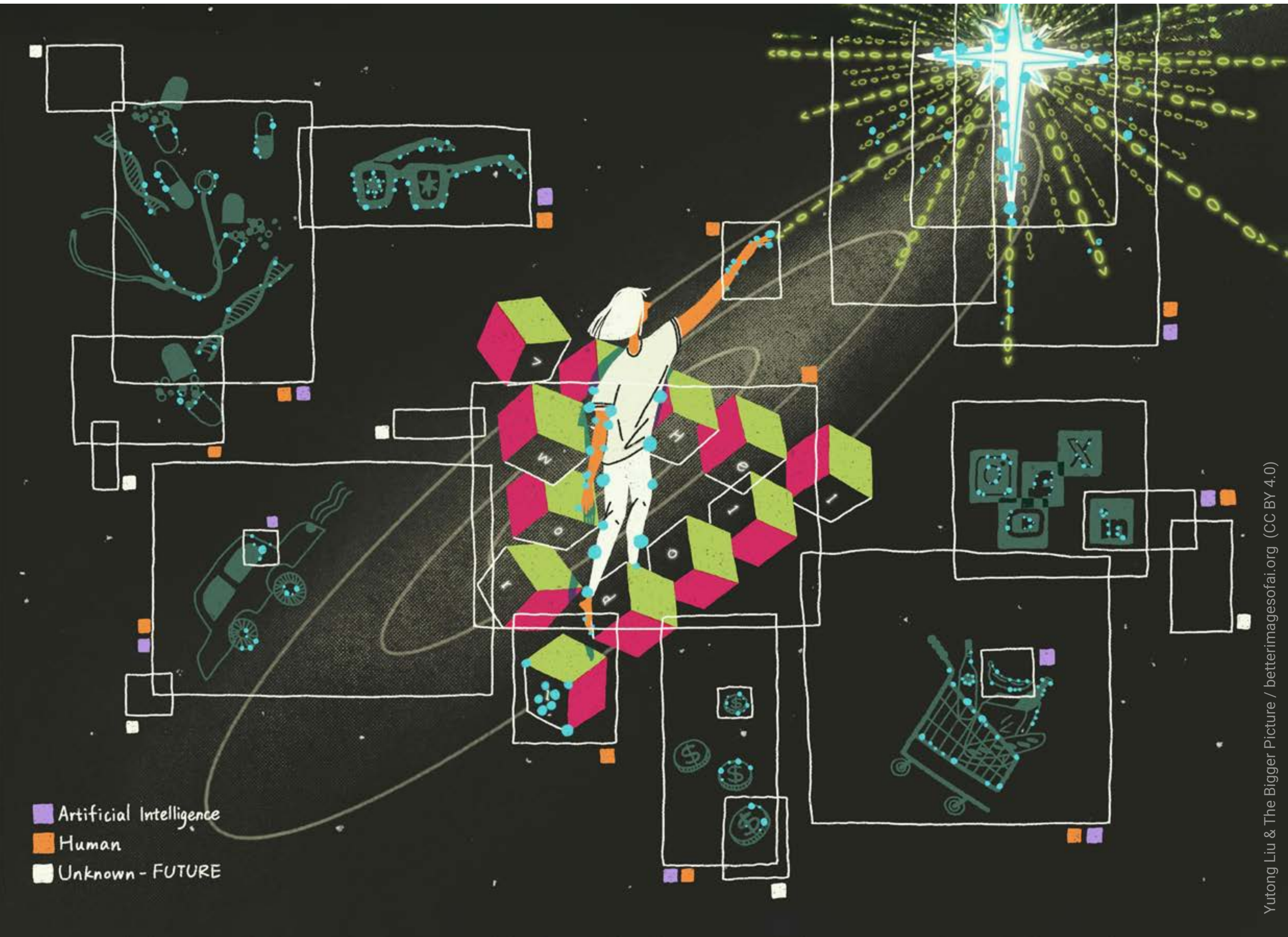

# Bibliographies

## Introduction

---

## Sustainability Risks and Allocative Harms

## A Taxonomy of AI for Sustainability Sciences

## AI for Sustainability Sciences—A Literature Review

## Preparing for a Future of Interconnected Shocks

## Foundation Modeling in Climate and Sustainability Science

## Understanding a Complex Earth System

# Using AI to Detect Earth System Tipping Points

## Stewarding Our Blue Planet

# Securing Freshwater for All

## Enhancing Nature's Contributions to People

## Prospering on an Urban Planet

## Improving Sustainability Science Communication

## Climate Emotions and AI

## Collective Decisions for a Planet Under Pressure

# Appendix

Scan the qr code for the full appendix:

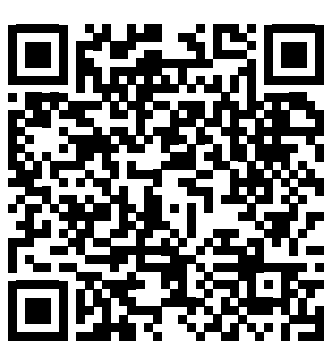

Or go to: https://stockholmuniversity.box.com/s/j7zkkh9c0nprou33tgsvq50g2tob5681

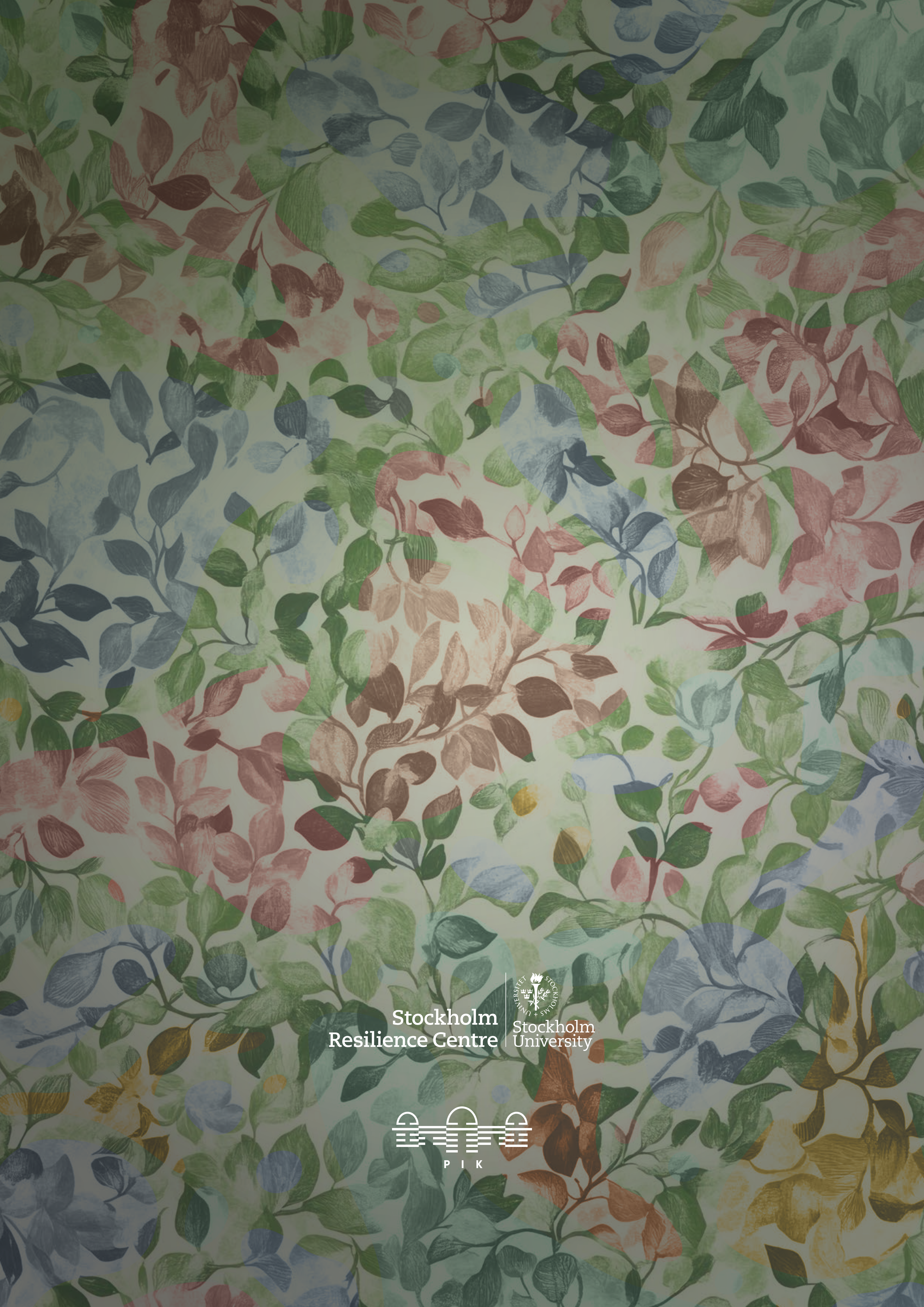